\documentclass[nohyper,11pt,letterpaper]{JHEP3}
\usepackage[dvips]{epsfig}
\usepackage{amsmath}
\usepackage{yfonts}
\usepackage{epsfig,latexsym}

\newcommand{\mo}{m_0}
\newcommand{\zb}{\bar{z}}
\newcommand{\bw}{\bar{\omega}}

\newcommand{\xb}{\bar{x}}
\newcommand{\cE}{{\cal E}}
\newcommand{\rn}{\textswab{R}}
\newcommand{\qn}{\textswab{q}}
\newcommand{\wn}{\textswab{w}}
\newcommand{\be}{\begin{equation}}
\newcommand{\ee}{\end{equation}}
\newcommand{\bea}{\begin{eqnarray}}
\newcommand{\eea}{\end{eqnarray}}
\newcommand{\ba}{\begin{eqnarray}}
\newcommand{\ea}{\end{eqnarray}}
\newcommand{\nn}{\nonumber \\}

\newcommand{\beq}{\begin{equation}}
\newcommand{\eeq}{\end{equation}}
\newcommand{\beqa}{\begin{eqnarray}}
\newcommand{\eeqa}{\end{eqnarray}}
\newcommand{\beqar}{\begin{eqnarray*}}
\newcommand{\eeqar}{\end{eqnarray*}}

\newcommand{\reef}[1]{(\ref{#1})}

\newcommand{\eg}{{\it e.g.,}\ }
\newcommand{\ie}{{\it i.e.,}\ }

\newcommand{\rinf}{{\rho_{\infty}}} 

\newcommand{\nq}{n_\mt{q}}
\newcommand{\dd}{\tilde{d}}

\newcommand{\tg}{\tilde{g}} 
\newcommand{\da}{\dot{A}} 
\newcommand{\de}{\dot{E}_x} 
\newcommand{\dde}{\ddot{E}_x} 

\newcommand{\tk}{\qn}
\newcommand{\tom}{\wn}

\def\nc {N_\mt{c}}
\def\nf {N_\mt{f}}

\def\t6 {T_\mt{D6}}

\def\gym {g_\mt{YM}}

\newcommand{\tf}{T_\mt{fund}}

\newcommand{\mq}{M_\mt{q}}      
\newcommand{\N}{{\cal N}} 
\newcommand{\mbar}{\bar{M}}

\newcommand{\gs}{g_\mt{s}}
\newcommand{\ls}{\ell_\mt{s}}

\newcommand{\mt}[1]{\textrm{\tiny #1}}

\newcommand{\om}{{r_0}}  

\def\ofo{ { {}_2 \! F_1 }}

\newcommand{\lam}{\lambda}

\newcommand{\labell}[1]{\label{#1}}

\newcommand{\bi}{\begin{itemize}}
\newcommand{\ei}{\end{itemize}}
\newcommand{\ben}{\begin{enumerate}}
\newcommand{\een}{\end{enumerate}}

\newcommand{\bmtx}{\left[ \begin{array}{cc}}
\newcommand{\emtx}{\end{array} \right]}
\newcommand{\bvec}{\left[ \begin{array}{c}}
\newcommand{\evec}{\end{array} \right]}

\newcommand{\bfig}{\begin{figure}}
\newcommand{\efig}{\end{figure}}

\newcommand{\ide}{I_\mt{E}}

\newcommand{\prt}{\partial}

\title{
Holographic spectral functions and diffusion constants for fundamental matter}
\author{
Robert C. Myers,$^{a,b,c}$ Andrei O. Starinets$^{a}$ and
Rowan M. Thomson$^{a,b}$\\
$^a$ Perimeter Institute for Theoretical Physics,
Waterloo, Ontario N2L 2Y5, Canada \\
$^b$ Department of Physics and Astronomy, University of Waterloo,
Waterloo, Ontario\\
\ \  N2L 3G1, Canada \\
$^c$ Kavli Institute for Theoretical Physics, University of
California, Santa Barbara, CA\\
\ \ 93106-4030, USA\\
\\E-mail: \email{
rmyers@perimeterinstitute.ca, starina@perimeterinstitute.ca,
 rthomson@perimeterinstitute.ca }}

\preprint{arXiv:0706.0162 [hep-th] }

\date{\today}

\abstract{The holographic dual of large-$\nc$ super-Yang-Mills
coupled to a small number of flavours of fundamental matter, $\nf
\ll \nc$, is described by $\nf$ probe D7-branes in the gravitational
background of $\nc$ black D3-branes. This system undergoes a first
order phase transition characterised by the `melting' of the mesons.
We study the high temperature phase in which the D7-branes extend
through the black hole horizon.  In this phase, we compute the
spectral function for vector, scalar and pseudoscalar modes on the
D7-brane probe. We also compute the diffusion constant for the
flavour currents.}


\begin{document}{\vskip 1cm}


\section{Introduction}

The gauge/gravity duality has created a powerful framework for the study
of strongly coupled gauge theories \cite{juan,adscft,itz,bigRev}.
Thermal properties of such theories are of considerable interest
both in their own right and in connection with the experimental
program on heavy ion collisions at RHIC and LHC.

In this paper, we study certain aspects of finite-temperature
behaviour of strongly coupled  ${\cal N}=2$ super-Yang-Mills theory
with dynamical quarks in Minkowski space. The holographic
description of the theory
 with gauge group $U(\nc)$ and $\nf$ hypermultiplets in the
fundamental representation (as well as one adjoint hypermultiplet)
is given by the well studied system of D3- and D7-branes. In the
limit of large $\nc$ and large 't Hooft coupling $\lam = g^2_{YM}
N_c$ with fixed $\nf$, the theory is described by  $\nf$ probe
D7-branes in the near-horizon geometry of $\nc$ D3-branes, \ie
AdS$_5\times S^5$ \cite{karchkatz}.
 At finite temperature, the
background geometry contains a black hole \cite{witten}. Although
the ${\cal N}=2$ theory is non-confining and thus no
confinement-deconfinement phase transition is expected at
 finite temperature, the presence of the
quark mass scale $M_q$ leads to a first order phase transition for
fundamental matter \cite{prl}. The transition occurs at a
temperature $T\sim M_q/\sqrt{\lam}$ and is characterized by the
dissociation or `melting' of the mesons, \ie the quark-antiquark
bound states\footnote{Recall that for these supersymmetric field
theories, the fundamental matter includes both fermions and scalars,
which we will refer to collectively as `quarks'.}.

Our goal is to study thermal dissociation of the bound states as
well as the flavour current diffusion by computing spectral
functions of mesonic operators in the framework of gauge/gravity
duality.
\FIGURE{
 \includegraphics[width=0.8 \textwidth]{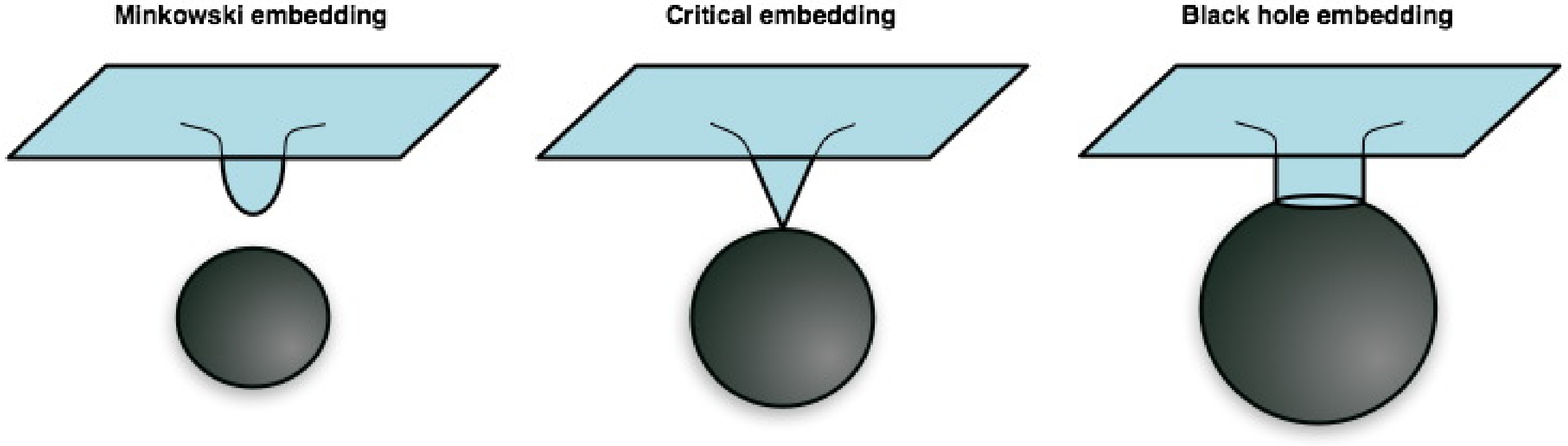}
\caption{Various possible D7-brane embeddings in the black D3-brane
geometry with temperature increasing from left to right.}
\label{embeddings2}}

The thermal behaviour of the ${\cal N}=2$ gauge theory with fundamental
matter has recently been actively studied
\cite{johanna,recent,prl,long}. The holographic description gives a
simple picture of the phase transition. If the D7-branes are
sufficiently far from the event horizon of the black hole, they are
gravitationally attracted towards the horizon but their tension is
sufficient to balance this attractive force. The probe branes then
lie entirely outside of the black hole in what were denoted
 `Minkowski' embeddings in \cite{prl,long} -- see
fig.~\ref{embeddings2}. As the temperature is raised, both the
radial position and the energy density of the event horizon
increase. Therefore, above some critical temperature, the
gravitational attraction of the black hole overcomes the brane
tension and the D7-branes are pulled into the horizon. These
configurations where the branes fall through the horizon are
referred to as `black hole' embeddings. In between these two
branches, there also exists a critical solution which just
`touches' the horizon. Thermodynamic considerations show the
latter is by-passed by a first order phase transition, in which the
probe branes jump discontinuously from a Minkowski to a black hole
embedding.

In the dual field theory, this phase transition is exemplified by
discontinuities in various physical quantities, \eg the quark
condensate. However, the most striking feature of the transition is
found in the spectrum of the mesons. The latter correspond to
excitations supported on the probe branes -- see, \eg
\cite{meson,rob-rowan}. In the low-temperature or Minkowski phase,
the mesons are stable (to leading order within the approximations
above) and the spectrum is discrete with a finite mass gap. In the
high-temperature or black hole phase, mesons are all destabilized
and rather one finds a continuous and gapless spectrum of
excitations\footnote{For a discussion of thermal resonances in the
context of holographic models see \eg  \cite{Caron-Huot:2006te}.}.
Accordingly, spectral functions of mesonic operators in the
low-temperature phase are characterized by $\delta$-function
peaks\footnote{A derivation of the scalar meson spectral function at
$T=0$ appears in appendix \ref{specT0}.}
(with the decay width of these particles and the continuum
contribution of multi-particle states both suppressed by factors of
$1/\nc$), whereas in the high-temperature phase spectral functions
are essentially featureless\footnote{The temperature-dependent part
of the spectral functions exhibits damped oscillations with the
period proportional to a Matsubara frequency, see section
\ref{simple} for details.} (fig.~\ref{schema}).
\FIGURE{
\begin{tabular}{cc}
 \includegraphics[width=0.45 \textwidth]{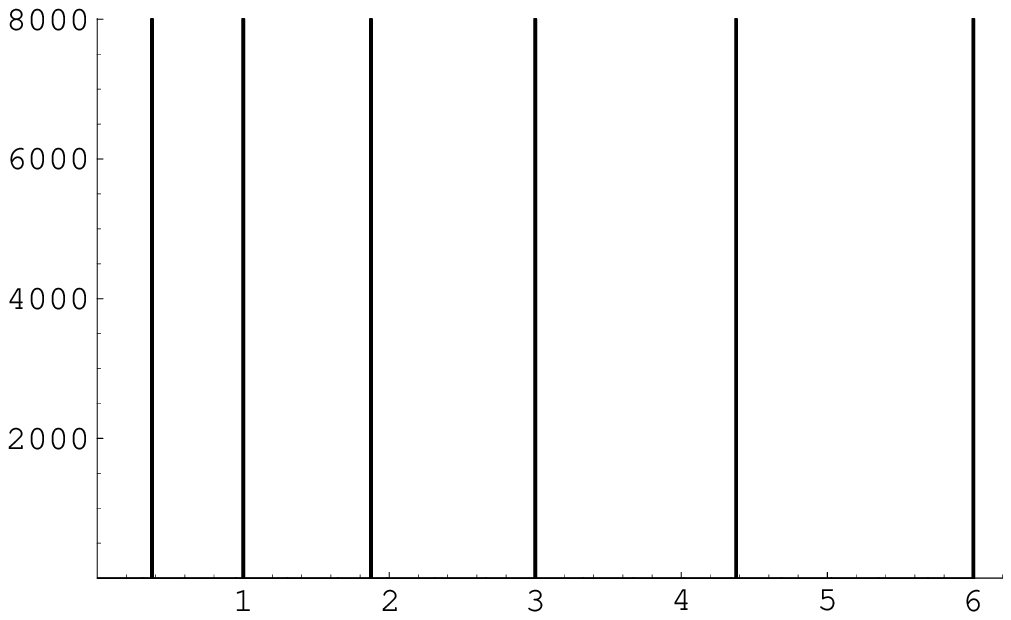}
&
 \includegraphics[width=0.45 \textwidth]{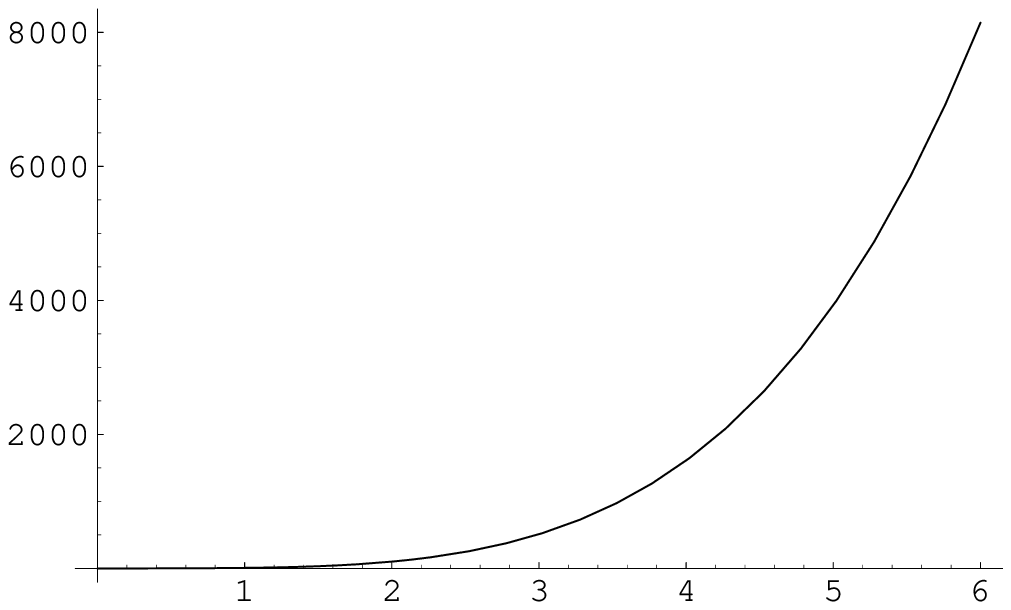} \\
 (a) & (b)
 \end{tabular}
\caption{Sketch of typical spectral functions in the (a)
low-temperature (Minkowski) and (b) high-temperature (black hole)
phases. } \label{schema} }
More interesting behaviour is observed when the system evolves from
the high-temperature phase  into the low-temperature phase through
the metastable `supercooled' phase. We show that in this case the
serene landscape of fig.~\ref{schema}b is distorted by peaks
corresponding to quasiparticle excitations, and these excitations
are eventually transformed into the stable resonances shown in
fig.~\ref{schema}a.

These features of the spectral functions are controlled by the
analytic structure of the corresponding retarded correlators in the
complex frequency plane. In the high-temperature phase, the poles of
the retarded correlators (with the exception of the poles
corresponding to hydrodynamic excitations) are located at a finite
distance from the real axis thus making the spectral function
featureless. As the temperature is lowered with respect to the quark
mass scale, the poles move closer to the real axis and the spectral
functions exhibit distinct peaks. Holographically, the poles of the
retarded correlators correspond to quasinormal modes of the
gravitational background  \cite{Birmingham:2001pj,Son:2002sd}.
Numerical investigation of the full
quasinormal spectrum of the D3-D7 system faces certain technical
difficulties. In this paper we focus on computing the spectral
functions for which the numerical methods are reliable.

In addition to characterizing quasiparticle excitations of a thermal
system, spectral functions also carry information about the medium's
transport properties. Adapting techniques from
\cite{Policastro:2002se, Kovtun:2003wp, Kovtun:2006pf}, we compute
the quark diffusion constant as a function of the parameter $M_q/T$
in the high-temperature phase, and attempt to give a qualitative
description of its dependence on the coupling for the full range of
temperatures.

Thermal dissociation of mesons as well as transport properties of
the quark-gluon plasma can be studied in lattice QCD with the help
of indirect methods such as the maximal entropy method
\cite{Umeda:2002vr,Asakawa:2003re,Datta:2003ww,Jakovac:2006bx,
Aarts:2006nr,Aarts:2007pk,Aarts:2006cq,Petreczky:2006zk,Aarts:2007wj,Meyer:2007ic}.
These studies suggest, in particular, that mesons survive as
relatively well-defined resonances at temperatures well above $T_c$
($2-3 T_c$). While the uncertainties of these lattice methods remain
large, the holographic approach used in this paper serves as a
source of quantitative and often analytically exact results for
qualitatively similar finite-temperature models.

An overview of the paper is as follows: in section \ref{prelude} we
review properties of thermal spectral functions in field theory and
outline methods of computing spectral functions from dual gravity.
These methods are illustrated by a simple example of  computing the
spectral function and diffusion constant for $R$ currents in ${\cal
N}=4$ SYM. For vanishing spatial momentum, the correlator,
quasinormal spectrum, and the spectral function can be computed
analytically. In section \ref{revGeom} we introduce the D3/D7-brane
framework and  review the D7-brane embeddings and thermodynamics. In
section \ref{goldfinger}, we turn to the calculation of the spectral
function for various mesonic operators in the high temperature phase
of the ${\cal N}=2$ gauge theory.  We consider a vector operator in
section \ref{vectors}.  In the special case of vanishing quark mass,
we determine the spectral function analytically. In general, for
arbitrary quark mass, the vector spectral function is computed
numerically. In section \ref{scalars}, we turn to spectral functions
for scalar and pseudoscalar operators, which are again bilinear in
the fundamental fields. Section \ref{diffusion} presents three
independent computations of the diffusion constant for `light'
quarks, using the membrane paradigm method \cite{Kovtun:2003wp}, the
Green-Kubo formula, and by calculating the lowest quasinormal
frequency for the vector field on the D7-brane. Section
\ref{discuss} contains  discussion and observations about our
results. Some details of our analysis are relegated to appendices:
appendix \ref{diction} provides a review of the holographic
dictionary relating the D7-brane worldvolume fields to their dual
operators in the gauge theory;
appendix \ref{specT0} contains a derivation of the scalar spectral
function at $T=0$; appendix \ref{high} provides a derivation of the
high frequency asymptotics for the spectral functions; appendix
\ref{schroe} gives a partial analysis of the quasinormal modes for
the pseudoscalar and scalar excitations; and finally, appendix
\ref{app-diffuse} extends the computation of the quark diffusion
constant, described in section \ref{membrane}, to the holographic
framework described by a Dq-brane probe in a near-extremal Dp-brane
throat.

\section{Prelude: spectral functions and holography}
\label{prelude}
In general, finite-temperature correlation functions of conserved
charge densities carry information about a medium's transport
properties and quasiparticle excitations. This  information is
given, roughly, by the poles and the corresponding residues of the
correlators, or, equivalently, by their spectral functions. Recently
the study of these objects has been used to great effect in a
holographic framework to study the thermal properties of various
strongly coupled field theories \cite{new}. In a holographic
setting, the spectral functions are often easier to compute than the
full correlators on the gravity side. According to the holographic
dictionary, the poles are determined by the quasinormal spectrum of
a dual bulk field fluctuation, whereas the spectral function is
given by the imaginary part of the retarded correlator which is
independent of the radial coordinate \cite{Son:2002sd,
Kovtun:2005ev}. In this section, we combine the necessary tools for
computing the spectral functions from dual gravity and analyzing
their properties, and then illustrate this technique using the
simple example of strongly coupled ${\cal N}=4$ supersymmetric
$SU(\nc)$ Yang-Mills (SYM) theory at large $\nc$. In this case, the
$R$-current spectral function has been analyzed elsewhere
\cite{Kovtun:2005ev,Teaney:2006nc,Caron-Huot:2006te} but we present
a new analytic result (for vanishing spatial momentum).

\subsection{Field theory picture}
\label{prelude_field}
A thermal spectral function  of an operator ${\cal O}$ is defined
as\footnote{Our metric convention is $(-,+,+,+)$. We assume
translation invariance to be an unbroken symmetry of the theory.}
\begin{equation}
\rn(\omega, {\bf q} ) =\int d^4 x\; e^{-i\omega t + i {\bf q} {\bf
x}} \langle [{\cal O}(t, {\bf x}),{\cal O}(0) ]\rangle\,,
\end{equation}
where the correlator is computed in thermal equilibrium at a
temperature $T$. The spectral function is proportional to
 the imaginary part of the retarded correlator,
\begin{equation}
 \rn(\omega, {\bf q} ) = -2\, {\rm Im}\, G^R (\omega, {\bf q})\,,
\end{equation}
where
\begin{equation}
  G^R(\omega, {\bf q} )
  = -i\!\int\!d^4x\,e^{-i\omega t + i {\bf q} {\bf x}}\,
  \theta(x^0) \langle
 [{\cal O}(t, {\bf x}),{\cal O}(0)] \rangle \,.
\label{retarded}
\end{equation}
If ${\cal O}$ is an operator of a
density of a conserved charge in a rotation invariant theory, the
retarded thermal two-point function
is determined by two independent scalar functions. In Fourier space,
the correlator can be decomposed into the transverse and
longitudinal parts  \cite{Kovtun:2005ev}
\begin{equation}
   G^R_{\mu\nu}(\omega, q) = P_{\mu\nu}^T\, \Pi^T(\omega,q) +
                   P_{\mu\nu}^L\, \Pi^L(\omega,q)\,,
\label{eq:C-rotation-inv}
\end{equation}
where the index structure is absorbed into two mutually
orthogonal projectors $P_{\mu\nu}^T$ and  $P_{\mu\nu}^L$.
Without loss of generality we can take the spatial momentum
oriented along the $x^3$ direction, so that $k_\mu=(-\omega,0,0,q)$,
with $k^2=-\omega^2+q^2$. Then one has  \cite{Kovtun:2005ev}
\begin{equation}
 G^R_{x^1x^1}(k) = G^R_{x^2x^2}(k) = \Pi^T(\omega,q)\ .
\end{equation}
Other components of the current-current correlator are
\begin{eqnarray}
   G^R_{tt}(k) = -\frac{q^2}{q^2-\omega^2}\, \Pi^L(\omega,q)\ ,
   \label{eq:Ctt}\\
   G^R_{tx^3}(k) = -\frac{\omega q}{q^2-\omega^2}\, \Pi^L(\omega,q)\ ,
   \label{eq:Ctz}\\
   G^R_{x^3x^3}(k) = -\frac{\omega^2}{q^2-\omega^2}\, \Pi^L(\omega,q)\ .
   \label{eq:Czz}
\end{eqnarray}
In the long-time, long-wavelength limit (\ie for $\omega/T \ll 1$,
$q/T \ll 1$) the functions  $\Pi^T(\omega,q)$ and $\Pi^L(\omega,q)$
have a universal behaviour dictated by hydrodynamics:
$\Pi^T(\omega,q)$ is nonsingular as a function of the frequency,
while $\Pi^L(\omega,q)$ has a simple pole at
\begin{equation}
\omega = - i D q^2\,,
\end{equation}
where $D$ is the charge diffusion constant.

The rotation invariance implies that in the limit of vanishing
spatial momentum at fixed $\omega >0$ the two scalar functions
coincide: $\Pi^T(\omega,0) = \Pi^L(\omega,0)= \Pi(\omega)$.
Correspondingly, at $q=0$ one can define
\begin{equation}
\rn (\omega) \equiv \rn_{x^1x^1} (\omega,0) =\rn_{x^2x^2} (\omega,0)
= \rn_{x^3x^3} (\omega,0)\,.
\end{equation}
The Green-Kubo formula relates the diffusion constant to the
zero-frequency limit of the spectral function $\rn (\omega)$:
\begin{equation}
D\, \Xi =
 \lim\limits_{\omega\rightarrow 0} {1\over 2 \omega} \rn (\omega) \,.
\labell{GKF}
\end{equation}
Here $\Xi$ is the charge susceptibility. The susceptibility is
determined by the thermodynamics of the system in a grand canonical
ensemble,
\begin{equation}
\Xi = {\partial n (\mu)\over \partial \mu}\, \Biggl|_{\mu=0}\,,
\label{suscep}
\end{equation}
where $n(\mu)$ is the charge density, $\mu$ is the corresponding
chemical potential.

In addition to  hydrodynamic poles, the retarded correlators may
have other singularities located in the lower half-plane of complex
frequency. Assuming one of these singularities is a simple pole,
$$
G^R \sim \frac{1}{\omega - \Omega (q,\alpha) + i\Gamma(q,\alpha)}\,,
$$
where $\alpha$ represents a set of parameters relevant for
a particular theory, for the spectral function one has
$$
\rn (\omega) \sim \frac{\Gamma}{(\omega - \Omega)^2 + \Gamma^2}\,.
$$
Thus in the vicinity of $\omega = \Omega$, the spectral function has
a peak characterized by a width $\sim \Gamma$ and a height
(`lifetime') $\sim 1/\Gamma$. The peak has a quasiparticle
interpretation if $\Gamma \ll \Omega$.

The spectral function $\rn (\omega)$ also has a characteristic form
in the high frequency limit. This behaviour is determined by the
leading short-distance singularity
 \beq
\lim_{(t^2-{\bf x}^2)\rightarrow0} \langle {\cal O}(t, {\bf
x})\,{\cal O}(0) \rangle=\frac{{\cal A}}{|t^2-{\bf
x}^2|^\Delta}+\cdots \, ,\labell{sing}
 \eeq
where $\Delta$ denotes the dimension of the operator $\cal O$ and
$\cal A$ is a dimensionless constant. A Fourier transform then leads
to the following contribution to the spectral function
 \beq
\rn (\omega)\sim  {\cal A}\, \omega^{2\Delta-4}\
.\labell{hightail}
 \eeq

\subsection{Gravity picture}

In the dual gravity picture, the conserved current $J_\mu$
couples to a boundary value of the gauge field fluctuation
$A_\mu$ propagating in a
specific gravitational background. One can form two gauge-invariant
 combinations of the fluctuation whose equations of motion
(supplemented with appropriate boundary conditions) contain (in the
limit where the gravity description is valid) full information about
the functions $\Pi^T(\omega,q)$ and $\Pi^L(\omega,q)$ introduced in
section \ref{prelude_field}. These gauge invariant combinations are
the transverse and longitudinal (with respect to a chosen direction
of the spatial momentum) components $E_T$, $E_L$ of the electric
field in curved space \cite{Kovtun:2005ev}. Quasinormal spectra of
the fluctuations  $E_T$ and $E_L$ determine the position of the
poles of $\Pi^T(\omega,q)$ and $\Pi^L(\omega,q)$ in the complex
$\omega$ plane.

\subsection{A simple example:  spectral function of
 $R$ currents in ${\cal N}=4$ SYM}\label{simple}

Correlators of $R$-currents in strongly coupled ${\cal N}=4$
$SU(\nc)$ supersymetric Yang-Mills (SYM) theory at large $\nc$ were
previously studied by means of the AdS/CFT correspondence both at
zero \cite{Freedman:1998tz, Chalmers:1998xr}  and finite temperature
\cite{Policastro:2002se, Nunez:2003eq, Kovtun:2005ev, Teaney:2006nc, Caron-Huot:2006te}.

In thermal  ${\cal N}=4$ SYM, the retarded two-point  correlators of
the $SU(4)_R$ R-symmetry currents $J^a_\mu$ are determined by two
independent scalar functions\footnote{In an equilibrium state
without chemical potentials for the $R$-charges, the correlation
function of $R$ currents $j^a_\mu$ has the form
$C_{\mu\nu}^{ab}=\delta^{ab} C_{\mu\nu}(x)$. In all expressions, the
factor $\delta^{ab}$ is omitted.}, $\Pi^T(\omega,q)$ and
$\Pi^L(\omega,q)$.

The holographic dual of  thermal  ${\cal N}=4$ SYM in flat space is
well known. The supergravity background describing the decoupling
limit of $\nc$ black D3-branes is (see, \eg \cite{bigRev})
\beq
ds^2 = \frac{r^2}{L^2} \left( -f(r)dt^2 +d{\bf x}^2\right) +
\frac{L^2}{r^2} \left( \frac{dr^2}{f(r)} +r^2 d\Omega_5^2\right)\, ,
\quad \quad C_{0123}=-\frac{r^4}{L^4}\, , \labell{D3geom-r}
\eeq
where $f(r) = 1-r_0^4/r^4$ and the dilaton is constant.  The horizon
lies at $r=r_0$ and the radius of curvature $L$ is defined in terms
of the string coupling constant $\gs$ and the string length scale
$\ls$ as $L^4 = 4\pi\, \gs \nc \, \ls^4$. According to the duality,
originally proposed by Maldacena
 \cite{juan}, type IIB string theory on these backgrounds is
dual to four-dimensional ${\cal N}=4$ super-Yang-Mills (SYM)
$SU(\nc)$ gauge theory.  The holographic dictionary between the
theories relates the Yang-Mills and string coupling constants
$\gym^2 = 2 \pi \gs$. The temperature of the gauge theory is
equivalent to the Hawking temperature of the black hole horizon:
\beq T = \frac{r_0}{\pi L^2}\, . \labell{Temper}\eeq

In the supergravity approximation (corresponding to the limit
$\nc\rightarrow \infty$, $\gym^2 \nc\rightarrow \infty$ in the field
theory), full information about functions $\Pi^T(\omega,q)$ and
$\Pi^L(\omega,q)$ can be obtained by solving the linearized Maxwell
equations for the bulk electric field components $E_T$, $E_L$
\cite{Kovtun:2005ev}
\begin{eqnarray}
E_{T} '' &+& {f'\over f} E_{T} ' + {\wn^2 - \qn^2 f\over (1-\xb) f^2} E_T =0\,,
\label{eqE_x} \\
E_L '' &+& {\wn^2 f'\over f (\wn^2 - \qn^2 f)} E_L' +
{\wn^2 - \qn^2 f\over (1-\xb) f^2} E_L = 0\,.
\label{eqE_z}
\end{eqnarray}
where $'$ indicates a derivative with respect to $\xb \equiv
1-r_0^2/r^2$. We have also introduced the dimensionless quantities
\begin{equation}
\wn = \frac{\omega}{2 \pi T}\,, \qquad \qn = \frac{q}{2 \pi T}\,.
\labell{dim}
\end{equation}
An analysis of eqs.~(\ref{eqE_x}), (\ref{eqE_z}) including the
perturbative solution for small $\wn$, $\qn$ can be found in
\cite{Kovtun:2005ev}. Here we shall focus on a particular case of
vanishing spatial momentum that admits analytic solution for
arbitrary frequency.

For $\qn =0$, the components $E_T = E_L \equiv E$ obey the same equation
\begin{equation}
E '' + {f'\over f} \, E ' + {\wn^2\over (1-\xb) f^2}\, E =0\,.
\label{eqE_i}
\end{equation}
Writing
\begin{equation}
E (\xb) = \xb^{-i \wn/2}\, (2-\xb)^{-\wn/2}\, F(\xb)\,,
\labell{gensole}
\end{equation}
where $F(\xb)$ is by construction regular at the horizon
$\xb=0$, we obtain the equation
\begin{equation}
F'' + {2 i \wn + 2 (\xb-1) - (1 + i) \wn \xb\over (\xb-2) \xb}\, F' +
{\wn
 ((1 + i) (1 - \xb) - i \wn ((1 + 2 i) - \xb))
\over 2 \xb (\xb-1)(\xb-2)}\, F =0\,. \label{eqF_x}
\end{equation}
Two linearly independent solutions of eq.~(\ref{eqF_x})
 are written in terms of the Gauss hypergeometric function
\begin{equation}
F_1(\xb) = (1-\xb)^{{(1+i)\wn \over 2}}
\ofo \left( 1 - { (1+i) \wn \over 2}\,,
 - { (1+i) \wn \over 2}\, ; 1- i \wn; {\xb\over 2 (\xb-1)}\right)\,,
\label{eqF_x_1}
\end{equation}
\begin{equation}
F_2(\xb) = \xb^{i \wn} \, (1-\xb)^{{(1-i)\wn \over 2}}
\ofo \left( 1 - { (1-i) \wn \over 2}\,,
 - { (1-i) \wn \over 2}\, ; 1+ i \wn; {\xb\over 2 (\xb-1)}\right)\,.
\label{eqF_x_2}
\end{equation}
To compute the retarded correlators, we need a solution obeying
the incoming wave boundary condition
at $\xb=0$ \cite{Son:2002sd}.
The correct solution is thus given by  eq.~(\ref{eqF_x_1}).

The retarded correlation functions can be computed from the boundary
supergravity action using the Lorentzian AdS/CFT prescription
\cite{Son:2002sd}. For vanishing spatial momentum, the result reads
\cite{Kovtun:2005ev, Caron-Huot:2006te}
\begin{equation}
\Pi (\omega)
 = {\nc^2 T^2\over 8}\, \lim\limits_{\xb\rightarrow 1}\;
 {E'(\xb)\over
E(\xb)} \,.
\label{cor_1}
\end{equation}
Substituting the solution  (\ref{eqF_x_1}) into  eq.~(\ref{cor_1})
we obtain
\begin{equation}
\Pi (\omega) = {\nc^2 T^2\over 8}\, \left\{
i \, \wn + \wn^2 \left[ \psi \left( {(1-i)\wn\over 2}\right)
+  \psi \left(- { (1+i)\wn\over 2}\right)\right]\right\}\,,
\label{correl_2}
\end{equation}
where $\psi(z)$ is the logarithmic derivative of the gamma-function.
The spectral function is given by
\begin{equation}
\rn (\omega) = - {\nc^2 T^2\over 4}\,\mbox{Im} \left\{
i \, \wn + \wn^2 \left[ \psi \left( {(1-i)\wn\over 2}\right)
+  \psi \left(- { (1+i)\wn\over 2}\right)\right]\right\}\,.
\label{sf}
\end{equation}
Using the property of the digamma function $\psi (z) -\psi (-z)
= - \pi \cot{\pi z}-1/z$, the spectral function (\ref{sf})
can be written in a more compact form
\begin{equation}
\rn (\omega) =  {\nc^2 T^2\over 4}\,
\frac{ \pi \wn^2 \sinh{\pi\wn}}{\cosh{\pi \wn} - \cos{\pi\wn}}\,.
\label{sfmod}
\end{equation}
These analytic results for the retarded Green's function
\reef{correl_2} and the spectral function \reef{sf},  \reef{sfmod}
 are new and
allow their various features to be easily examined. The asymptotics
of the spectral function for large and small frequency can be easily
computed
\begin{eqnarray}
\rn (\omega) &=&
 {\pi \nc^2 T^2 \wn^2\over 4}\, \left( 1 + 2 e^{-\pi \wn}\,
\cos{ \pi \wn} + \cdots \right)\,,
\qquad  \wn \rightarrow \infty\,, \label{lfa}\\
\rn (\omega) &=&
 {\nc^2 T^2 \wn\over 4}\,\left( 1 + {\pi^2 \wn^2 \over 6} +\cdots\right)\,,
 \qquad  \wn \rightarrow 0 \,. \labell{sfa}
\end{eqnarray}
As expected, the high frequency asymptotic coincides with the
zero-temperature result for the spectral function  \cite{Son:2002sd}
\begin{equation}
\rn^{T=0} (\omega) =  { \nc^2 \omega^2\over 16 \pi}\,.
\end{equation}
The retarded correlator (\ref{correl_2}) is a meromorphic function of $\wn$
with poles located at\footnote{The exact
 location of the poles was previously found in
 \cite{Nunez:2003eq} using the continued fraction method. }
\begin{equation}
\wn = \pm n - i\, n\,, \qquad n=1,2,...\,.
\end{equation}
The position of the poles coincides with the quasinormal spectrum of
the fluctuations $E(\xb)$ determined by the Dirichlet condition
$E(1)=0$. For each pole, the imaginary part has the same magnitude
as the real one, and thus none of the singularities can be given a
`quasiparticle' interpretation. Indeed, as shown in
fig.~\ref{chiVsOmGaugem0}a, the spectral function is quite
featureless, although
 not  monotonic: its finite-temperature part, $\rn(\omega)-
\rn^{T=0} (\omega)$, exhibits damped oscillations reflecting
the diminishing influence of the sequence
of poles receding farther and farther away from the real axis
 in the complex frequency plane (fig.~\ref{chiVsOmGaugem0}b).
The oscillatory behaviour of the finite-temperature part of
 the spectral function is evident from eq.~(\ref{lfa}).
\FIGURE{
\begin{tabular}{cc}
 \includegraphics[width=0.45 \textwidth]{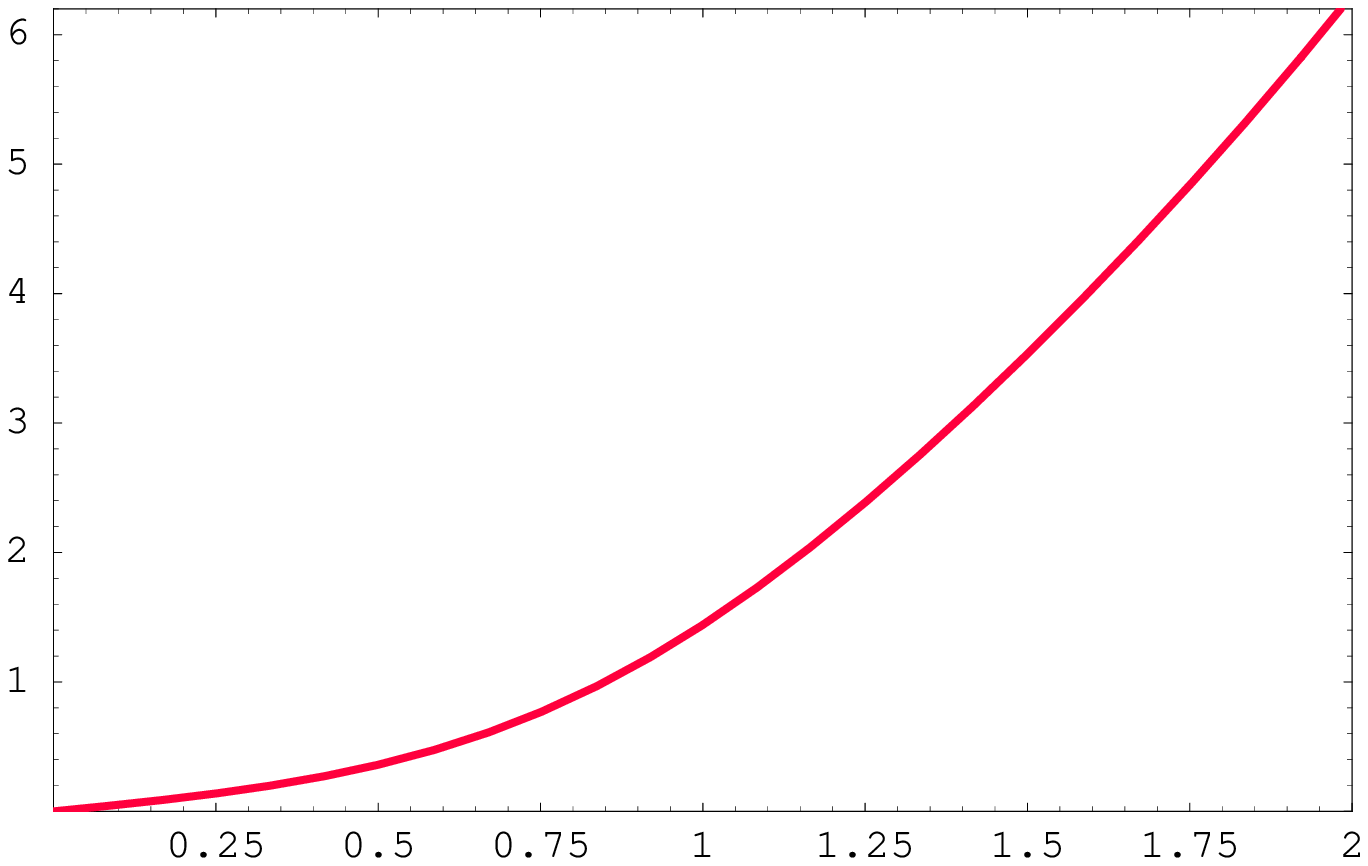} &
 \includegraphics[width=0.45 \textwidth]{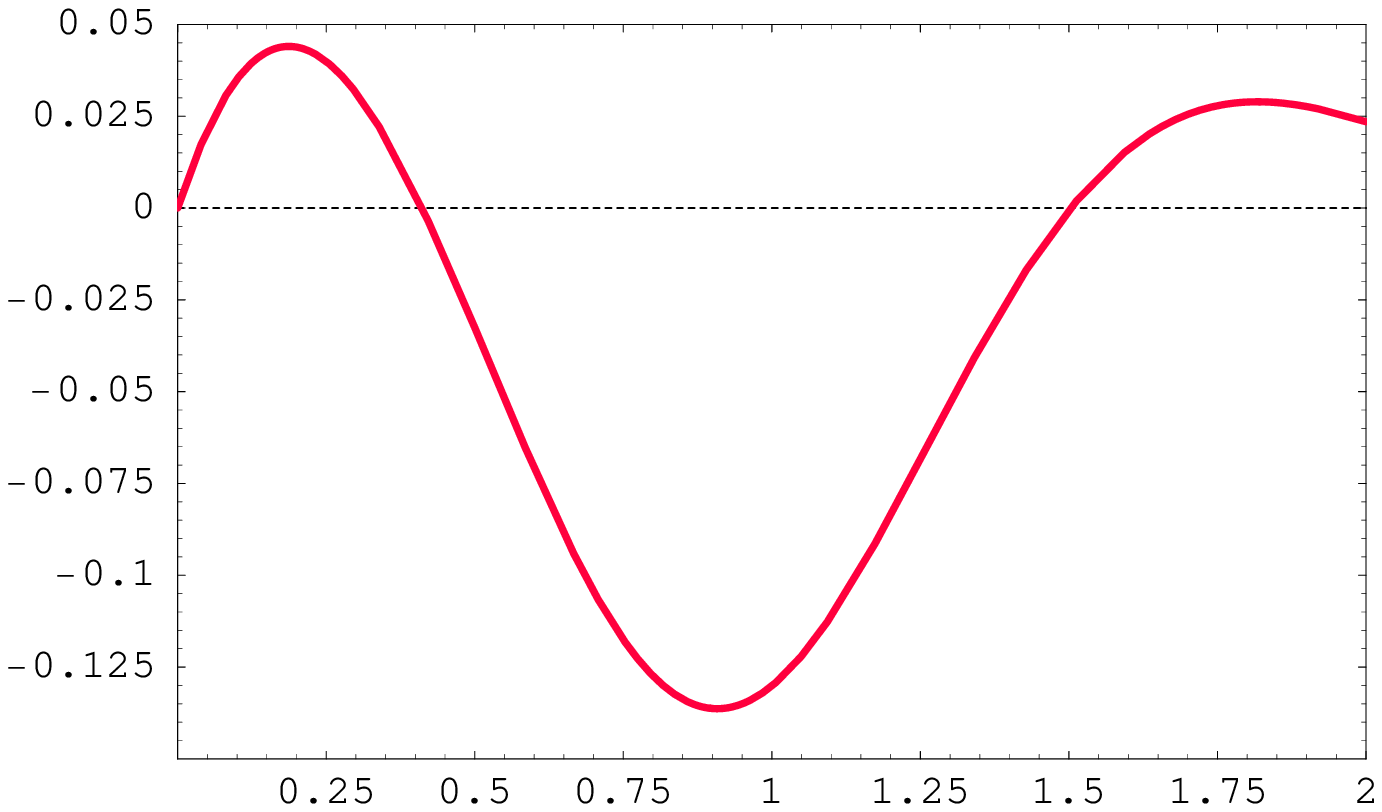} \\
 (a) & (b)
 \end{tabular}
\caption{The  ${\cal N}=4$ SYM $R$-current spectral function
 at zero spatial momentum
$\rn(\omega)$ (a) and its finite temperature
 part $\rn(\omega) - \nc^2 \omega^2/16 \pi$ (b)
 in units of $\nc^2T^2/2$ as
 a function of $\tom = \omega/2\pi T$.} \label{chiVsOmGaugem0} }

Using the Green-Kubo formula (\ref{GKF})
and the low frequency limit (\ref{sfa})
of the spectral function at zero spatial momentum
one finds the product of the $R$-charge diffusion constant and the
 charge susceptibility
 \begin{equation}
D \Xi = {\nc^2 T\over 16\pi}\,.
\end{equation}
The susceptibility is determined from thermodynamics according to
eq.~(\ref{suscep}). The dependence of the charge density on the
chemical potential was found in \cite{Son:2006em}. For small $\mu$,
one has
 \begin{equation}
n(\mu) = {\nc^2 T^2\over 8}\, \mu +\cdots\,,
\end{equation}
and thus from eq.~(\ref{suscep}), $\Xi = \nc^2 T^2/8$. We conclude
that the $R$ charge diffusion constant  is given by $D = 1/2\pi T$,
in agreement with the result of an earlier calculation
\cite{Policastro:2002se}, where the value of $D$ was determined from
the hydrodynamic pole of the longitudinal part of the correlator at
small but nonvanishing spatial momentum.

\section{Adding flavour: D7-brane embedding and thermodynamics}\label{revGeom}

All fields in the ${\cal N}=4$ SYM theory transform in the adjoint
representation of the $SU(\nc)$ gauge group. Fields transforming in
the fundamental representation of the gauge group can be introduced
in the gravity dual by inserting a second set of D-branes in the
supergravity background \cite{karchkatz}. In particular, we consider
the decoupling limit of the intersection of $\nc$ black D3-branes
and $\nf$ D7-branes as described by the array:
\begin{equation}
\begin{array}{ccccccccccc}
   & 0 & 1 & 2 & 3 & 4& 5 & 6 & 7 & 8 & 9\\
D3 & \times & \times & \times & \times & & &  &  & & \\
D7 & \times & \times & \times & \times & \times  & \times & \times & \times &  &   \\
\end{array}
\labell{array}
\end{equation}
The dual field theory is now an ${\cal N}=2$ gauge theory consisting
of the original SYM theory coupled to $\nf$ fundamental
hypermultiplets. Taking the decoupling limit with $\nf \ll \nc$, the
D7-branes may be treated as probes in the black D3-brane geometry
\reef{D3geom-r}. The holographic framework has been used extensively
to study the thermal properties of the ${\cal N}=2$ gauge theory at
large $\nc$ \cite{johanna,recent,prl,long}. In particular, it was
found that the fundamental matter undergoes a phase transition
characterised by the dissociation of the mesonic bound states, as
will be reviewed below.

Following \cite{prl,long}, it is helpful to introduce a
dimensionless radial coordinate $\rho$, related to the coordinate
$r$ via
\beq
(\om \rho)^2 = r^2 + \sqrt{r^4-\om^4}\, . \labell{rhor}
\eeq
In this case, the metric for the black D3-brane geometry
\reef{D3geom-r} becomes
\beq
ds^2(g) = \frac{1}{2} \left(\frac{\om \rho}{L}\right)^2
\left[-{f^2\over \tilde f}dt^2 + \tilde{f} d{\bf x}^2 \right] +
\frac{L^2}{\rho^2}\left[ d\rho^2 +\rho^2 d\Omega_5^2  \right]\, , \labell{D3geom}
\eeq
where $f(\rho)=1-1/\rho^4$, $\tilde{f}(\rho)=1+1/\rho^4$ and
\beq d\Omega_5 ^2 = d\theta ^2 + \sin ^2 \theta d\Omega_3 +\cos^2
\theta \, d\phi^2 \, .\label{spheroid} \eeq
The worldvolume directions of the D3-branes by the coordinates $\{
t,x^i\}$. The probe D7-branes fill these coordinates, as well as
wrapping the $S^3$ in \reef{spheroid} and extending in the radial
direction $\rho$. In general then, the D7-brane embedding is
specified by giving its profile in the angular directions, $\theta$
and $\phi$. For simplicity, we fix $\phi=0$ for our background
embeddings. Then requiring translational symmetry in the 0123-space
and rotational symmetry in the 4567-directions motivates us to
consider $\theta = \theta (\rho)$. In fact, it is more convenient to
work with $\chi(\rho) \equiv \cos \theta(\rho)$ rather than
$\theta$. The induced metric on the D7-branes is then
\beq
ds^2 = \frac{1}{2} \left(\frac{\om \rho}{L}\right)^2
\left[-{f^2\over \tilde f}dt^2 + \tilde{f} d{\bf x}^2 \right] +
 \left(\frac{L^2}{\rho^2}+\frac{L^2 \dot{\chi}^2}{1-\chi^2} \right)
 d\rho^2 +L^2(1-\chi^2) d\Omega_3^2\, ,
\labell{induce}
\eeq
where $\dot{\chi} = d \chi /d\rho$.
The D7-brane action follows as
\beq
\frac{S_\mt{D7}}{\N} = -\int d\rho \left( 1-\frac{1}{\rho^8}\right)
\rho^3 (1-\chi^2) \sqrt{1-\chi^2+\rho^2 \dot{\chi}^2} \, ,\labell{act2s}
\eeq
where the normalisation constant is $\N=N_f T_\mt{D7} \om^4 \Omega_3
/ 4 T$ with $\Omega_3 = 2\pi^2$ and $T_\mt{D7}=2\pi/(2\pi\ls)^8
g_s$. The equation of motion for $\chi(\rho)$ is then
\beq
\partial_\rho \left[\left( 1-\frac{1}{\rho^8}\right) \frac{\rho^5
(1-\chi^2) \dot{\chi}}{\sqrt{1-\chi^2+\rho^2 \dot{\chi}^2}} \right]
+\rho^3 \left( 1-\frac{1}{\rho^8}\right)\frac{3 \chi (1-\chi^2)+2
\rho^2 \chi \dot{\chi}^2}{\sqrt{1-\chi^2+\rho^2 \dot{\chi}^2}}=0 \labell{psieom}
\eeq
which implies that the field $\chi$ asymptotically approaches zero as
\beq
\chi = \frac{m}{\rho} + \frac{c}{\rho^3}+\cdots\, . \labell{asympD7}
\eeq
The operator dual to $\chi$ is the supersymmetric extension of the
quark mass term, defined in \reef{mass}. Holography then relates the
dimensionless constants $m$ and $c$ to the quark mass and condensate
via \reef{Mm} and \reef{Oc}. Eq.~\reef{Mm} implies the relationship
$m=\mbar/T$ between the dimensionless quantity $m$, the temperature
$T$ and the mass scale
\beq \mbar= \frac{2 M_\mt{q}}{\sqrt{\lambda}} =
\frac{M_\mt{gap}}{2\pi}\,. \labell{mbar} \eeq
Here $M_\mt{gap}$ is the meson mass gap in the D3/D7 brane theory at
zero temperature \cite{meson}.

In the limits of large and small $m$ it is possible to find
approximate analytic solutions for the embeddings -- see
\cite{long}.  However, for arbitrary $m$ we numerically integrated
\reef{psieom} -- see \cite{prl,long}. In the present case, we are
studying the high temperature phase and so we are interesting in the
black hole embeddings, which are found by imposing the following
boundary conditions on the event horizon $\rho_\mt{min}=1$:
$\chi=\chi_0$ and $d\chi/d\rho =0$ for $0 \leq \chi_0 <1$. In order
to compute the constants $m,\, c$ corresponding to each value of
$\chi_0$, we fitted the numerical solutions to the asymptotic form
\reef{asympD7}.

\subsection{Thermodynamics of the brane}\label{thermoBrane}

As described above, introducing D7-brane probes into the black
D3-brane background corresponds to adding dynamical quarks to the
gauge theory.  The resulting theory has a rich spectrum of
quark-antiquark bound states or mesons.  As these mesons are dual to
strings with both ends on D7-branes, the mesons can be studied by
examining the fluctuations of the D7-branes, \eg the spectrum of the
lowest-lying mesons can be found by computing the spectrum of
fluctuations of the worldvolume fields on the D7-branes -- see \eg
\cite{meson, rob-rowan, ramallo}. At finite temperature, there are
two classes of embeddings for the D7-branes and a first-order phase
transition that goes between these classes. For temperatures below
the phase transition $T < T_\mt{fund}$, the D7-branes close off above
the black hole horizon (Minkowski embeddings), while above the
transition ($T > T_\mt{fund}$), the D7-branes extend through the
horizon (black hole embeddings). In the gauge theory, the most
striking feature of this transition is the change in the meson
spectrum \cite{prl,long}. Below the phase transition, the spectrum
of mesons has a mass gap and is discrete while above, the mesonic
excitations are unstable and are characterised by a discrete
spectrum of quasinormal modes \cite{hoyos}.

The above phase transition has been studied in detail in
\cite{johanna, recent, prl, long} and we review a few salient facts
here needed for later discussions. The Minkowski and black hole
embeddings are separated by a critical solution that just touches
the black hole horizon, as depicted in fig.~\ref{embeddings2}.
Plotting the quark condensate $c$ as a function of $m=\mbar/T$
reveals that $c$ is not a single-valued function of $m$ and that the
two families of solutions spiral around the critical solution -- see
fig.~4 of \cite{long}. The physical solution corresponds to the
embedding which minimizes the free energy density of the D7-branes.

A plot of the free energy $F$ as a function of temperature near the
phase transition is given in fig.~\ref{action} and this shows the
`swallow tail' which is typical of first order phase transitions.
Starting from low temperatures, we follow the blue dotted line
depicting Minkowski embeddings to the point where this line
intersects the solid red line for black hole embeddings.   The phase
transition occurs at this point, and the physical embedding jumps
from a Minkowski embedding, a finite distance from the black hole
horizon, to the black hole embedding with $\chi_0 \simeq 0.94$. At
this temperature, the quark condensate, entropy, and energy density
each exhibit a finite discontinuity, indicating that the phase
transition is first order.
\FIGURE{
\includegraphics[width = 0.8 \textwidth]{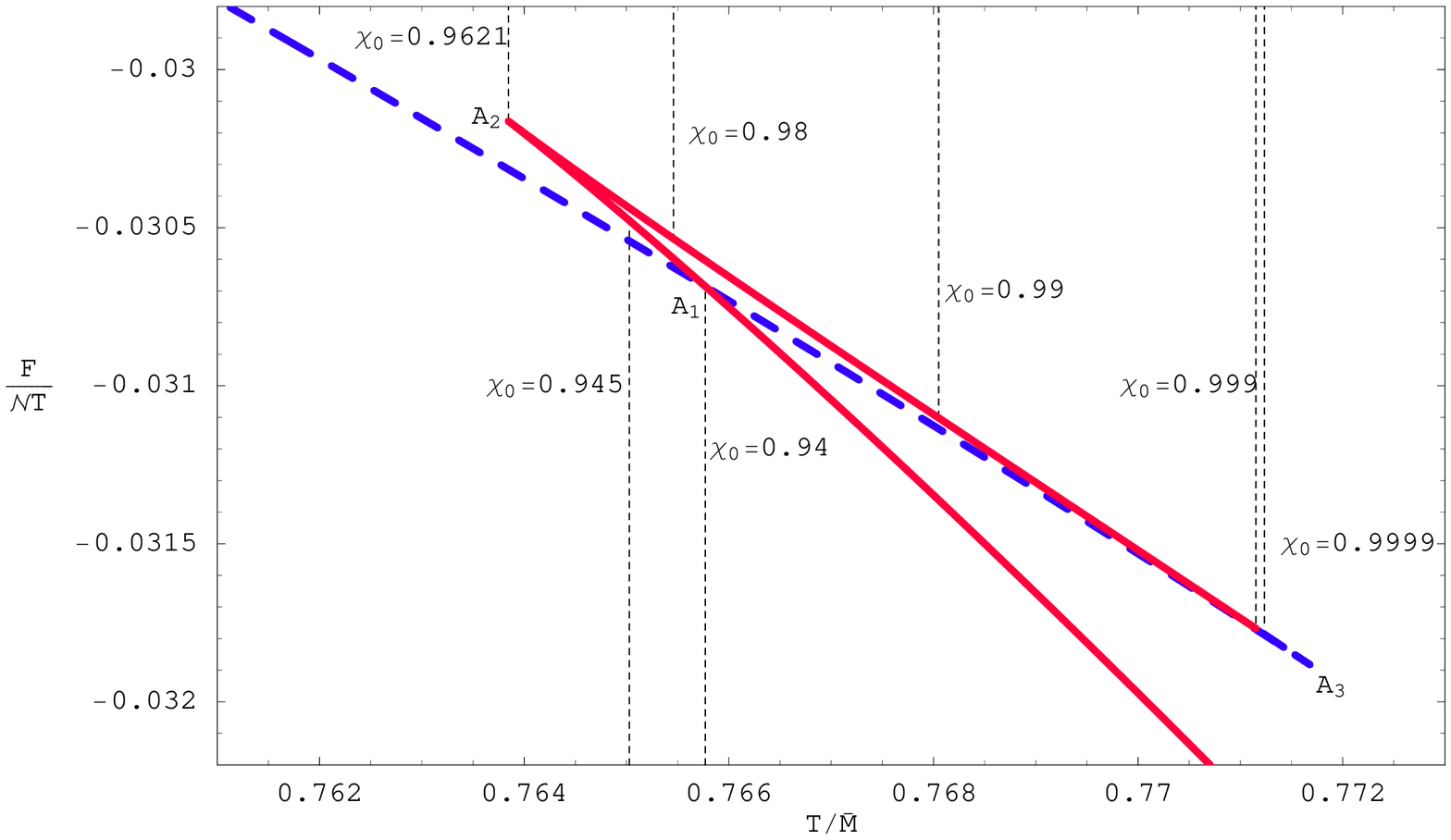}
\caption{The free energy $F/{\cal N}T$  versus temperature $T/\mbar$
for a D7-brane in the black D3-brane geometry where ${\cal N} =
\lambda \nf \nc T^3/32$.  The blue dashed (red continuous) curves
correspond to the Minkowski (black hole) embeddings.  The values of
$\chi_0$ for certain black hole D7-brane embeddings are noted for
future reference. The phase transition is indicated by the vertical
dashed line labelled $\chi_0=0.94$.  } \label{action} }

It is interesting to ask whether the D7-brane embeddings beyond the
phase transition, \eg between $A_1$ and $A_2$ on the black hole
branch, still represent metastable configurations. If so, the
corresponding states of the gauge theory might be accessed by a
process analogous to `super-cooling' the system. Examining the
specific heat $c_\mt{V}=\partial E/\partial T$ reveals that
$c_\mt{V}$ becomes negative as the curves (\eg of the condensate,
entropy or energy density as a function of $T$) spiral around the
critical solution, indicating that the system should be unstable for
these embeddings. In particular, the specific heat first becomes
negative at $A_2$ on the black hole branch and $A_3$ on the
Minkowski branch. Examining the scalar fluctuation spectrum of the
D7-brane Minkowski embeddings (corresponding to the meson spectrum
in the low temperature phase of the dual gauge theory) reveals that
a dynamical instability appears precisely at $A_3$: At the first
kink in the free energy, the lowest-lying scalar mode on the
D7-branes becomes tachyonic. In fact, at the second kink, the second
lowest-lying scalar mode becomes tachyonic and new tachyonic modes
seem to appear at each such kink \cite{long}.

Hence the behaviour on the Minkowski branch  is clear: Continuing
along the Minkowski branch past the phase transition, the system
exhibits a dynamical instability, which matches the thermodynamic
prediction, at the point $A_3$ in fig.~\ref{action}, which is the
first kink in the free energy on the Minkowski branch and
corresponds to the first turn-around in the spiral. Hence while
these configurations remain metastable between $A_1$ and $A_3$, all
of the embeddings beyond $A_3$ are simply unstable. In fact, more
and more instabilities appear as the embeddings approach
the critical solution, as described above.

On the black hole branch we expect similar phenomena. At point $A_2$
in fig.~\ref{action}, the specific heat becomes negative, indicating
a thermodynamic instability.  Though the full calculation of the
quasinormal spectrum remains to be performed -- preliminary results
appear in \cite{hoyos} -- we will see in section \ref{scalar} that
the scalar spectral function provides evidence that new tachyonic
modes again appear at each turn in the spiral along the black hole
branch. Appendix \ref{schroe} presents a complementary analysis
which also supports the appearance of tachyons in the quasinormal
spectrum of the scalar fluctuations. Hence we expect that only the
configurations between $A_1$ and $A_2$  on the black hole branch
represent metastable states of the gauge theory.


\section{Spectral functions for excitations of fundamental fields}
\label{goldfinger}

In this section, we compute spectral functions for excitations of
fundamental fields in the high temperature phase of the theory,
$T>\tf$, by studying vector and scalar fluctuations of the D7-brane
probes.  The details of the holographic dictionary relating the
fluctuations of the probe branes to the hypermultiplet operators of
the gauge theory are described in appendix \ref{diction}. In each
case, we begin by considering modes that are constant on the
internal $S^3$ but then extend the analysis to include modes with nontrivial
angular momentum on this space. The latter modes are dual to higher
dimension operators which are `charged' under the global SO(4)
symmetry, as outlined in appendix \ref{diction}.

\subsection{Vector}\label{vectors}

In the gravity dual, the vector is one of several possible
excitations of the worldvolume gauge field on the D7-branes. These
modes are characterised as having only $A_{t,x,y,z}$ nonzero with
$A_\rho=A_{S^3}=0$ \cite{rob-rowan}. The holographic dictionary
outlined in appendix \ref{diction} reveals that the vector is dual
to the current $J_\mt{q}^\mu$  which is the conserved current
corresponding to the diagonal $U(1)_\mt{q}$ of the global flavour
symmetry.

The full action for the gauge fields on a D7-brane contains the
Dirac-Born-Infeld (DBI) action plus a Wess-Zumino term, however, for
gauge fields with $A_\rho=A_{S^3}=0$ only the DBI portion of the
action is relevant. Further since we only study linearized
fluctuations about the background, the gauge field action is only
needed to quadratic order, which is simply
\beq S = -\frac{(2 \pi \ls^2)^2}{4} T_\mt{D7} \nf \int d^8 \sigma
\sqrt{- g}\, g^{cd}g^{ef} F_{fc}F_{de}\, , \labell{gaugeAction} \eeq
where the Latin indices run over the D7 worldvolume directions and
$g_{ab}$ is the induced metric on the D7-brane given in
\reef{D3geom}.

Assuming that the gauge field is independent of the coordinates on
the $S^3$, we can easily reduce \reef{gaugeAction} to an effective
action in five-dimensions. The induced metric in these directions is
\beq ds^2(\tilde{g}) = \frac{1}{2} \left(\frac{\om \rho}{L}\right)^2
\left[-{f^2\over \tilde f}dt^2 + \tilde{f} dx^2_3 \right] +
\frac{L^2}{\rho^2}\left(\frac{1-\chi^2+\rho^2
\dot{\chi}^2}{1-\chi^2} \right) d\rho^2 \, ,\labell{metricTilde} \eeq
and so the determinant of the full induced metric \reef{D3geom} can
be written as
\beq \sqrt{-g} =
\frac{\sqrt{-\tilde{g}}}{g^2_{eff}(\rho)}\sqrt{h_3}\, , \quad \quad
\frac{1}{g_{eff}^2(\rho)} \equiv L^3 (1-\chi^2)^{\frac{3}{2}}\,.
\labell{geff} \eeq
Here $h_3$ is the determinant of the metric on the $S^3$ of unit
radius and $g_{eff}$ is a radially-dependent `effective coupling'.
Integrating over the three-sphere, the action \reef{gaugeAction}
reduces to
\beq S = -\frac{(2 \pi \ls^2)^2}{4} \Omega_3 T_\mt{D7} \nf \int dt
d^3x d\rho \sqrt{- \tilde{g}}\, \frac{F^{\alpha \beta}F_{\alpha
\beta}}{g_{eff}^2(\rho)} \, ,\labell{gaugeAction2} \eeq
where $\alpha, \beta=t,x,y,z,\rho$. Of course, Maxwell's equations
follow as
\beq
\partial_\alpha \left(\frac{\sqrt{-
 \tilde{g}}}{g_{eff}^2}\, F^{\alpha \beta}\right) =0. \labell{maxwell}
\eeq

Using the equation of motion \reef{maxwell}, the action
\reef{gaugeAction2} can be written as
\beq S= -\frac{(2 \pi \ls^2)^2}{2}T_\mt{D7} \Omega_3 \nf \int dx^4
d\rho\,
\partial_\alpha \left[\frac{\sqrt{-
\tilde{g}}}{g_{eff}^2}A_\beta F^{\alpha \beta} \right]\,. \eeq
Retaining only the terms at the $\rho$-boundaries and using the
metric \reef{metricTilde}, this becomes
\beq S = -\frac{(2 \pi \ls^2)^2}{2}T_\mt{D7} \Omega_3
\frac{\om^2}{2} \nf \int d^4 x \left[\frac{f \rho^3
(1-\chi^2)^2}{\sqrt{1-\chi^2+\rho^2 \dot{\chi}^2}} \left(A_i
\partial_\rho A_i-\frac{\tilde{f}^2}{f^2}A_t \partial_\rho A_t
\right) \right]^{\rho \to \infty}_{\rho \to 1} \, , \eeq
where $i$ is summed over $x,y,z$. Following \cite{Kovtun:2005ev}, we
take the Fourier transform of the gauge field and with
$k_\mu=(-\omega,q,0,0)$,
\beq A_\mu = \int \frac{d\omega \, dq}{(2\pi)^2} e^{-i \omega t +
iqx} A_\mu(k,\rho) \eeq
(with $A_\rho=0$, as discussed earlier), and the boundary action can
be written as
\begin{small}
\[ S=-\frac{\nf \nc T^2}{2^4}\int \frac{d\omega dq}{(2 \pi)^2} \left[\frac{f
\rho^3 (1-\chi^2)^2}{\sqrt{1-\chi^2+\rho^2 \dot{\chi}^2}}
\left(A_i(\rho,-k)  \partial_\rho
A_i(\rho,k)-\frac{\tilde{f}^2}{f^2}A_t(\rho,-k) \partial_\rho
A_t(\rho,k) \right) \right]^{\rho \to \infty}_{\rho \to 1}\, .
\]
\end{small}
We construct gauge-invariant components of the electric field: $E_x
\equiv q A_t+{\omega} A_x$ and $E_{y,z}\equiv\omega A_{y,z}$.  Note
that in the language of section \ref{prelude}, $E_x$ corresponds to
the longitudinal electric field $E_L$ while $E_{y,z}$ correspond to
the transverse electric field $E_T$.  With these gauge-invariant
fields, the action can be rewritten as (using eq.~\reef{gauge1}
below)
\beqa S &=& -\frac{\nf \nc T^2}{2^4} \int \frac{d\omega dq}{(2
\pi)^2} \left[\frac{f \rho^3 (1-\chi^2)^2}{\sqrt{1-\chi^2+\rho^2
\dot{\chi}^2}} \left( \frac{E_x(\rho,-k) \partial_\rho
E_x(\rho,k)}{\omega^2 -q^2
f^2/\tilde{f}^2} \right. \right. \labell{fire1}\\
&& \quad \quad \quad \quad \quad \quad \quad\quad \quad \left.
\left.- \frac{1}{\omega^2} \left( E_y(\rho,-k) \partial_\rho
E_y(\rho,k)+E_z(\rho,-k) \partial_\rho E_z(\rho,k)\right)\right)
\right]^{\rho \to \infty}_{\rho \to 1}\,. \nonumber\eeqa

Focusing on the longitudinal electric field, we write
\beq E_x(k,\rho) = E_0(k) \frac{E_k(\rho)}{E_k(\rinf)} \, , \eeq
where it is understood that eventually the limit $\rinf \to \infty$
will be taken.  We can then define the flux factor for $E_x$ as
\cite{Son:2002sd}:
\beqa \mathcal{F} = -\frac{\nf N_c T^2}{2^4}  \left[\frac{f \rho^3
(1-\chi^2)^2}{\sqrt{1-\chi^2+\rho^2 \dot{\chi}^2}}
\frac{E_{-k}(\rho) \partial_\rho E_{k}(\rho)}{(\omega^2 -q^2
f^2/\tilde{f}^2)E_{-k}(\rinf) E_k(\rinf)} \right]  . \eeqa
The usual AdS/CFT prescription tells us to evaluate it at the
boundary $\rho \to \infty$ to find the retarded Green's function for
$E_x$ \cite{Son:2002sd}:
\beqa G = -2 \mathcal{F} = \frac{\nf N_c T^2}{2^3} \left[\frac{f
\rho^3 (1-\chi^2)^2}{\sqrt{1-\chi^2+\rho^2 \dot{\chi}^2}}
\frac{E_{-k}(\rho) \partial_\rho E_{k}(\rho)}{(\omega^2 -q^2
f^2/\tilde{f}^2)E_{-k}(\rinf) E_k(\rinf) }\right]_{\rho \to \infty}.
\eeqa
The retarded Green's function for $A_x$ is the above expression
times $\omega ^2$, which for $q=0$ gives
\beq G_{xx} = \frac{\nf N_c T^2}{8} \left[\rho^3 \frac{\partial_\rho
E_k(\rho)}{E_k(\rho)} \right]_{\rho \to \infty}\, \labell{greens}
\eeq
upon using the asymptotic expansion \reef{asympD7} for $\chi$. Of
course, this is the analogue of the expression in eq.~\reef{cor_1} for
the example discussed in section \ref{simple}. The spectral function
for $q=0$ is then
\beq \rn_{xx}(\omega,0)=-2 \textrm{Im} G_{xx}(\omega,0)= - \frac{\nf
N_c T^2}{4} \textrm{Im} \left[\rho^3 \frac{\partial_\rho
E_k(\rho)}{E_k(\rho)} \right]_{\rho \to \infty}\, .
\labell{spectralGauge} \eeq

In order to evaluate the spectral function, we must solve the
equations of motion \reef{maxwell}.  For $A_\rho = A_{S^3}=0$ and
$A_\mu$ an s-wave on the $S^3$, the equations for $A_t$ and $A_x$
are
\beqa
&&\tg^{tt} \omega \dot{A}_t - \tg^{xx}q \dot{A}_x=0 \, ,\labell{gauge1}\\
&&\partial_\rho \left(\frac{\sqrt{-\tg}}{g_{eff}^2}\tg^{tt}\tg^{\rho\rho}
\da_t \right)- \frac{\sqrt{-\tg}}{g_{eff}^2}\tg^{tt}\tg^{xx}
\left(\omega q A_x+q^2 A_t \right)=0 \,  ,\labell{gauge2}\\
&&\partial_\rho \left(\frac{\sqrt{-\tg}}{g_{eff}^2}\tg^{\rho \rho}\tg^{xx}
\da_x \right)- \frac{\sqrt{-\tg}}{g_{eff}^2}\tg^{tt}\tg^{xx}
\left(\omega q A_t+\omega^2 A_x \right)=0 \,. \labell{gauge3}
\eeqa
Given the longitudinal field $E_x = qA_t+\omega A_x$ as above, the
system of equations \reef{gauge1}-\reef{gauge3} yields
\beq \dde + \left[\frac{4 \omega^2 \tilde{f}\dot{f}}{f (\omega^2
\tilde{f}^2-q^2 f^2)}+\partial_\rho \ln\left(
\frac{\sqrt{-\tg}}{g_{eff}^2} \tg^{tt}\tg^{\rho \rho}\right)
\right]\de
 +\frac{\tg^{xx}}{\tg^{\rho\rho}}\left(\frac{\tilde{f}^2}{f^2}\omega^2-q^2
 \right) E_x =0. \labell{Eeq}
\eeq
Substituting for the induced metric in \reef{Eeq}, the equation of
motion for $E_x$ is:
\beqa \ddot{E}_x &&+ \left[\frac{4 \tom^2 \tilde{f}\dot{f}}{f
(\tom^2 \tilde{f}^2-\tk^2 f^2)}+\frac{f}{\tilde{f}^2}
\frac{\sqrt{1-\chi^2+\rho^2 \dot{\chi}^2}}{\rho^3
(1-\chi^2)^2}\partial_\rho \left(  \frac{\tilde{f}^2 \rho^3
(1-\chi^2)^2}{f\sqrt{1-\chi^2+\rho^2 \dot{\chi}^2} } \right)
\right]\dot{E}_x \nn && +8 \frac{1-\chi^2+\rho^2
\dot{\chi}^2}{\rho^4
\tilde{f}(1-\chi^2)}\left(\frac{\tilde{f}^2}{f^2}\tom^2-\tk^2
\right) E_x =0 \, .  \labell{Eeq2} \eeqa
Here we use the dimensionless frequency $\tom$ and momentum $\tk$
are defined in eq.~\reef{dim}.

Returning to the transverse electric field $E_{T}=E_{y,z}$, the
equation of motion is
\beq
\partial _\rho \left[\frac{\sqrt{-\tilde{g}}}{g_{eff}^2} g^{\rho \rho}
 g^{yy} \partial_\rho E_T \right]-\frac{\sqrt{-\tilde{g}}}{g_{eff}^2}
 g^{yy} \left(\omega^2 g^{tt}+q^2 g^{xx} \right)E_T=0\, .
\eeq
For vanishing spatial momentum $q=0$, this equation and that for
$E_x$ (eq.~\reef{Eeq2}) coincide. Thus, as discussed in section
\ref{prelude}, for $q=0$, the spectral functions  are identical \ie
$\rn_{xx}(\omega,0)=\rn_{yy}(\omega,0)=\rn_{zz}(\omega,0)$ and we
denote these by $\rn(\omega)$ henceforth.

We proceed to compute the spectral function $\rn(\omega)$ by solving
the equation of motion \reef{Eeq2} with $\tk =0$. First we note that
the case of massless quarks corresponds to the equatorial embedding
of the D7-branes for which $\chi (\rho) =0$.  Hence $g^2_{eff}$ in
eq.~\reef{gaugeAction2} is constant and the induced metric
\reef{metricTilde} matches precisely that of the AdS$_5$ black hole.
Hence except for an overall normalization, the calculation of
$\rn(\omega)$ is identical to that in the example discussed in
section \ref{simple} and so in this case, it is possible to solve
\reef{Eeq2} exactly. We leave this exercise for the following
subsection as it is a special case of the general analysis of
charged vector operators, for which the case $\mq=0$ is also exactly
soluble.

For massive quarks ($m\ne 0$), the embedding equation \reef{psieom}
must be solved numerically and hence it was necessary to numerically
integrate \reef{Eeq2} to solve for $E_x$. Near the horizon ($\rho
\to 1$), eq.~\reef{Eeq2} implies that $E_x \sim (\rho-1)^{\pm i
\tom}$.  Choosing the negative sign enforces incoming wave boundary
conditions at the horizon.  Thus, for each choice of quark mass, we
solved \reef{Eeq2} numerically, taking $E_x (\rho) = (\rho-1)^{-i
\tom} F(\rho)$ and where $F(\rho)$ is regular at the horizon with
$F(1)=1$ and $\partial_\rho F (1) = i \tom/2$ for real $\tom$.

\FIGURE{
\begin{tabular}{c}
 \includegraphics[width=0.8 \textwidth]{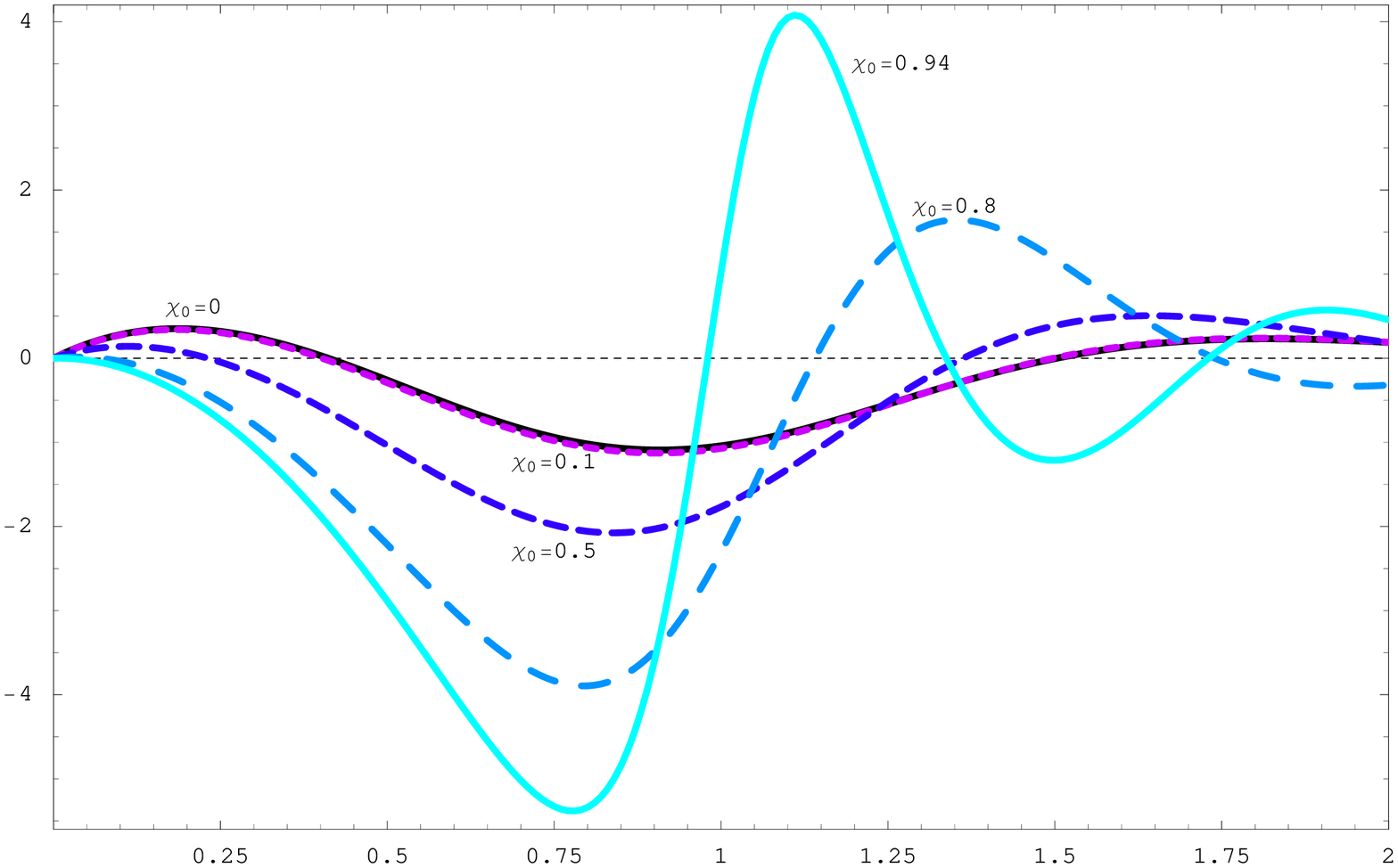} \\
 \includegraphics[width=0.8 \textwidth]{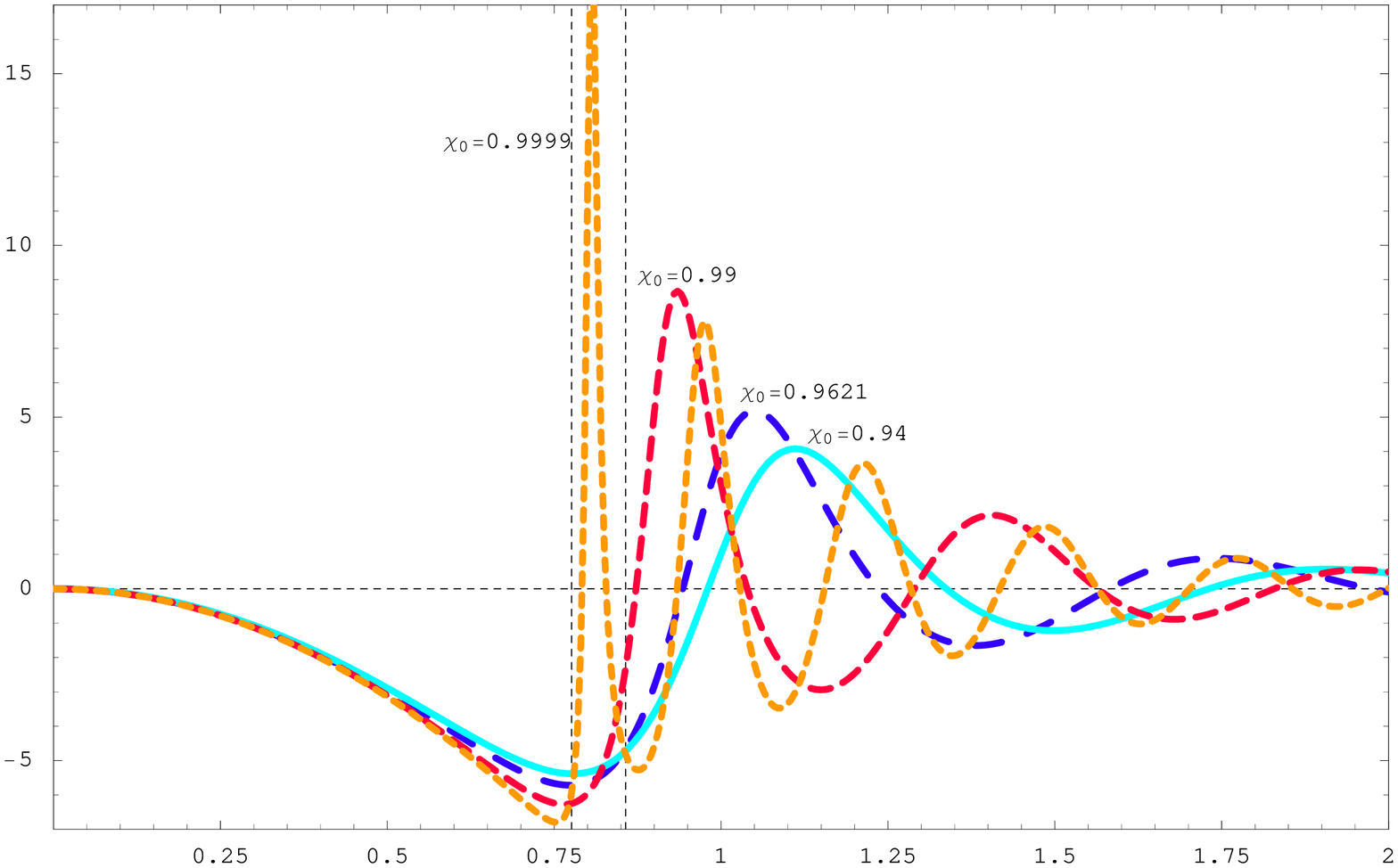}
 \end{tabular}
\caption{The finite temperature part of the vector spectral
function, \ie $\rn-\nf \nc \omega^2/4\pi$, in units of $\nf \nc
T^2/4$,  versus $\tom=\omega/2\pi T$ for various values of $\chi_0$
corresponding to different values of $m=\mbar/T$. The upper plot
shows values of $\chi_0$ corresponding to temperatures above the
phase transition while the lower plot is for values of $\chi_0$ past
the transition. The vertical dotted lines represent the mass of the
lowest and first excited vector mesons in the low temperature
(Minkowski) phase for a near-critical Minkowski embedding.}
\label{gaugePlots} }
The spectral function was evaluated using the numerical solutions
for $E_x$ and eq.~\reef{spectralGauge}.  In the high frequency
limit, the spectral function asymptotes to $\nf \nc \omega^2/4\pi$
-- see appendix \ref{high}. Figure \ref{gaugePlots} provides plots
of the finite-temperature part of the spectral 
function,\footnote{In using the wording `finite temperature part of
the spectral function,' we are adopting the language used previously
for ${\cal N}=$ SYM in section \ref{simple} and
ref.~\cite{Kovtun:2005ev}.  In the present case, this refers to the
spectral function minus its high frequency asymptotics.}
$\rn(\omega)-\nf \nc \omega^2/4\pi$, for various D7-brane
embeddings, specified by $\chi_0=\chi(\rho =1)$ (or equivalently by
$m$).  The upper plot shows the finite temperature part of the
spectral function for temperatures above the phase transition:
$\chi_0=0$ ($m=0$), $\chi_0=0.1$ ($m=0.1667$), $\chi_0=0.5$
($m=0.8080$), $\chi_0=0.8$ ($m=1.2026$), $\chi_0=0.94$ ($m=1.3059$)
-- the last of these corresponds to $T/\mq$ for the phase
transition. Note that the $\chi=0$ and $\chi_0=0.1$ lines are
virtually coincident. The lower plot shows the finite temperature
part of the spectral function for values of $\chi_0$ corresponding
to black hole embeddings after the phase transition, \ie along the
lines $A_1$ to $A_2$ and $A_2$ to $A_3$ on the black hole branch in
fig.~\ref{action}. Note that as $\chi_0$ approaches 1, the finite
temperature part of the spectral function displays high peaks. It is
interesting to note that as $\chi_0 \to 1$, as well as growing
sharper, the peaks in the spectral function become more closely
spaced and move towards lower frequencies.  For example, the peaks
in the $\chi_0=0.9999$ line are much more closely spaced than those
in the $\chi_0=0.99$ line.

It is interesting to compare the positions of these peaks to the
masses of the lowest vector mesons in the Minkowski phase
\cite{rowanThesis}. The vertical dotted line at $\tom \simeq 0.776$
represents the mass of the lowest vector meson for a Minkowski
embedding very close to the critical embedding. It seems that the
position of the first peak of the spectral function is converging to
a very similar value as $\chi_0\rightarrow1$ -- certainly for
$\chi_0 =0.9999$, the first peak is very close to $\tom \simeq
0.776$. The second vertical dotted line at $\tom \simeq 0.857$
represents the mass of the first excited vector meson ($n=1,\ell=0$)
as the Minkowski embedding approaches the critical solution. In this
case, it is likely that the second peak in the spectral function is
approaching this value, but certainly it is not converging on this
position as rapidly as the first peak. 
From the supergravity perspective, it is natural that these peaks
are converging on the Minkowski phase spectrum, as described above,
because near the critical embedding both the Minkowski and black
hole embeddings of the D7-brane will be nearly identical except for
very near the event horizon. The small horizon that still appears in
the induced geometry of the near-critical black hole embeddings
ensures, however, that there is a small imaginary frequency
component of the quasinormal fluctuations.

Note that in the Minkowski phase, $\tom \simeq 0.98$ corresponds to
the mass of the lowest vector meson just at the phase transition.
This is significantly above both masses quoted above near the
critical solution -- recall that in general the meson masses
decreased as the critical embedding was approached \cite{long}.
Further then, this mass does not seem to be correlated with the
positions of the spectral peaks for the black hole embedding at the
phase transition, beyond being in the same general range.

The high peaks in the spectral function for $\chi_0 \to 1$ may be
interpreted in terms of quasiparticle states because their width
$\Gamma$ and is much less than their frequency $\Omega$: $\Gamma \ll
\Omega$. Appendix \ref{schroe} presents a complementary discussion
which reaches the same conclusion. The appendix only explicitly
discusses the pseudoscalar channel but the results for the vector
are almost identical. In particular, the effective potential shown
in appendix C develops a finite barrier at intermediate values of
the radius as $\chi_0 \to 1$. This suggests the existence of
metastable states in the corresponding Schroedinger problem which,
as discussed in the appendix, would correspond to quasinormal
frequencies with $\Gamma \ll \Omega$ in this regime.

\subsubsection{Charged vectors}\label{chargev}

The ${\cal N}=2$ gauge theory under study here has an internal
$SO(4)=SU(2)\times SU(2)$ global symmetry, which is dual to
rotations on the D7-brane's internal $S^3$. The vector modes which
are considered above are all singlets under this symmetry. However,
these operators only correspond to the lowest dimension operators in
an infinite family of vector operators transforming in the
($\ell/2,\ell/2$) representation of the internal symmetry
\cite{meson}. As outlined in appendix \ref{diction}, these operators
are built up by combining the adjoint hypermultiplet fields
(scalars) with the fundamental fields appearing in the singlet
operators.

Evaluating the spectral function for these vectors with $\ell\ne0$
follows closely the analysis in the previous section and so we only
present the salient steps here. Of course, the first step is to
consider an expansion of the world-volume vector in terms of
spherical harmonics on the $S^3$,
\beq A_\mu = \sum_\ell \mathcal{Y}^\ell (S^3) \, A_\mu^\ell (\rho,
x^\mu)\,, \labell{newsum}\eeq
with
\beq \nabla^2_\mt{[3]}\mathcal{Y}^\ell
=-\ell(\ell+2)\,\mathcal{Y}^\ell \,, \labell{why}\eeq
where $\nabla^2_\mt{[3]}$ is the Laplacian on the unit
three-sphere.\footnote{Of course, the spherical harmonics for a
given $\ell$ are also labeled by two further $SU(2)$ quantum
numbers, but we drop these as they are irrelevant in the following.
Implicitly, our normalization is such that
$\mathcal{Y}^{\ell=0}_{m=0,n=0} (S^3)=1$ and so $ \int d^3\Omega
\sqrt{h_3}\, \mathcal{Y}^{*\ell'}_{m'n'}
\mathcal{Y}^{\ell}_{mn}=2\pi^2\, \delta_{\ell \ell'}\,
\delta_{mm'}\,\delta_{nn'}$.} Examining the eight-dimensional
Maxwell equations arising from eq.~\reef{gaugeAction}, one finds
that with $\ell\ne0$ (and $T\ne0$) $A^\ell_\rho$ cannot be set to
zero in general. Hence the general analysis becomes somewhat more
elaborate. However, if we focus on spatially independent
fluctuations, \ie the $q\to0$ limit above, then both
$A^\ell_{\rho,t}$ decouple and the calculations are greatly
simplified. In this case then, the analog of eq.~\reef{fire1}
becomes
\beq S_\ell = -\frac{\nf \nc T^2}{2^6\pi^2} \int
\frac{d\omega}{\omega^2} \left[\frac{f \rho^3
(1-\chi^2)^2}{\sqrt{1-\chi^2+\rho^2 \dot{\chi}^2}}
E^\ell_{i}(\rho,-k) \partial_\rho E^\ell_{i}(\rho,k)\right]^{\rho
\to \infty}_{\rho \to 1}\, , \labell{fire2}\eeq
where $i$ is summed over $x,y,z$ and $E_{i}^\ell \equiv{\omega}
A^\ell_{i}$.  Recall that with vanishing spatial momenta, there is
no distinction between longitudinal and transverse electric fields
in the language of section \ref{prelude}.

Examining the asymptotic behaviour of any of the electric field
components, we write
\beq E^\ell_{i}(\omega,\rho) = E^\ell_0(\omega) \frac{(\pi
T)^\ell}{2^{\ell/2}} \rho_\infty^\ell  \frac{E_{\ell,
\omega}(\rho)}{E_{\ell,\omega}(\rinf)} \, , \labell{outer}\eeq
where it is understood that eventually the limit $\rinf \to \infty$
will be taken.  Note the factor of $\rho_\infty^\ell$ required to
obtain the correct asymptotic behaviour -- see appendix
\ref{diction}. As above, taking variations of $E^\ell_0(k)$ then
yields the the flux factor $\mathcal{F}_\ell$ for $E^\ell_{i}$. This then
leads to the following expression for the spectral function for
$A_i$:
\beq \rn_{ii}^\ell(\omega)\equiv-2 \textrm{Im} G_{ii}^\ell(\omega)=
- \frac{\pi^{2\ell}}{2^{\ell+2}} \nf \nc T^{2\ell+2} \textrm{Im}
\left[\rho^{2\ell+3} \frac{\partial_\rho
E_{\ell,\omega}(\rho)}{E_{\ell,\omega}(\rho)} \right]_{\rho \to
\infty}\,  \labell{spectralGauge5} \eeq
with no sum on $i$. Instead for any value of $i=x,y,z$, the spectral
functions are identical, \ie
$\rn_{xx}^\ell(\omega)=\rn_{yy}^\ell(\omega)=\rn_{zz}^\ell(\omega)$,
because we are limiting our analysis here to the case of vanishing
spatial momentum. Hence in the following, we denote these spectral
functions by $\rn_\ell(\omega)$.

In order to evaluate the spectral functions, we must solve the
Maxwell equations arising from the eight-dimensional action
\reef{gaugeAction}. Expressed in terms of the electric field
components, the relevant equation of motion is
\beqa \ddot{E}_{\ell,\omega} &&+
\left[\frac{4\dot{f}}{\tilde{f}\,f}+\frac{f}{\tilde{f}^2}
\frac{\sqrt{1-\chi^2+\rho^2 \dot{\chi}^2}}{\rho^3
(1-\chi^2)^2}\partial_\rho \left( \frac{\tilde{f}^2 \rho^3
(1-\chi^2)^2}{f\sqrt{1-\chi^2+\rho^2 \dot{\chi}^2} } \right)
\right]\dot{E}_{\ell,\omega} \nn
&& +\frac{1-\chi^2+\rho^2
\dot{\chi}^2}{\rho^2(1-\chi^2)^2}\left[\frac{8(1-\chi^2)\tilde{f}}{\rho^2
f^2} \tom^2-\ell (\ell+2)\right] E_{\ell,\omega} =0 \, .
\labell{Eeq22} \eeqa
We now proceed to compute the spectral function $\rn_\ell(\omega)$,
first for massless quarks ($m=0$) and then for quarks with a finite
mass.

Recall that the case of massless quarks corresponds to the
equatorial embedding of the D7-branes for which $\chi (\rho) =0$. In
the previous section, we noted that for the $\ell=0$ vector, the
calculation of $\rn(\omega)$ then becomes the same as that in our
example of section \ref{simple}. In particular, an analytic solution
can be found for $q=0$ because it is possible to solve \reef{Eeq2}
exactly when $\chi=0$. Here we show that in fact the general
equation \reef{Eeq22} for any $\ell$ has an analytic solution in
this case of massless quarks.

Setting $\chi=0$ and making the change of variables $\xb= 1-2/\rho^2
\tilde{f} = 1-2\rho^2/(1+\rho^4)$, the equation for the fluctuation
$E_{\ell,\omega}(\xb)$ is
\begin{equation}
E_{\ell,\omega} '' + {f'\over f} \, E_{\ell,\omega} ' + \left[
{\wn^2\over (1-\xb) f^2}\, - {\ell(\ell+2)\over 4 (1-\xb)^2
f}\right] \, E_{\ell,\omega} = 0\, , \label{eqE_il}
\end{equation}
where the prime denotes a derivative with respect to $\xb$. As in
eq.~\reef{gensole}, the solution is given by\footnote{Note that near
the horizon $\xb\simeq2(\rho-1)^2$ and so the small $\xb$ behaviour
here is consistent with the boundary condition at the horizon
discussed for the numerical solution.}
\begin{equation}
E_{\ell,\omega}(\xb) = \xb^{-i \wn/2}\, (2-\xb)^{-\wn/2}\, F(\xb)\,,
\labell{gensole2}
\end{equation}
where the regular function $F(\xb)$ is a straightforward
generalization of the result \reef{eqF_x_1}:
\begin{equation}
F(\xb) = (1-\xb)^{{(1+i)\wn \over 2}} \ofo \left( 1 +{\ell\over 2} -
{ (1+i) \wn \over 2}\,, -{\ell\over 2} - { (1+i) \wn \over 2}\, ; 1-
i \wn; {\xb\over 2 (\xb-1)}\right)\,. \label{eqF_x_1_l}
\end{equation}
The spectral function is then given by
\begin{equation}
 \rn_\ell (\omega)= - \lim_{\epsilon
\rightarrow 0} \, \frac{\pi^{2\ell}}{2^\ell}{\nf \nc T^{2\ell+2}} \,
\textrm{Im} f(\xb) \left( {1+\sqrt{f}\over 1-\xb}\right)^\ell
{E(\xb,-\wn)\over E(1-\epsilon,-\wn)}
 {E'(\xb,\wn)\over E(1-\epsilon,\wn)}\,.
\label{sxxx}
\end{equation}
The right hand side of eq.~(\ref{sxxx}) is independent of the radial
coordinate \cite{Son:2002sd} and thus can be computed at any value
of $\xb$, \eg at $\xb=0$. We obtain
\begin{equation}
 \rn_\ell (\omega)= {2^{\ell}\pi^{2\ell-1}\over (\ell!)^2}\, {\nf \nc T^{2\ell+2}}\,
 {\sinh{\pi \wn}}\,
\Biggl| \Gamma \left( 1 + {\ell\over 2} - {\wn \over 2} - {i \wn
\over 2}\right) \Gamma \left( 1 + {\ell\over 2} + {\wn \over 2} -
 {i \wn \over 2}\right)\Biggr|^2\,.
\label{sxxl}
\end{equation}
Eq.~(\ref{sxxl}) shows that the poles of the retarded correlator
corresponding to  $\rn_\ell (\omega)$ are located at
\begin{equation}
\wn = \pm \left( n + 1 + {\ell\over 2}\right) \left( 1 \mp
i\right)\,, \qquad n= 0,1,...\,.
\end{equation}
Note that there is an interesting degeneracy in the positions of
these quasinormal modes in that their position only depends on
$n+\ell/2$. This is reminiscent of the unexpected degeneracy found
in \cite{meson}, where the meson masses only depended on the
combination $n+\ell$ (at $T=0$). For $\ell=0$, eq.~(\ref{sxxl})
reduces (up to the normalization)
 to the result (\ref{sfmod}).
For odd and even $\ell>0$, respectively, eq.~(\ref{sxxl}) can be
written in the form
\beqar
 \rn_{\ell=2n-1} (\omega) &=& \frac{\pi^{2\ell}}{2^\ell} {\nf \nc T^{2\ell+2}}\, { 2^{4n}\Gamma^4 (n+1/2)
\over 2 \pi  [(2n-1)!]^2}\, {  \sinh{\pi \wn}\over \cosh{\pi \wn} +
\cos{\pi \wn}}
\prod_{k=1}^n \left( 1 + {4\wn^4\over  (2k-1)^4}\right), \\
 \rn_{\ell=2n} (\omega) &=&  \frac{\pi^{2\ell}}{2^\ell}{\nf \nc T^{2\ell+2}}\, { 2^{4n}(n!)^4\over [(2n)!]^2}\,
{ \pi \wn^2 \sinh{\pi \wn}\over \cosh{\pi \wn} - \cos{\pi \wn}}\,
\prod_{k=1}^n \left( 1 + {\wn^4\over 4 k^4}\right),
\eeqar
where $n=1,2,...$. The asymptotics of the spectral function for
large and small frequency are
\begin{eqnarray}
\rn_\ell (\omega) &=& \frac{\pi^{2 \ell+1}}{(\ell!)^2}
 {\nf \nc T^{2\ell+2}}\, \wn^{2\ell+2} \left( 1 + (-1)^\ell 2
 e^{-\pi \wn} \cos{\pi \wn} \right)\left( 1+ O(1/\wn^4)\right)\,, \labell{lfal} \\
&& \qquad \qquad \qquad \qquad \qquad \qquad \qquad \qquad \qquad \qquad \qquad \qquad \wn \rightarrow \infty\,, \nonumber\\
\rn_\ell (\omega) &=& \frac{2^\ell \pi^{2 \ell}}{(\ell !)^2} \nf \nc
T^{2\ell+2} \Gamma^4(1+{\ell\over 2}) \, \wn \,,
 \qquad  \wn \rightarrow 0 \,. \labell{sfal}
\end{eqnarray}

In particular, we have the $\ell=1$ spectral function:
\begin{equation}
\rn_1 (\omega) = \frac{\pi^3}{4}\nf \nc T^4 {(1+4\wn^4)\sinh{\pi
\wn}\over \cosh{\pi \wn} + \cos{\pi \wn}}\,. \label{sfl=1}
\end{equation}
The large frequency asymptotics of  $\rn_1 (\omega)$ is
\begin{equation}
\rn_1 (\omega)\rightarrow \bar{\rn}_1(\omega) = \frac{\pi^2}{2}\nf
\nc T^4 \left[ 2\pi \wn^4 \left(  1 - 2 e^{-\pi \wn}
 \cos{\pi \wn} \right) + {\pi\over 2} \right]\,,
\label{asymwl}
\end{equation}
where we have dropped $\mathcal{O}(e^{-\pi \wn})$ terms. Thus for
sufficiently large values of $\wn$ the finite temperature part of
the spectral function, $\rn_1 (\omega) - \pi^3 \nf \nc T^4 \wn^4$,
exhibits damped oscillations around $\pi^3 \nf \nc T^4 /4$ -- see
fig.~\ref{gaugePlotslll}. Note that for $\ell\geq 2$ the  finite
temperature part of the spectral function asymptotes $\wn^{2\ell+2}$
for large $\wn$ and thus the oscillatory behaviour again becomes a
subdominant effect.
\FIGURE{
 \includegraphics[width=0.8 \textwidth]{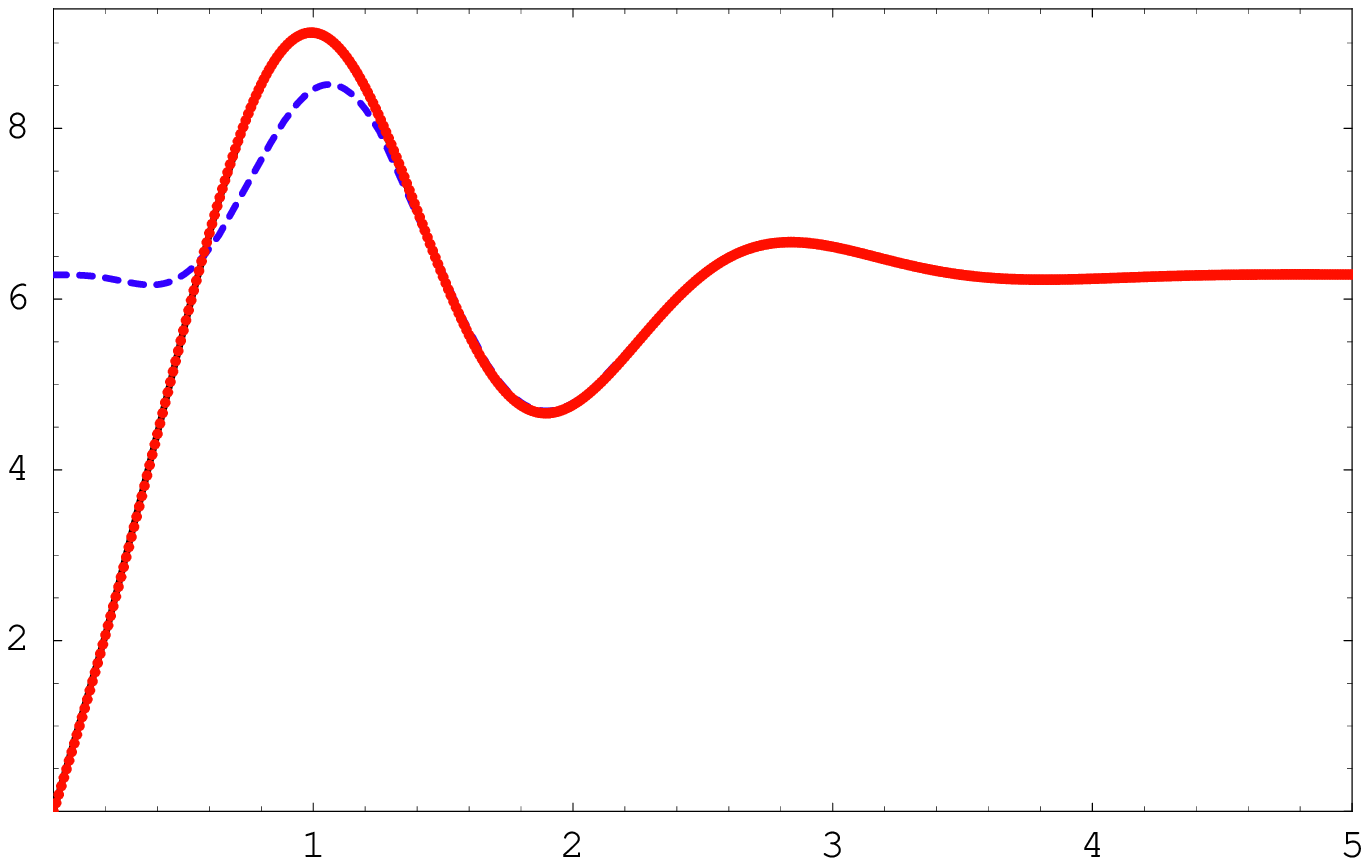}
\caption{The finite temperature parts of the $\mq=0$ ($\chi_0=0$),
$q=0$, $\ell=1$ vector spectral function ($\rn_1-\nf \nc \omega^4/16
\pi $) and of its high frequency asymptotics \reef{asymwl}
($\bar{\rn}_1-\nf \nc \omega^4/16 \pi $) (dashed blue line), in
units of $\pi^2 \nf \nc T^4/8$,  versus $\tom=\omega/2\pi T$. Note
the figure also demonstrates the precise agreement between the
numerical results (red dots) and the exact result (solid black line,
which is essentially invisible above).} \label{gaugePlotslll} }

As in the previous section, for massive quarks ($\chi_0\ne 0$), both
the embedding equation \reef{psieom} and the vector equation of
motion \reef{Eeq22} must be solved numerically. Solving for
$E_{\ell,\omega}$ requires special attention to the boundary
conditions near the horizon ($\rho \to 1$). As for the $\ell=0$
case, the appropriate incoming wave conditions are imposed by taking
$E_{\ell,\omega} (\rho) = (\rho-1)^{-i \tom } F(\rho)$ with $F(1)=1$
and $\partial_\rho F (1) = i\tom/2$.

The vector spectral function for $\ell=1$ is shown for various
values of $\chi_0$ in fig.~\ref{vecPlotl1}. For all values of
$\chi_0$, the $\ell=1$ spectral functions approach $\nf \nc
\omega^4/ 16 \pi$  at large $\omega$ -- see appendix \ref{high}.
While this common behaviour is not clear in fig.~\ref{vecPlotl1}, it
can be seen by going to larger $\tom$. Note that the spectral
functions in the upper plot, which correspond to values of the
$\mq/T$ above the phase transition, seem to be essentially
featureless. In contrast, the lower plot shows that as the critical
embedding is approached with $\chi_0 \to 1$ some peaks are appearing
in the spectral function. The masses of the lowest two $\ell=1$
vector mesons in the low temperature phase for a near-critical
Minkowski embedding have been included in this plot as well. While
these lines lie close to the first peak for the $\chi_0=0.9999$
spectral function, the peaks do not seem to be converging to these
positions nearly as rapidly as was seen for $\ell=0$.
\FIGURE{
\begin{tabular}{c}
 \includegraphics[width=0.8 \textwidth]{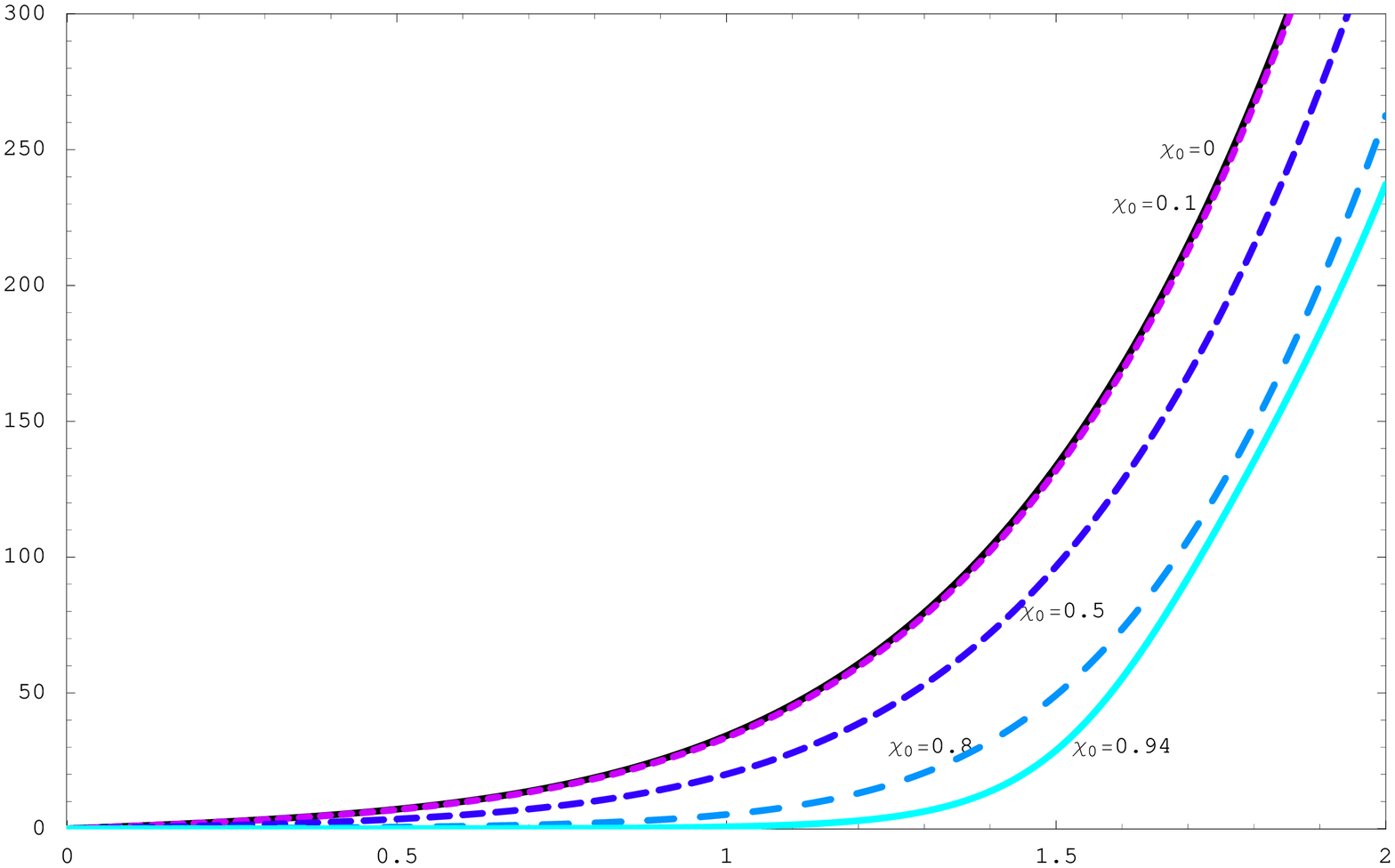}\\
 \includegraphics[width=0.8 \textwidth]{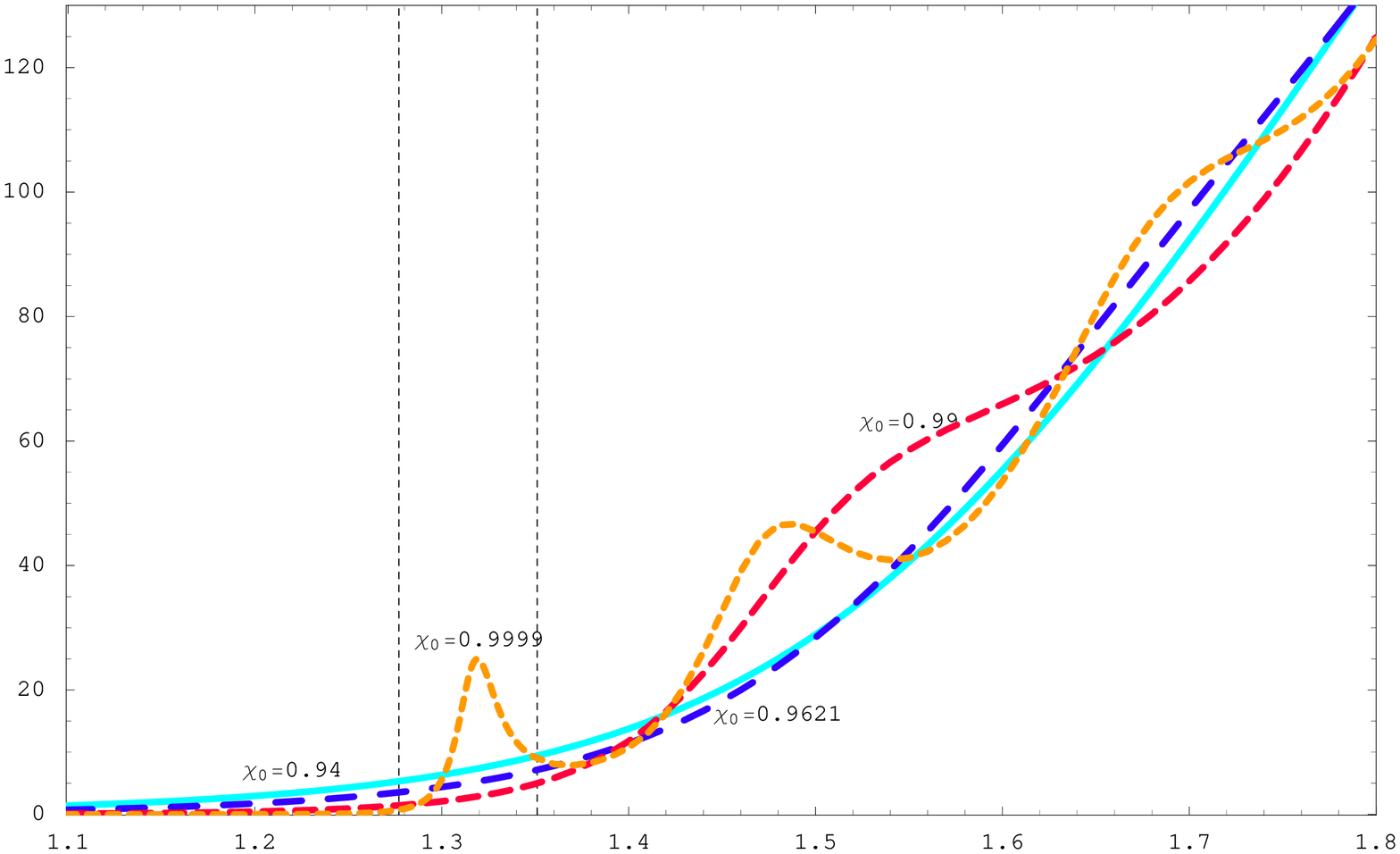}
\end{tabular}
\caption{The vector spectral function for $\ell=1$ in units of
$\pi^2 \nf \nc T^4/8$ versus $\tom$.  The upper plot shows values of
$\chi_0$ corresponding to temperatures above the phase transition
while the lower plot is for values of $\chi_0$ past the phase
transition.  In the lower plot we focus on values of $\tom$ for
which the spectral function shows structure.  As usual, the vertical
dotted lines represent the mass of the lowest and first excited
vector mesons for $\ell=1$ in the low temperature (Minkowski) phase
for a near-critical Minkowski embedding.} \label{vecPlotl1}}

The peaks in the spectral function for $\chi_0 \to 1$ may again be
interpreted in terms of quasiparticle states when their width
$\Gamma$ and is much less than their frequency $\Omega$: $\Gamma \ll
\Omega$. Hence, as discussed for $\ell=0$, it appears that the
quasinormal frequencies are approaching the real axis in this
regime. However, we stress that this approach is occurring much more
slowly for the $\ell>0$ modes. In particular, the spectral function
remains essentially structureless for $\chi_0=0.94$, which
corresponds to the phase transition between the black hole and
Minkowski embeddings. Therefore the mesonic states corresponding to
the higher-$\ell$ vector operators dissociate immediately at the
phase transition.

Note the complementary discussion in Appendix \ref{schroe} would
lead to similar conclusions. In particular, a barrier in the
effective potential develops as in the $\ell=0$ analysis but one
must go to values of $\chi_0$ much closer to one when $\ell>0$.
Hence metastable states in the corresponding Schroedinger problem
would only appear for $\chi_0$ in this regime very close to one.


\subsection{Scalars}\label{scalars}

We now turn to scalar and pseudoscalar excitations of the fundamental
fields. In the dual gravity picture, these correspond to scalar
fluctuations of the D7-brane probes in the black D3 geometry
\reef{D3geom} about the fiducial embedding given by
$\theta_v(\rho)$:
\beq
\theta (\sigma^a) = \theta_v(\rho) + \delta \theta (\sigma^a)\, ,\qquad \phi
= 0 + \delta \phi (\sigma^a)\,
\eeq
where $\sigma^a$ denotes the D7-branes' worldvolume coordinates.
Appendix \ref{diction} describes the holographic dictionary relating
the scalar $\delta \theta$ and the pseudoscalar $\delta \phi$ to the
corresponding gauge theory operators. We present the analysis for
general $\ell$ modes but our results will focus on the spectral
functions of lowest dimension operators with $\ell=0$.

The pull-back of the bulk metric \reef{D3geom} to the D7 worldvolume is:
\beq
ds^2 = ds^2 (g) - {2 L^2 \dot{\chi} \over \sqrt{1-\chi^2}}
\partial_a (\delta \theta) dx^a d\rho+L^2 \left[\partial_a
(\delta \theta)\partial_b (\delta \theta) + \chi^2 \partial_a
(\delta
\phi)\partial_b (\delta \phi) \right]dx^a dx^b \eeq
where
\beq ds^2(g)= \frac{1}{2}\left({\om \rho \over L} \right)^2 \left[-
{f^2 \over \tilde{f}} dt^2+\tilde{f}dx_3^2 \right] +{L^2 \over
\rho^2} \left[\left( 1+{\rho^2 \dot{\chi}^2 \over 1-\chi^2}\right)
d\rho^2 +\rho^2 \sin^2 (\theta_v+\delta \theta) d\Omega_3^2 \right].
\labell{aMet} \eeq
As before, we've put $\chi(\rho) = \cos \theta_v(\rho)$. Using the
DBI action and retaining terms only to quadratic order in the
fluctuations, the Lagrangian density is
\beqa \mathcal{L} &=& \mathcal{L}_0 + -{\nf T_\mt{D7} \om^4\over 4}
\sqrt{h_3}\left(\tilde{\mathcal{L}}_1+\tilde{\mathcal{L}}_2 \right)
-{\nf T_\mt{D7} \om^4\over 4} \sqrt{h_3} \,
\rho^3 f \tilde{f} \sqrt{1-\chi^2+\rho^2 \dot{\chi}^2}\,   \labell{scalarLag} \\
&& \times \left[ -{3
\over 2} {1-\chi^2 \over
1-\chi^2+\rho^2 \dot{\chi}^2}\left( \delta \theta \right)^2+{L^2 \over 2}(1-\chi^2) g_v^{ab} \left( {(1-\chi^2)
\partial_a(\delta \theta) \partial_b(\delta \theta) \over 1-\chi^2
+\rho^2 \dot{\chi}^2}+\chi^2 \partial_a (\delta \phi)\partial_b (\delta \phi)
 \right) \right] \nonumber
\,
\eeqa
where $g_v^{ab}$ is the metric \reef{aMet} with $\delta \theta =0$,
$\mathcal{L}_0$ is the Lagrangian density for the fiducial embedding
$\chi$ (given in equation \reef{act2s}), and the boundary terms $
\tilde{\mathcal{L}}_1$ and  $\tilde{\mathcal{L}}_2$ are
\beqa
\tilde{\mathcal{L}}_1 &=& \partial_\rho
\left[-\frac{\rho^5 f\tilde{f}(1-\chi^2)^{3/2}
\dot{\chi}}{\sqrt{1-\chi^2+\rho^2 \dot{\chi}^2}}
\delta \theta \right]\\
\tilde{\mathcal{L}}_2 &=& \partial_\rho
\left[-\frac{3}{2}\frac{\rho^5 f\tilde{f}(1-\chi^2)\chi
\dot{\chi}}{\sqrt{1-\chi^2+\rho^2 \dot{\chi}^2}}
(\delta \theta)^2 \right]\,.
\eeqa
We eliminated terms linear in $\delta \theta$ by integrating by
parts and using the equation of motion \reef{psieom} for $\chi$.

The equations of motion for the fluctuations follow from \reef{scalarLag} as
\beq
\partial_a\left[\sqrt{h_3} \,\rho^3 f \tilde{f} (1-\chi^2) \chi^2
\sqrt{1-\chi^2+\rho^2 \dot{\chi}^2} \, g_v^{ab} \, \partial_b
(\delta \phi) \right]=0 \labell{phieom}\eeq
for $\delta \phi$ and
\beq L^2 \partial_a\left[ {\sqrt{h_3} \,\rho^3 f \tilde{f}
(1-\chi^2)^2 \over \sqrt{1-\chi^2+\rho^2 \dot{\chi}^2}} \, g_v^{ab}
\, \partial_b (\delta \theta)  \right] + 3 {\sqrt{h_3} \,\rho^3 f
\tilde{f} (1-\chi^2) \over \sqrt{1-\chi^2+\rho^2
\dot{\chi}^2}}\delta \theta =0 \labell{thetaeom}\eeq
for $\delta \theta$.

\subsubsection{Pseudoscalar $\delta \phi$}\label{pseudo}

The relevant
portion of the action \reef{scalarLag} for the pseudoscalar $\delta \phi$ is
\beq
S_{\delta \phi} = -\frac{T_\mt{D7} \nf \om^4 L^2 }{8} \int d^8 \sigma
\, \partial_a \left[\sqrt{h_3}\,\rho^3 f \tilde{f} (1-\chi^2) \chi^2
\sqrt{1-\chi^2+\rho^2 \dot{\chi}^2} \, g^{ab}_v \, \delta \phi \,
\partial_b \delta \phi \right] \labell{quadLphi}
\eeq
where we've integrated by parts and used the equation of motion
\reef{phieom}.  To evaluate the spectral function we only need the
complex part of \reef{quadLphi} and hence in the following we retain
only the term involving the $\rho$ derivative.  Expanding the
fluctuation in terms of spherical harmonics on the $S^3$ of unit
radius,
\beq
\delta \phi = \sum_\ell \mathcal{Y}^\ell (S^3) \delta \phi_\ell (\rho, x^\mu)\,
, \labell{pseudoLs}
\eeq
the term needed to evaluate the spectral function for the $\ell$th mode is
\beq
S_{\delta \phi_\ell} = -\frac{T_\mt{D7} \nf \om^4 \Omega_3}{8}  \int d^4 x   \,
\left[\frac{\rho^5 f \tilde{f} (1-\chi^2)^2 \chi^2}{
\sqrt{1-\chi^2+\rho^2 \dot{\chi}^2}}  \, \delta \phi_\ell
\, \partial_\rho \delta \phi_\ell \right]_{\rho \to \infty} . \labell{boundphi}
\eeq
We take the Fourier transform of $\delta \phi_\ell $ with $k=(-\omega, q,0,0)$,
\beq \delta \phi_\ell (\rho,x^\mu) = \int \frac{d\omega dq}{(2\pi)^2} e^{-i\omega t +i
qx} \delta \phi_{\ell } (\rho,k)\, , \labell{fourierPhi} \eeq
and write
\beq
\delta \phi_{\ell }(\rho, k) = \delta \phi_\ell ^0 (k) \frac{(\pi T)^\ell}{2^{\ell/2}}\rho_\infty^\ell
\frac{\mathcal{P}_{\ell, k }(\rho)}{\mathcal{P}_{\ell , k}(\rinf)} \labell{normPhi}
\eeq
where the limit $\rinf\to \infty$ will eventually be taken.  Note
the factor of $\rho_\infty^\ell$ required to obtain the correct
asymptotic behaviour $\delta \phi_\ell (\rinf,k)=\delta \phi_\ell
^0(k) \rho_\infty^\ell$ -- see appendix \ref{diction}. We can then
define the flux factor for the $\ell$th mode as
\beq
\mathcal{F}_{\phi_\ell} = -\frac{\pi^{2\ell}}{2^{\ell+6}} \lambda \nf  \nc T^{2 \ell+4} \left[\frac{\rho^5 f
\tilde{f}(1-\chi^2)^2 \chi^2}{ \sqrt{1-\chi^2+\rho^2 \dot{\chi}^2}}
\, \frac{\rho_\infty^{2\ell} \, \mathcal{P}_{\ell,-k}(\rho)\, \partial_\rho
\mathcal{P}_{\ell,k}(\rho)}{\mathcal{P}_{\ell,-k}(\rho_\infty)\, \mathcal{P}_{\ell,k}(\rho_\infty)}  \right]_{\rho \to
\infty}.
\eeq
The retarded Green's function is then $G = -2 \mathcal{F}$ \cite{Son:2002sd} from
which we obtain the spectral function $\rn=-2 \textrm{Im }
G$ for $q=0$ as
\beq \rn_{\phi_\ell}(\omega,0)
= -\frac{\pi^{2\ell}}{2^{\ell+4}}  \lambda\nf \nc T^{2\ell+4} m^2
\lim_{\rho \to \infty}\
\textrm{Im }\!\left[\rho^{3+2\ell} \,  \frac{\partial_\rho
\mathcal{P}_{\ell,k}(\rho)}{\mathcal{P}_{\ell,k}(\rho)} \right] \, ,
\labell{gamble1}\eeq
where we have simplified using eq.~\reef{asympD7}.

For the $\ell=0$ mode, the holographic dictionary in appendix
\ref{diction} describes how the variation $\delta \phi^0(k)$
introduced an insertion of the operator $\mq\,{\cal O}_{\phi}$.
Hence to get the spectral function for the dimension-three operator
${\cal O}_\phi$, we should normalize the spectral function by an
extra factor $1/\mq^2$. We expect that this should also hold for
$\ell >0$, in which case the operator ${\cal O}_{\phi_\ell}$ has
dimension $\ell+3$.  Recalling that $m^2 = 4 \mq^2/\lambda T^2$, we
arrive at:
\beq \tilde{\rn}_{\phi_\ell}(\omega)=\frac{1}{\mq^2}\rn_{\phi_\ell}(\omega,0) = -\frac{\pi^{2\ell}}{2^{\ell+2}} \nf \nc T^{2 \ell+2}
 \lim_{\rho \to \infty}\ \textrm{Im
}\!\left[\rho^{3+2 \ell}\, \frac{\partial_\rho
\mathcal{P}_{k}(\rho)}{\mathcal{P}_{k}(\rho)} \right]\ .
\labell{gamble2} \eeq

Using \reef{fourierPhi} and \reef{normPhi}, the
equation of motion \reef{phieom} becomes
\beqa &&
\partial_\rho \left[{ \rho^5 f\tilde{f} (1-\chi^2)^2 \chi^2 \over
\sqrt{1-\chi^2+\rho^2 \dot{\chi}^2}} \partial_\rho \mathcal{P}_{\ell, k} \right] \nonumber\\
&&\quad \quad \quad
+\rho^3 f\tilde{f} \chi^2 \sqrt{1-\chi^2+\rho^2 \dot{\chi}^2}
\left[\frac{8(1-\chi^2)}{\rho^2 \tilde{f}}\left( \frac{\tilde{f}^2 }{f^2}
\tom^2 - \tk^2\right) -\ell (\ell+2) \right]\mathcal{P}_{\ell, k}=0\, . \labell{phieom2}
\eeqa
Near the horizon ($\rho \to 1$) we impose incoming wave boundary
conditions so that taking
\beq
\mathcal{P}_{\ell, k}(\rho) \simeq (\rho-1)^{- i \tom}\left[1+\frac{i
\tom}{2}(\rho-1) + \mathcal{O}(\rho-1)^2 \right] \quad \textrm{ for } \rho \to 1 \, ,
\eeq
we were able to solve \reef{phieom2} numerically to evaluate the
spectral function \reef{gamble2}.  The high frequency asymptotics of
the spectral function are described in appendix \ref{high}.
\FIGURE{
\begin{tabular}{c}
\includegraphics[width = 0.8 \textwidth]{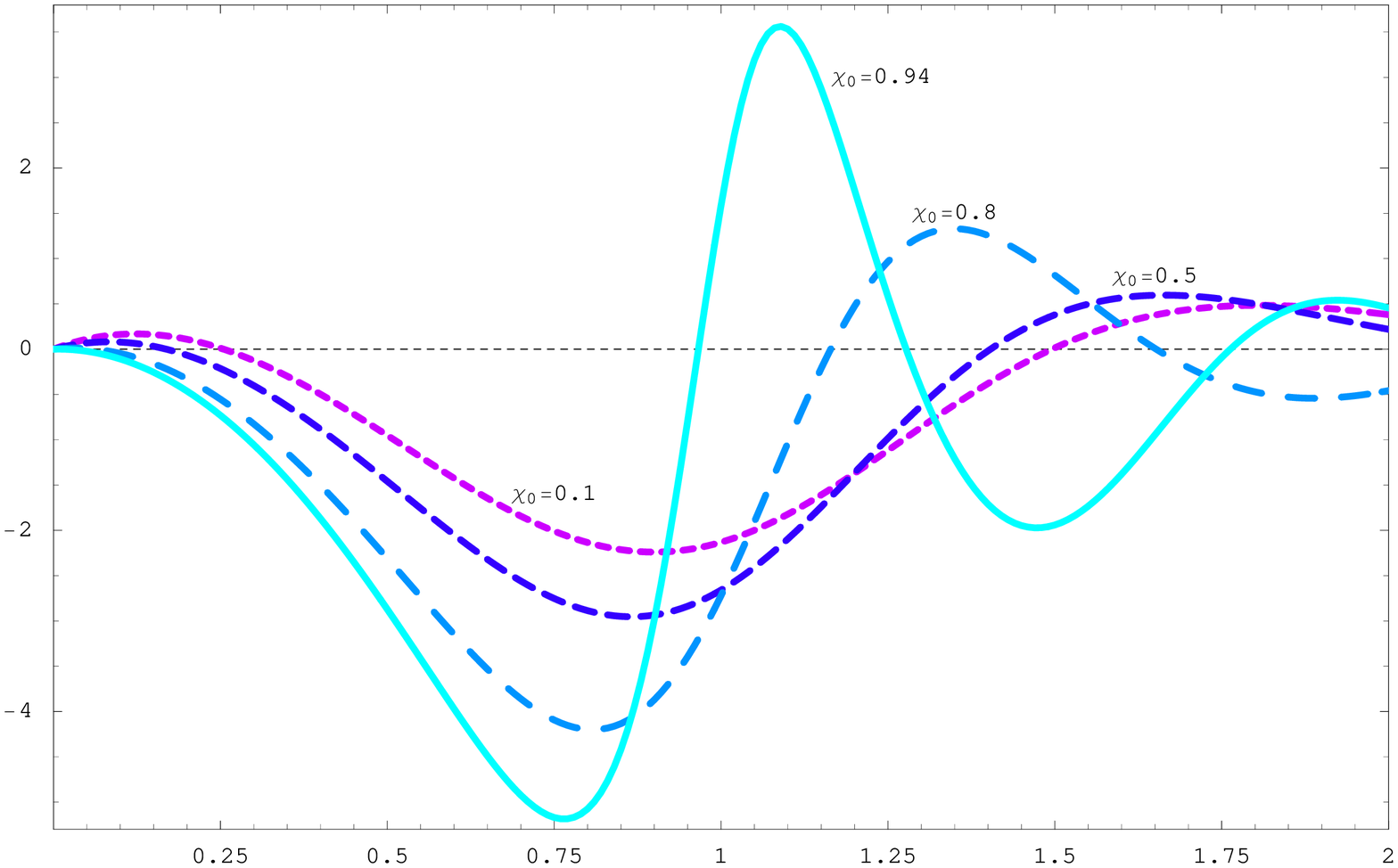} \\
\includegraphics[width =0.8 \textwidth]{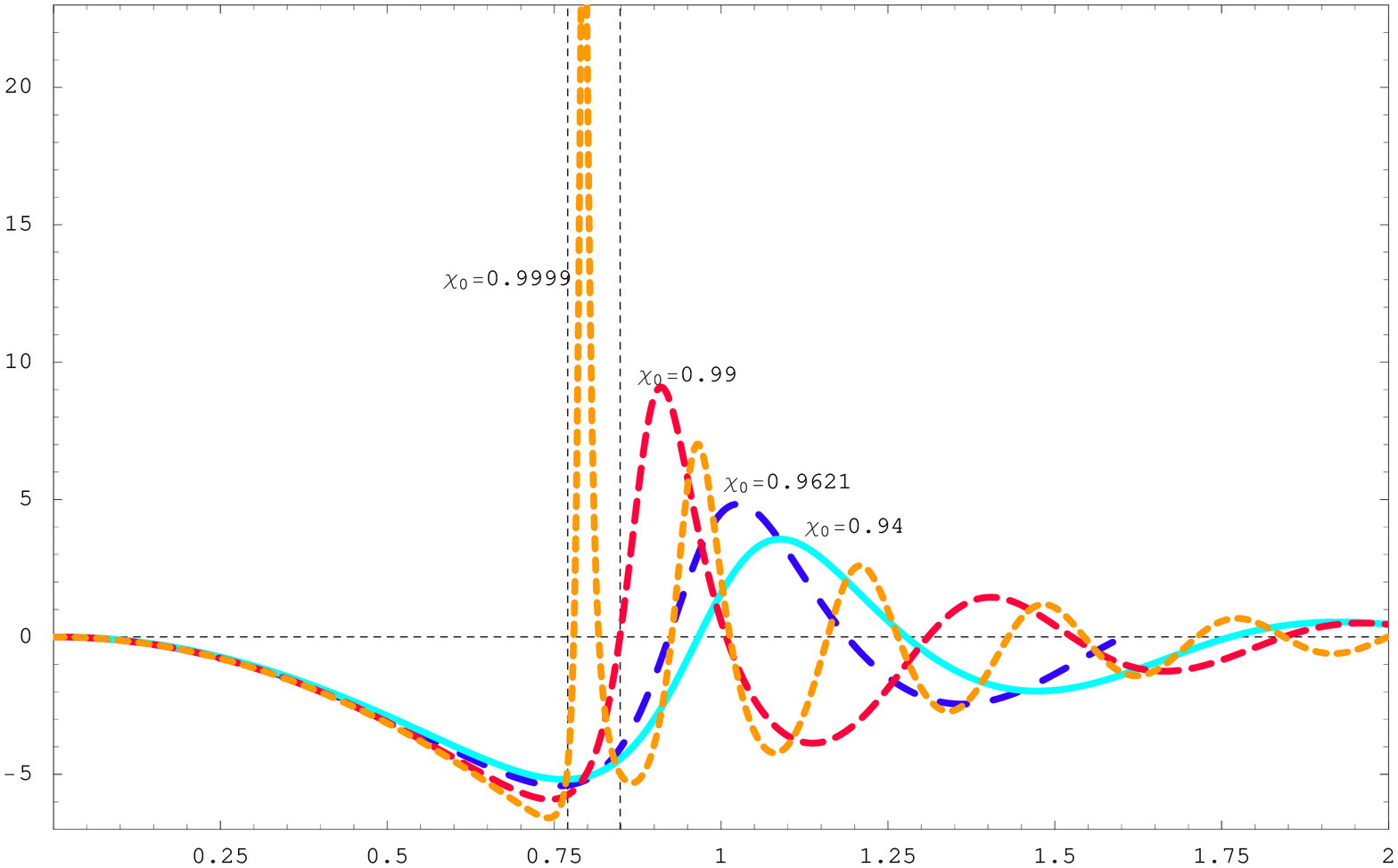}
\end{tabular}
\caption{The finite temperature part of the $\delta \phi$, $\ell=0$, spectral
function, $\tilde{\rn}_\phi - \nf \nc \omega^2/4 \pi$, in units of
$\nf \nc T^2/4$ versus $\tom=\omega/2\pi T$ for various values of
$\chi_0$. The upper plot shows the spectral function for values of $\chi_0$ corresponding to temperatures above the phase transition while the lower plot is for values of $\chi_0$ past the transition.  The vertical dotted lines represent the masses of the lowest two  pseudoscalar mesons for a near-critical Minkowski embedding.} \label{pseudoPlots} }

Figure \ref{pseudoPlots} provides plots of the finite temperature
part of the spectral function, $\tilde{\rn}_\phi - \nf \nc
\omega^2/4 \pi$, for the pseudoscalar $\delta \phi$, $\ell=0$, for
various values  of $\chi_0$. Qualitatively the results are the same
as for the vector spectal function shown in figure
\ref{gaugePlots}.\footnote{In the interests of space we do not
include plots of the pseudoscalar spectral function for $\ell>0$
here, however, the $\ell =1$ plot closely ressembles that for the
vector, shown in fig.~\ref{vecPlotl1}. } The quasiparticle peaks in
spectral function quickly dissipate above the phase transition, \ie
for $\chi_0 < 0.94$. High sharp peaks develop as $\chi_0 \to 1$. As
before, the position of these peaks may be compared with the masses
of the lowest pseudoscalar mesons on the Minkowski branch. The
vertical dotted lines mark the masses ($\tom
\simeq 0.770$ and $0.849$) of the lowest two $\delta \phi$ mesons (with $\ell=0$) for
a near-critical Minkowski embedding.  Note that the first peak in
the $\chi_0=0.9999$ line is nearly centred on the first value of
$\tom$. The second peak of this spectral function also seems to be
converging towards the mass of the next meson.
Hence as in the vector channel, we see that the spectra of these
pseudoscalar fluctuations in the Minkowski and black hole phases
seem to converge as we approach the critical embedding.

As before, the sharp peaks which develop in the spectral function as
$\chi_0$ approaches 1 may be interpreted in terms of quasiparticle
states. Again the complementary discussion of the quasinormal
spectrum, given in Appendix \ref{schroe}, leads to the same
conclusion.

\subsubsection{Scalar $\delta \theta$}\label{scalar}

The derivation of the spectral function for the scalar $\delta
\theta$ is entirely analogous to that for the pseudoscalar.  The
portion of the action for the $\delta \theta$ fluctuations is, from
\reef{scalarLag},
\beqar S_{\delta \theta} &=& -\frac{\nf T_\mt{D7} \om^4 }{4}
\int d^8 \sigma \, \left\{ \sqrt{h_3} \partial_\rho \left[ - \frac{\rho^5 f \tilde{f} (1-\chi^2)^{3/2}
\dot{\chi}} {\sqrt{1-\chi^2 +\rho^2 \dot{\chi}^2}} \delta \theta
-\frac{3}{2} \frac{\rho^5 f \tilde{f} (1-\chi^2)
\dot{\chi}\chi}{\sqrt{1-\chi^2
+\rho^2 \dot{\chi}^2}} (\delta \theta)^2 \right] \right. \\
& &\quad \quad \quad \quad\quad \quad \quad \quad \quad \quad\quad
\quad+ \left. \frac{L^2}{2} \partial_a \left[\frac{\sqrt{h_3} \rho^3 f
\tilde{f}(1-\chi^2)^2}{\sqrt{1-\chi^2 +\rho^2 \dot{\chi}^2}} g_v^{ab}\delta
\theta \partial_b \delta \theta \right] \right\}\,,\eeqar
where we've integrated by parts and used the equation of motion
\reef{thetaeom}. As discussed above, to evaluate the spectral
function we only need the imaginary part of the Green's function and
hence of this action.  Thus, only the $\rho$ derivative term from
the second line is needed.  We expand the scalar in terms of
spherical harmonics on the $S^3$ (as in \reef{pseudoLs}), take the
Fourier transform (as in \reef{fourierPhi}), and express the
$\ell$th mode as
\beq
\delta \theta_\ell (\rho,k) = {\delta \theta^0_\ell (k)}\, \frac{(\pi T)^\ell}{2^{\ell/2}}\rho_\infty^{\ell-1}
\frac{\mathcal{R}_{\ell,k}(\rho)}{\mathcal{R}_{\ell,k}(\rho_\infty)},
\labell{protract1}
\eeq
where we will eventually take the limit $\rho_\infty\to \infty$.
Note that the factor of $\rho_\infty^{\ell-1}$ is inserted to obtain
the correct asymptotic behaviour -- see appendix \ref{diction}.

Following the same procedure as with the pseudoscalars, we identify
\beq \mathcal{F}_{\theta_\ell} = -\frac{\pi^{2 \ell}}{2^{\ell+6}}\lambda\nf \nc T^{2 \ell+4} \left[
\frac{\rho^5 f \tilde{f} (1-\chi^2)^3}{(1-\chi^2+\rho^2
\dot{\chi}^2)^{3/2}} \frac{\rho_\infty^{2\ell-2}\mathcal{R}_{\ell,-k}(\rho)\partial_\rho
\mathcal{R}_{\ell,k}(\rho)}
{\mathcal{R}_{\ell,-k}(\rho_\infty)\mathcal{R}_{\ell,k
}(\rho_\infty)}
 \right]_{\rho
\to \infty}. \eeq
The spectral function then follows as
\beq \rn_{\theta_\ell} (\omega, 0) = - \frac{\pi^{2 \ell}}{2^{\ell+4}}\lambda
\nf \nc T^{2 \ell+4} \lim_{\rho \to
\infty}\ \textrm{Im }\! \left[\rho^{3+2\ell}
\frac{\partial_\rho
\mathcal{R}_{\ell,k}(\rho)}{
\mathcal{R}_{\ell,k}(\rho)}\right]\,,
\labell{thetaSpectral} \eeq
where we've used \reef{asympD7} to simplify.

Now recall $\chi=\cos\theta$ and asymptotically $\chi\simeq m/\rho$
where $m$ is determined by gauge theory quantities in eq.~\reef{Mm}.
Note that asymptotically we can relate a variation in $\theta$ with
a variation in $\chi$: $\delta\chi=-\delta\theta$.  Hence, a
variation of the coefficient of the operator ${\cal O}_m$ in the
gauge theory action (\ie figuratively we might say $\delta\mq(k)$)
corresponds to $(\sqrt{\lambda}\,T/2)\delta \theta_0(k)$ in
eq.~\reef{protract1}.  In the correlator \reef{thetaSpectral} two
factors of $\delta \theta_0$ have been stripped off, so in order to
normalize the correlator so that only the variations of the gauge
theory coefficient are removed, we should multiply by a factor of
$4/\lambda T^2$:
\beq
 \tilde{\rn}_\theta (\omega) =\frac{4}{\lambda T^2}\,\rn_\theta(\omega, 0)
 = - \frac{ \pi^{2\ell}}{2^{\ell+2}} \nf \nc T^{2\ell+2}  \lim_{\rho \to \infty}\ \textrm{Im }\!
\left[ \rho^{3+2 \ell}\frac{ \partial_\rho \mathcal{R}_k(\rho)}{
\mathcal{R}_k(\rho)}\right]\, . \labell{thetaSpectral2}
\eeq

With the Fourier transform of $\delta \theta$ and using the notation
\reef{protract1}, the equation of motion \reef{thetaeom} for $\delta
\theta$ becomes
\beqa &&
\partial_\rho \left[\frac{\rho^5 f \tilde{f} (1-\chi^2)^3}
{(1-\chi^2+\rho^2 \dot{\chi}^2)^{3/2}} \partial_\rho \mathcal{R}_{\ell, k}
\right]\nonumber\\
&&\quad \quad \quad
+\frac{\rho^3 f \tilde{f}(1-\chi^2)}{\sqrt{1-\chi^2+\rho^2
\dot{\chi}^2}} \left[\frac{8(1-\chi^2)}{\rho^2 \tilde{f}}\left( \frac{\tilde{f}^2}{f^2}\tom^2
- \tk^2\right)-(\ell+3)(\ell-1) \right]\mathcal{R}_{\ell, k}=0\,. \labell{thetaeom2}
\eeqa
As with the
vector and pseudoscalar, we set $\tk=0$ and impose incoming wave
boundary conditions at the horizon, requiring that the field behave
as
\beq
\mathcal{R}_{\ell,k}(\rho) \simeq (\rho-1)^{- i \tom}\left[1+\frac{i
\tom}{2}(\rho-1) + \mathcal{O}(\rho-1)^2 \right] \labell{horizonBCTheta}
\eeq
near $\rho =1$.
\FIGURE{
\begin{tabular}{c}
\includegraphics[width =  \textwidth]{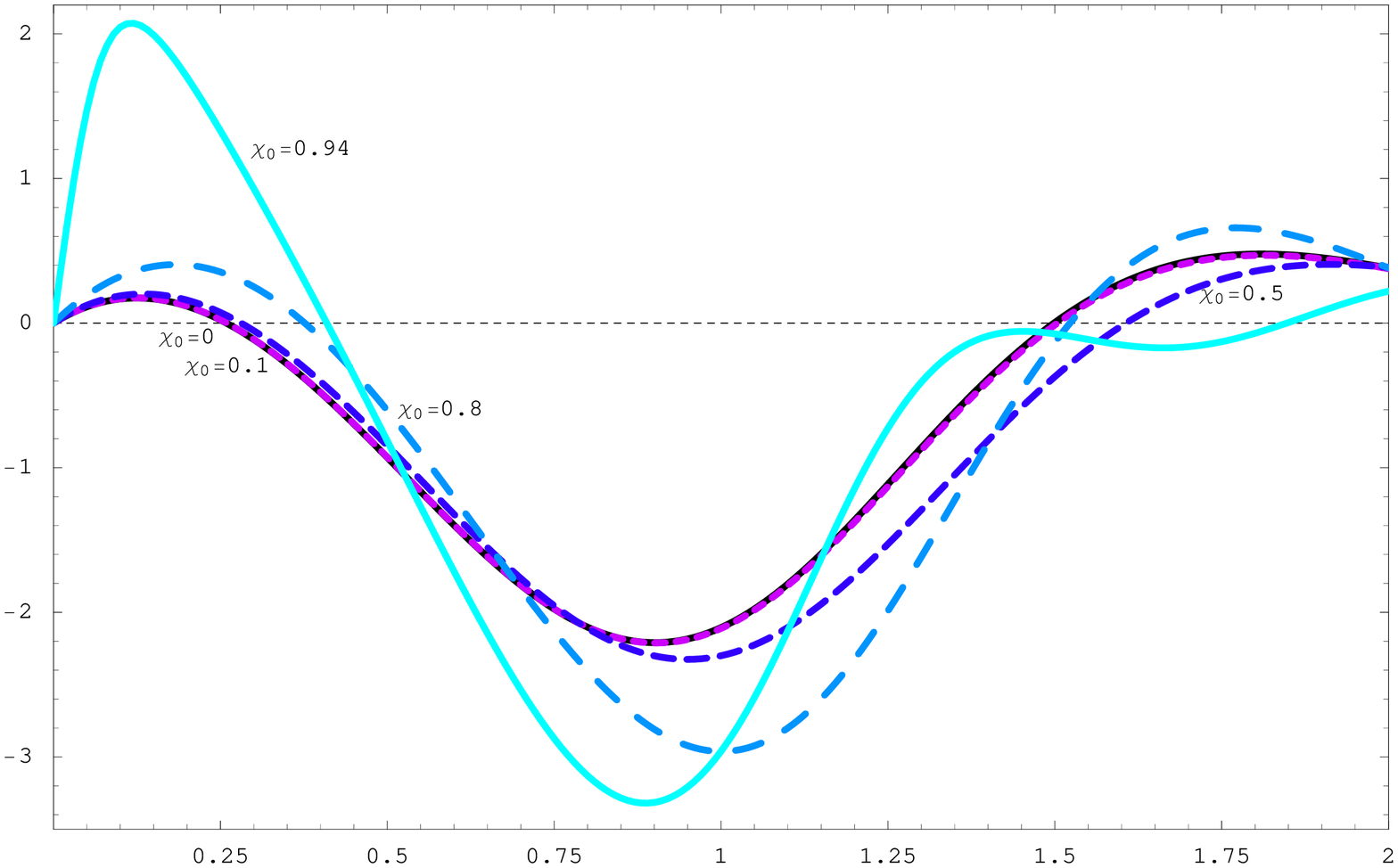}
\end{tabular}
\caption{The finite temperature part of the $\ell=0$ scalar ($\delta
\theta$) spectral function, $\tilde{\rn}_\theta - \nf \nc \omega^2/4
\pi $, in units of $\nf \nc T^2 /4$ for $\chi_0 \leq 0.94$,
corresponding to temperatures above the phase transition.}
\label{scalarLow} }

We solved \reef{thetaeom2} numerically and evaluated the spectral
function using \reef{thetaSpectral2}.  The high frequency
asymptotics of the spectral function appear in appendix \ref{high}.
Plots of the finite temperature part of the s-wave spectral function,
$\tilde{\rn}_\theta - \nf \nc \omega^2/4 \pi $ are provided in
fig.~\ref{scalarLow} for D7-brane embeddings corresponding to
temperatures above the phase transition.  The spectral function
shows no high peaks and little structure at temperatures above the
phase transition.

\FIGURE{
\begin{tabular}{c}
\includegraphics[width = 0.8 \textwidth]{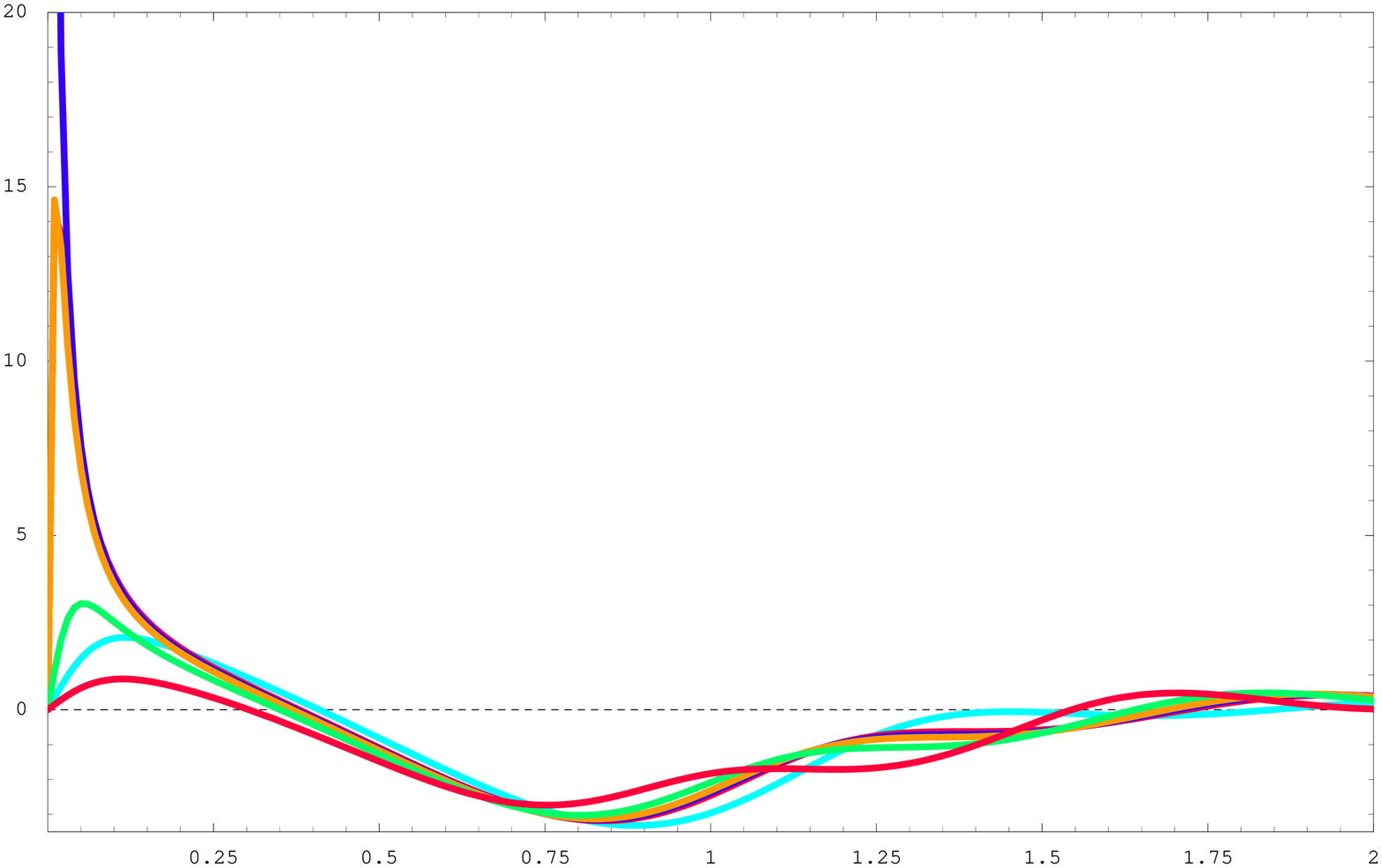} \\
\includegraphics[width = 0.8\textwidth]{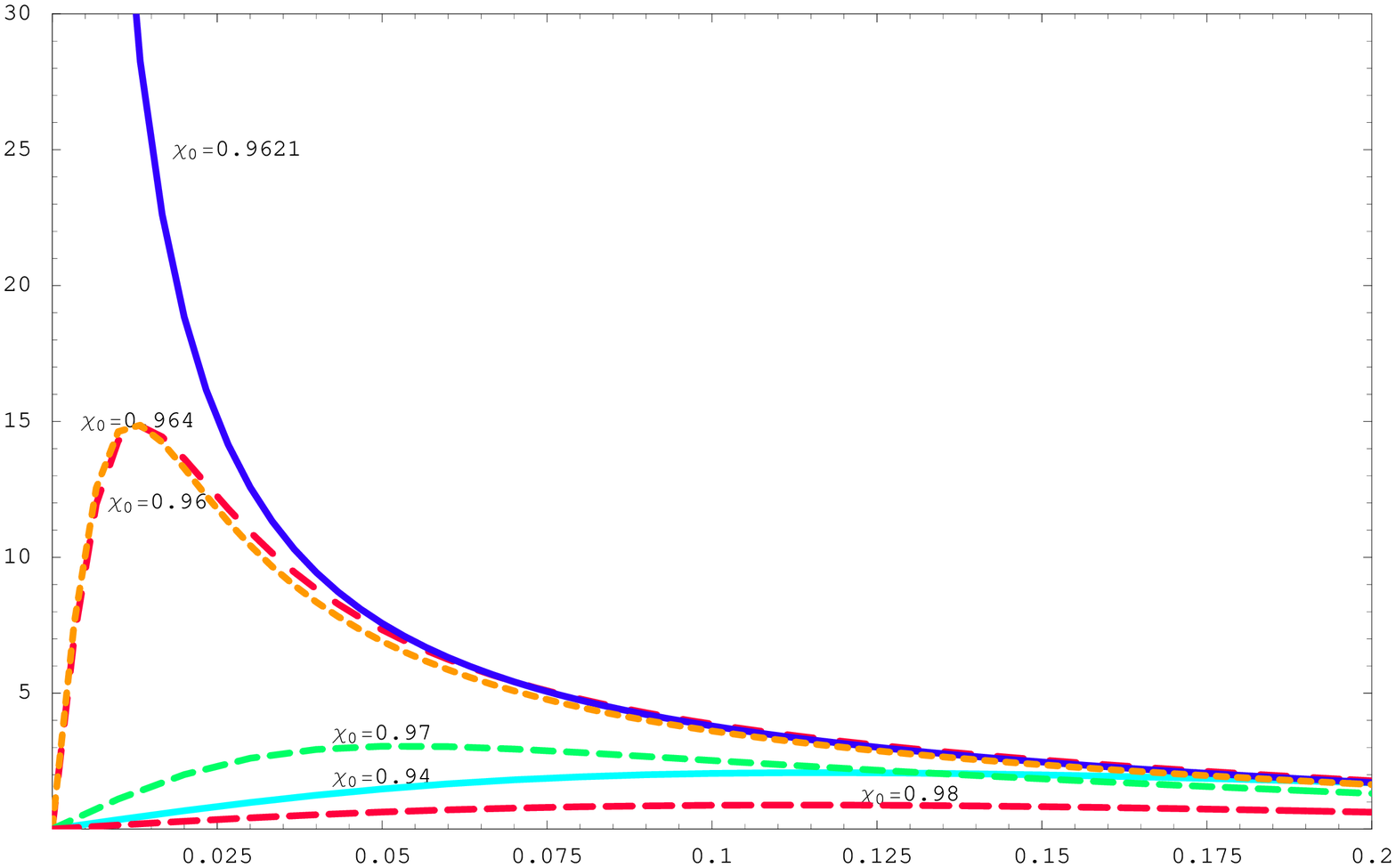}
\end{tabular}
\caption{The finite temperature part of the scalar ($\delta \theta$)
spectral function for $\ell=0$ in units of $\nf \nc T^2 /4$ for
values of $\chi_0$ past the phase transition.  The lower plot
focusses on the region near $\tom =0$ where a peak appears in the
spectral for $\chi_0 = 0.9621$, corresponding to the first kink in
the plot of the free energy versus temperature. }
\label{scalarKink1} }
Figure \ref{scalarKink1} provides plots of the spectral function for
values of $0.94<\chi_0 <1$, corresponding to black hole embeddings
past the phase transition, \ie continuing along the black hole
branch in fig.~\ref{action} past point $A_1$.  For $0.94 < \chi_0 <
0.96$, prior to the first kink in the free energy (between $A_1$ and
$A_2$ in fig.~\ref{action}), no striking peaks appear in the
spectral function.  However, for $\chi_0 = 0.9621$, point $A_2$ in
fig.~\ref{action}, a very high peak appears in the spectral
function, centred on $\omega = 0$.  Taking a value of $\chi_0$
slightly larger (smaller), say $\chi_0 = 0.964$ ($\chi_0 = 0.96$),
fig.~\ref{scalarKink1} shows that the peak is diminishing and is
centred on a small but nonzero value of $\omega$. A bit further away
from the first kink, \eg $\chi_0=0.97\, ,0.98$ no peak is evident.
Following the D7-brane embeddings to the second kink, which occurs
for $\chi_0 = 0.99973885$, the same behaviour is evident: Near this
value of $\chi_0$ a small peak starts to appear in the spectral
function and at $\chi_0 = 0.99973885$ a high peak, centred on
$\omega=0$ appears.  As we will discuss in section \ref{discuss},
this behaviour is a result of quasinormal eigenfrequencies crossing
the real axis from the lower to upper half of the complex
$\omega$-plane. As a result, these black hole embeddings become
unstable beyond $\chi_0 = 0.9621$, in precise agreement with the
thermodynamic discussion of section \ref{thermoBrane}.
Ref.~\cite{hoyos} examines the quasinormal modes in this channel
directly and find qualitative evidence of this behaviour, as well.

At first sight, the qualitative difference in the behaviour of the
$\ell=0$ scalar spectral function from the previous cases may seem
to be at odds with the expectation that the fluctuation spectra of
Minkowski and black hole phases should converge as they approach the
critical embedding. However, we must recall that it was precisely
the $\ell=0$ scalar modes that also realized an infinite family of
instabilities on the Minkowski branes \cite{long}. Hence we expect
that the spectra of the two phases are again converging but now on
the imaginary frequency axis. Hence the absence of a series of
quasiparticle peaks should in fact be the expected result.

Note, however, that in the analysis of the Minkowski phase
\cite{long}, the $\ell>0$ scalar modes remained stable. Hence we
might expect to see the appearance of quasiparticles in the spectral
functions for these channels on the black hole embeddings.
Fig.~\ref{scalarL1} provides a plot of the scalar spectral function
for the $\ell=1$ mode. The spectral function shows very little
structure for any values of $\chi_0$, or, equivalently, $m$.
However, we certainly found no evidence for $\ell=1$ of quasinormal
eigenfrequencies crossing the real axis from the lower to upper half
of the complex $\omega$-plane.  Instead from the general arguments
above, we expect that if this $\ell=1$ spectral function was studied
more intensively that in fact quasiparticle peaks would appear very
close to $\chi_0=1$.
\FIGURE{
\begin{tabular}{c}
 \includegraphics[width=0.8 \textwidth]{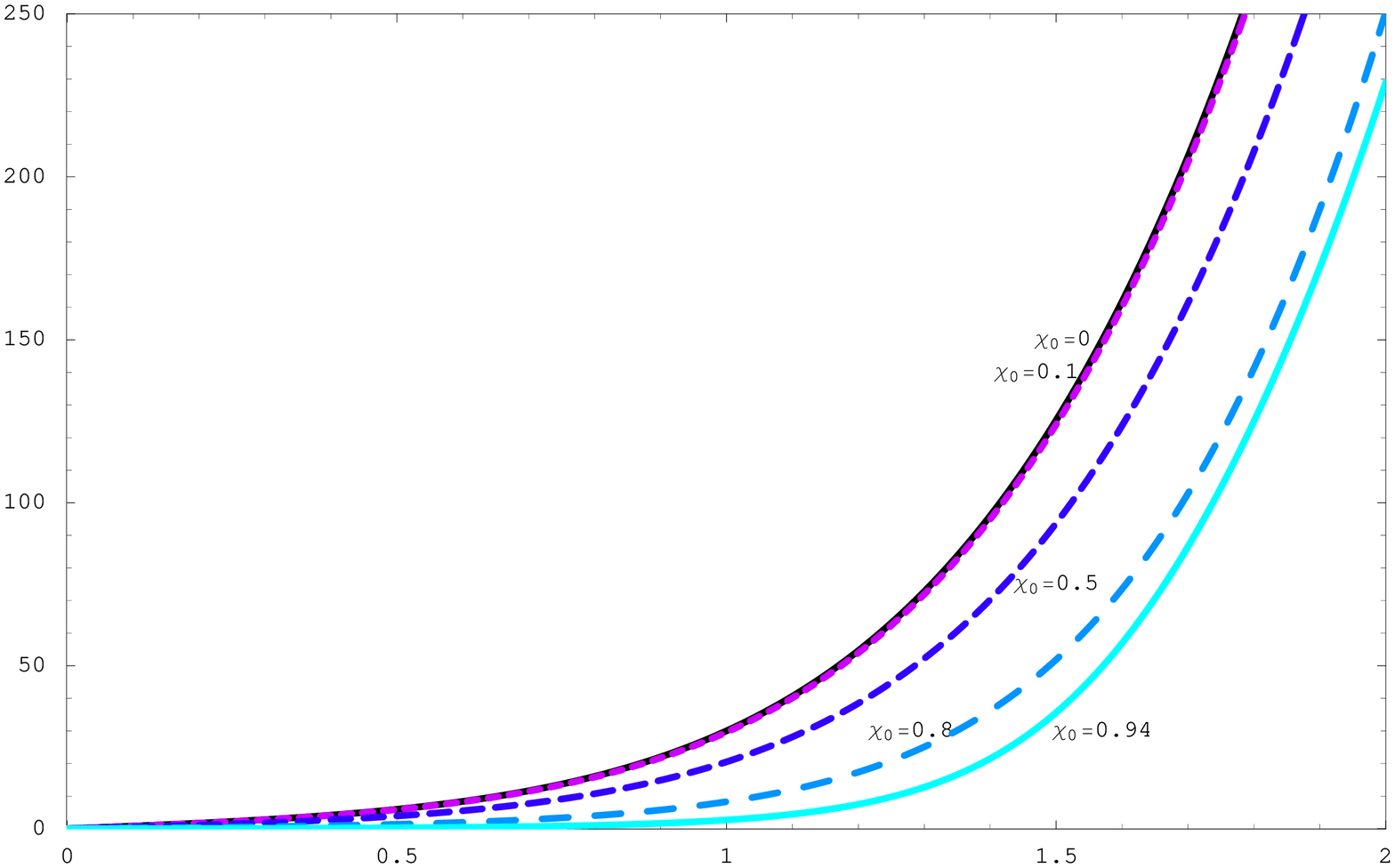}\\
 \includegraphics[width=0.8 \textwidth]{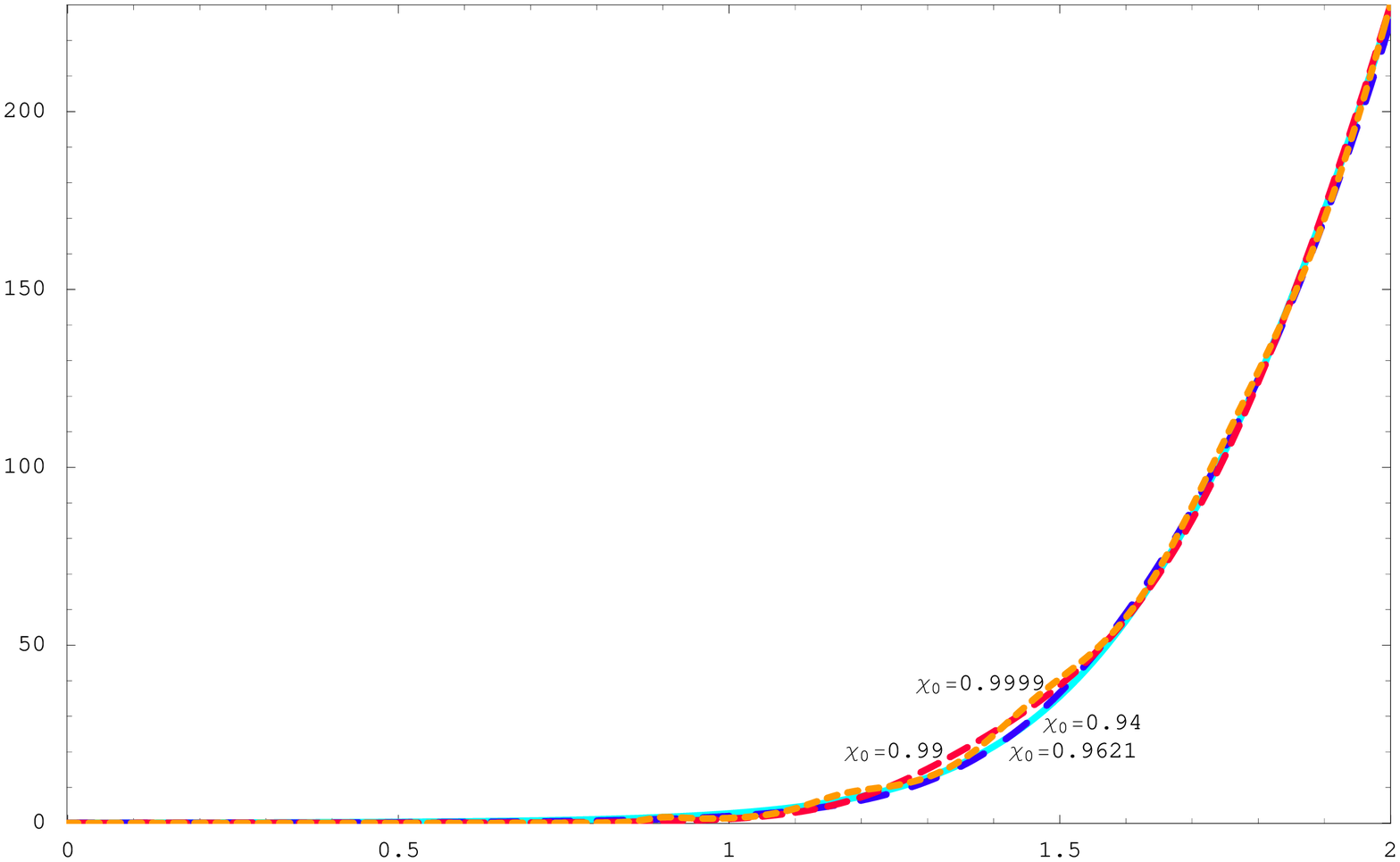}
\end{tabular}
\caption{The scalar spectral function for $\ell=1$ in units of
$\pi^2 \nf \nc T^4/8$ versus $\tom$.  The upper plot is for values
of $\chi_0$ corresponding to temperatures above the phase transition
while the lower plot is for $\chi_0$ past the phase transition.
Note in the lower plot that the lines  for
$\chi_0=0.9621,0.99,0.9999$ roughly coincide with that for
$\chi_0=0.94$ and that there is no structure suggesting the
existence of quasiparticle states.}\label{scalarL1}}

We close this section with one final observation. While the
pseudoscalar equation of motion \reef{phieom2} is singular in the
massless limit, \ie $\chi_0\to0$, the spectral function should have
a well defined limit. Further if comparing the $\chi\rightarrow0$
limits in the pseudoscalar and scalar channels in figures
\ref{pseudoPlots} and \ref{scalarLow}, we find that they converge on
the same spectral function in this limit. The fact that these
spectral functions coincide in this limit is a reflection of the
restoration of an additional $U(1)$ global symmetry, corresponding
to rotations in the 89-plane in the array \reef{array} when the
quark mass vanishes. In the massless limit, this symmetry relates
the two scalar operators.

\section{Diffusion constant for `light' quarks}\label{diffusion}

The worldvolume gauge field is dual to a conserved current in the dual
gauge theory. One then expects to see the diffusion of the conserved
charge, \ie quark charge, according to Fick's law with a certain
diffusion constant. This expectation can be confirmed in a
holographic context \cite{Kovtun:2003wp,Policastro:2002se} and in
fact, the computation of the diffusion constant can be performed in
a number of different ways. In the present D3/D7 brane system, we
have explicitly computed the diffusion constant in three different
ways: (a) using the membrane paradigm \cite{Kovtun:2003wp};  (b) the
Green-Kubo formula; and (c) the lowest quasinormal frequency in the
diffusion channel.  In this section, we describe these different
computations and our results confirm the internal consistency of the
holographic framework, in that we show these different methods all
give the same result.

\subsection{Membrane paradigm method}\label{membrane}

The computation of the diffusion constant via the membrane paradigm
was discussed in \cite{Kovtun:2003wp} where explicit formulae for
various transport coefficients in terms of metric components for a
wide class of metrics were derived.  There, the authors considered
perturbations of a black brane background and a formula for the
diffusion constant (eq.~(2.27) in \cite{Kovtun:2003wp}, also quoted
here in eq.~\reef{genMem}) resulted from a derivation of Fick's law.
An analogous computation can be performed for the D7-branes' vector
field for black hole embeddings and it gives\footnote{Note that the
same method can be applied to compute the diffusion constant for the
gauge theory corresponding to the supergravity configuration of a
Dq-brane probe in the near-horizon black Dp-brane geometry -- see
appendix \ref{app-diffuse}.}
\beqa
D &=& \left. \frac{\sqrt{-g}}{\sqrt{h_3}}\frac{1}{g_{xx}\sqrt{-g_{tt}g_{\rho\rho}}}
 \right|_{\rho=1} \int d\rho \left( -g_{tt}\right) g_{\rho \rho}\frac{\sqrt{h_3}}{\sqrt{-g}} \nonumber \\
&=& \frac{2 (1-\chi_0)^{3/2}}{\pi T}  \int_1 ^\infty d\rho \frac{f
\sqrt{1-\chi^2+\rho^2 \dot{\chi}^2}}{\tilde{f}^2 \rho^3
(1-\chi^2)^2} \labell{membraneD} \eeqa
where, in the first expression, the metric $g$ is the induced metric
on the D7-branes \reef{induce} and $h$ is the determinant of the
metric on the $S^3$ of unit radius.
\FIGURE{
\includegraphics[width =  0.8 \textwidth]{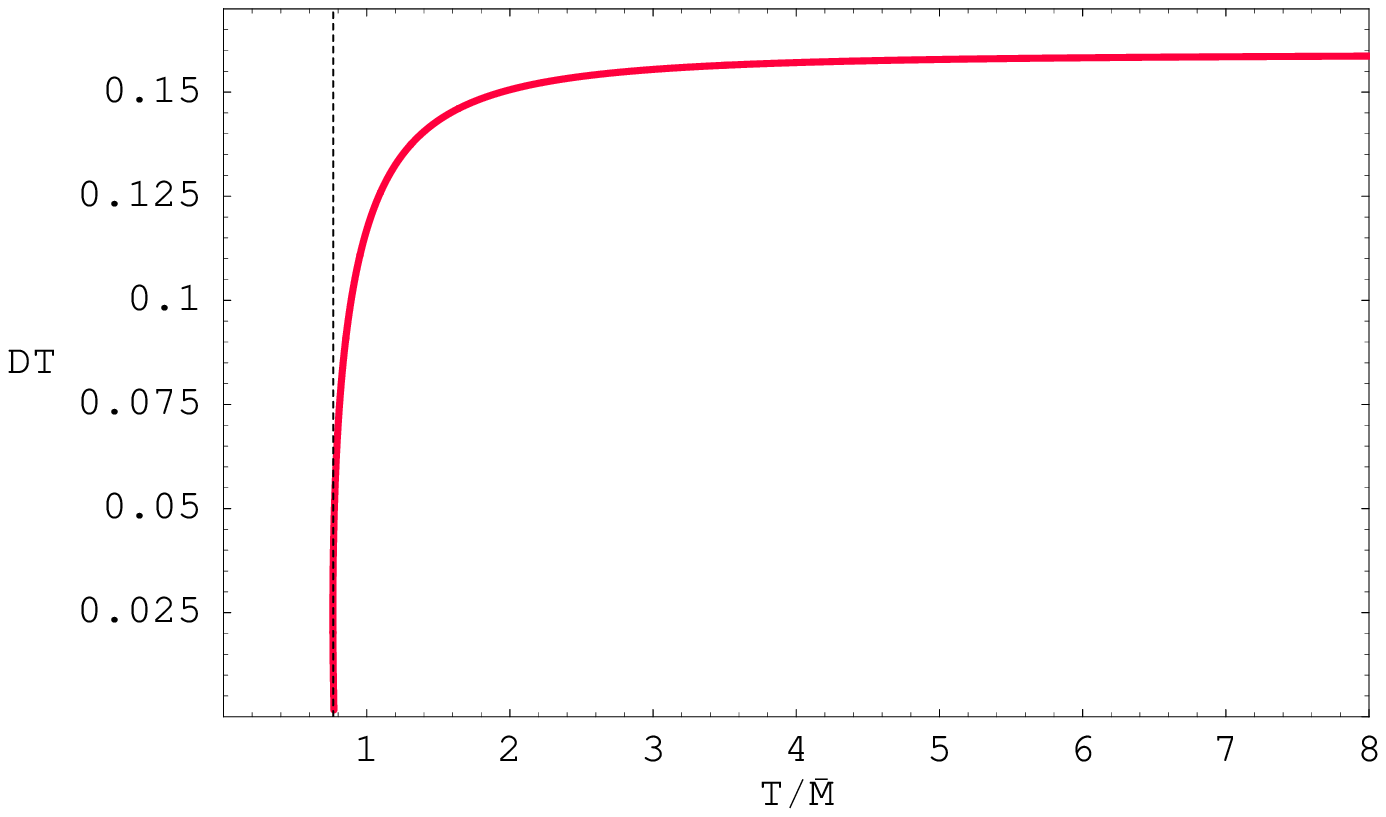}
\caption{The diffusion constant $D$ times the
  temperature $T$ versus temperature $T/\bar{M}$ for D7-brane probes in
  the black D3-brane geometry. The dotted vertical line marks the
  temperature of the phase transition. }
  \label{d3d7a}
}

Using the numerical solutions for the embedding $\chi$, we
numerically integrated \reef{membraneD} to find $DT$.  The results
are plotted in figs.~\ref{d3d7a} and \ref{d3d7b}. Fig.~\ref{d3d7a}
clearly shows that asymptotically at high temperatures, $DT$
approaches $1/2 \pi$. This coincides with the result for the
diffusion constant of R-charges in ${\cal N}=4$ SYM
\cite{Policastro:2002se}. At a
pragmatic level, this coincidence arises because both results are
constructed from correlators of a Maxwell field in a AdS$_5$ black
hole background. As the quark mass is increased, the induced
geometry on the D7-brane deviates from that of the background
geometry. Hence one finds a departure of $DT$ away from $1/2 \pi$ as
the ratio $T/\mbar$ decreases. Close to the phase transition, there
is a rapid decrease and $DT=0.036\simeq0.226/2\pi$ at the phase
transition. If we continue following the black hole branch beyond
the phase transition, $DT$ continues to fall and it also becomes a
multi-valued function of temperature, as shown in fig.~\ref{d3d7b}.
The latter simply reflects the fact that multiple embeddings can be
found for a single temperature in the vicinity of the critical
solution.
\FIGURE{
\begin{tabular}{cc}
\includegraphics[width = 0.5 \textwidth]{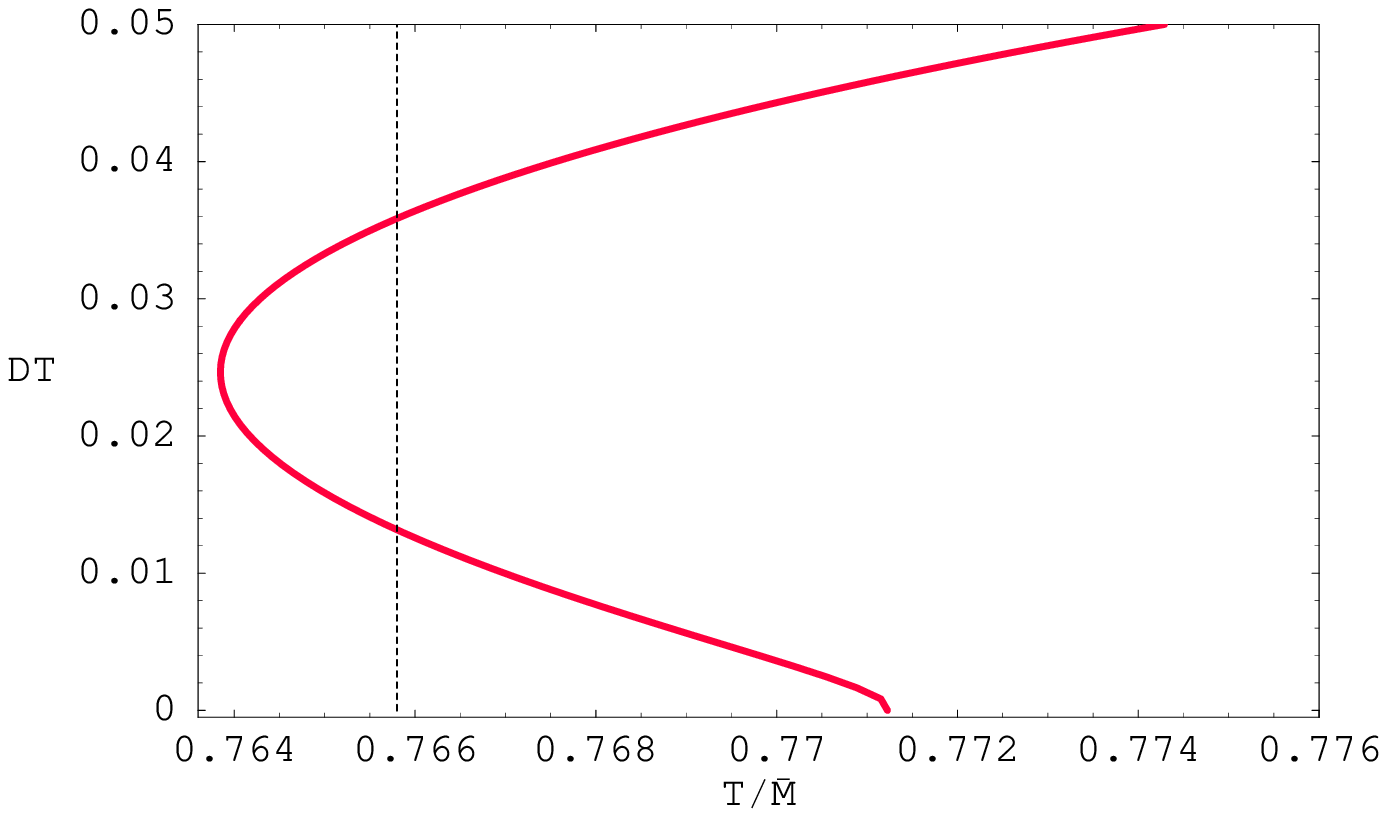} &
\includegraphics[width =  0.5 \textwidth]{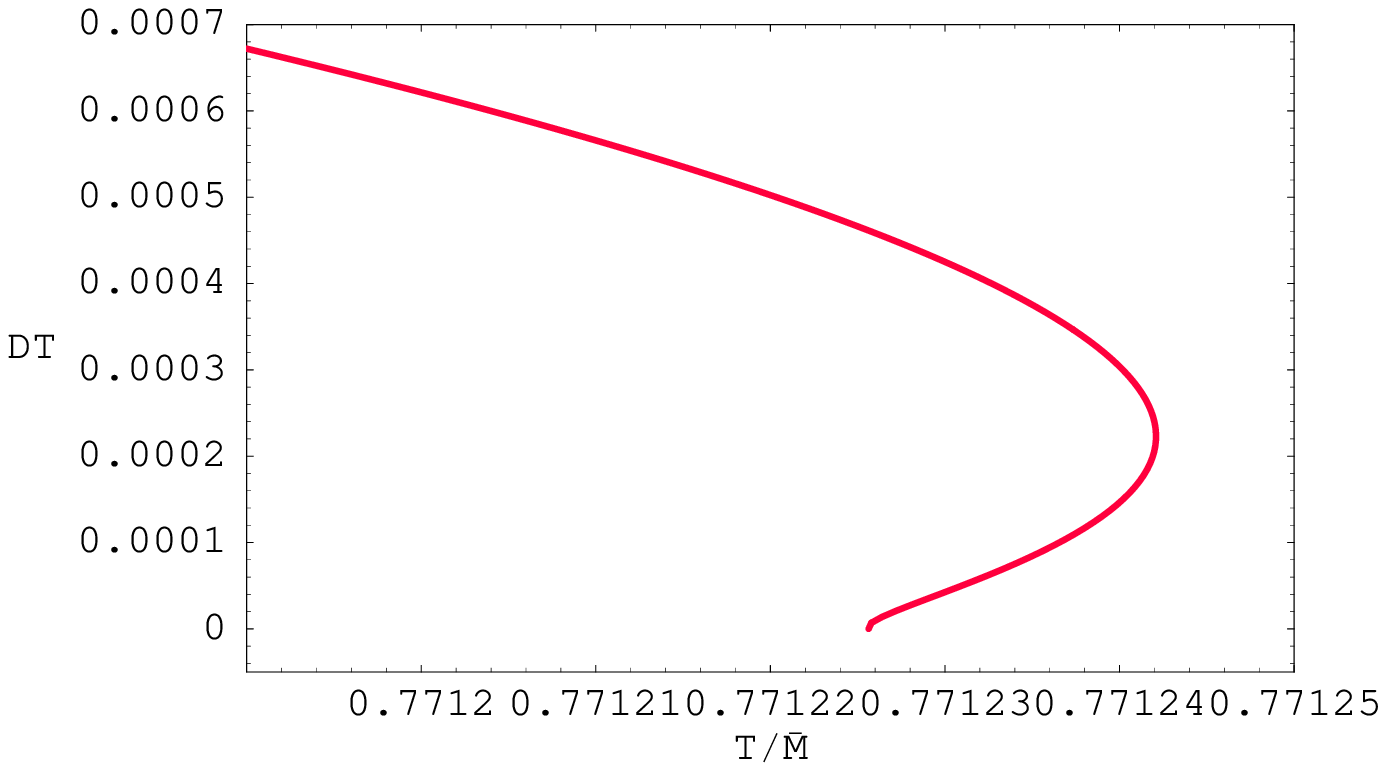}
  \end{tabular}
\caption{Plots of the diffusion constant $D$ times the temperature
$T$ versus temperature $T/\bar{M}$ for D7-brane probes in the black
D3-brane geometry, zooming in on the spiral  behaviour for
temperatures near the phase transition.} \label{d3d7b} }

\subsection{Green-Kubo formula}\label{green}

As discussed in section \ref{prelude_field}, the diffusion constant
may also be computed using the Green-Kubo formula \reef{GKF} which
relates the product of the diffusion constant $D$ and the
susceptibility $\Xi$ to the slope of the vector spectral function
($\ell=0$) for $\omega \to 0$: $D \Xi = \lim_{\omega \to 0}
\rn(\omega)/2 \omega$. The susceptibility is $\Xi = \partial
n_q/\partial \mu |_{\mu =0}$ where $n_q$ is the charge density and
$\mu$ is the chemical potential, both for fundamental matter (quarks
or their supersymmetric generalization).  In order to compute the
susceptibility, one must consider the D3/D7 brane system with a
finite chemical potential  \cite{findens}. The susceptibility can be
computed directly from eq.~(2.22) of \cite{findens}
\beq \tilde{\mu}  = 2\tilde{d} \int_1^\infty d\rho\,
\frac{f\sqrt{1-\chi^2+\rho^2\dot{\chi}^2}}
{\sqrt{\tilde{f}(1-\chi^2)[\rho^6 \tilde{f}^3 (1-\chi^2)^3 +8
\tilde{d}^2]}}\,,\labell{moo} \eeq
which applies for any black hole embeddings. From the appendix of
that paper, $\tilde{\mu}$ and $\tilde{d}$ are related to $\mu$ and $\nq$ via
\beq \tilde{\mu} =\sqrt{\frac{2}{\lam}}\,\frac{\mu}{T}\,, \qquad
\tilde{d} = \frac{2^{5/2}}{\nf\nc\lam^{1/2}}\,\frac{\nq}{T^3}\,.\labell{for2}
 \eeq
Combining these definitions, we
have
\beq \frac{\prt\dd}{\prt\tilde{\mu}}=
\frac{4}{\nf\nc T^2}\,\frac{\prt\nq}{\prt\mu}
\labell{derivative} \eeq
which, interestingly, is independent of the
't Hooft coupling $\lam$.

Note from eq.~\reef{moo} that $\tilde{\mu}=0$ is equivalent to
$\dd=0$ which means that we can calculate $\prt\tilde{\mu}
/\prt\dd|_{\dd=0}$ from this equation and take the inverse for the
desired derivative. Hence a straightforward calculation yields
\beq \frac{\prt\tilde{\mu}}{\prt\dd}  = 2 \int_1^\infty d\rho\,
\frac{\rho^6f\tilde{f}^4(1-\chi^2)^4\sqrt{1-\chi^2+\rho^2\dot{\chi}^2}}
{\left(\tilde{f}(1-\chi^2)[\rho^6 \tilde{f}^3 (1-\chi^2)^3 +8
\tilde{d}^2]\right)^{3/2}}\,,\labell{moo2} \eeq
and evaluating at $\dd=0$ gives
\beq \left.\frac{\prt\tilde{\mu}}{\prt\dd}\right|_{\dd=0}  = 2
\int_1^\infty d\rho\, \frac{f\sqrt{1-\chi^2+\rho^2\dot{\chi}^2}}
{\tilde{f}^2\rho^3(1-\chi^2)^2}\, .\labell{moo3} \eeq
Combining eqs.~\reef{derivative} and \reef{moo3}, our final
result is
\beq \Xi\equiv\left.\frac{\prt\nq}{\prt\mu}\right|_{\mu=0}=
\frac{\nf\nc}{4}\,T^2\, \left\lbrace\ \left.
\frac{\prt\tilde{\mu}}{\prt\dd}\right|_{\dd=0}\ \right\rbrace^{-1}
\,. \labell{final}\eeq
Note that in limit of massless quarks (\ie $\chi=0$), these
expressions have a simple form
\beq \left.\frac{\prt\tilde{\mu}}{\prt\dd}\right|_{\dd=0}  =
\frac{1}{2}\,,\qquad
\Xi= \frac{\nf\nc}{2}\,T^2\,. \labell{mqzero}\eeq

Numerically evaluating the low frequency limit of the spectral
function $\rn$ and the susceptibility (using \reef{final} and
\reef{moo3}), we computed the diffusion constant using \reef{GKF},
and the results confirm those displayed in figure \ref{d3d7a}.

Further, if we combine \reef{final} and \reef{membraneD}, we are
lead to write
\beq
D\,\Xi=\frac{(1-\chi_0^2)^{3/2}}{4\pi}\,\nf\nc\,T\, ,
\labell{prod}
\eeq
which, in view of \reef{GKF}, provides a simple analytic expression
for the low frequency ($\omega \to 0$) limit of the vector spectral
function:
\beq \rn (\omega) = \frac{(1-\chi_0^2)^{3/2}}{2 \pi} \nf \nc T
\omega +\cdots\, . \eeq
In a related analysis, the electric conductivity $\sigma$ of this
system with fundamental quarks was recently computed from Ohm's law
using AdS/CFT techniques \cite{andyandreas}. In the limit of
vanishing quark density and small external electric field, the
result of \cite{andyandreas} reduces to our formula \reef{prod}
above, as it should according to the generalized Einstein relation
$\sigma/e^2 = D\, \Xi$. (Note that $e^2$ is set to one in the
conventions of \cite{andyandreas}.)

\subsection{Lowest quasinormal frequency (in the diffusion channel)}\label{lowqnm}

The final computation of the diffusion constant comes from examining
the hydrodynamic dispersion relation, corresponding to the lowest
quasinormal frequency. At small three-momentum $q$, the diffusion
constant can be extracted from: $\omega = -iD q^2 +\mathcal{O}(q^4)$
\cite{Kovtun:2003wp}. In principle, the calculation of the
quasinormal mode spectrum from eq.~\reef{Eeq2} proceeds as follows:
For $\rho \to \infty$, eq.~\reef{Eeq2} implies that  $E_x \simeq
A+B\rho^{-2}$ for some constants $A,B$. Normalizable modes will be
those with $A=0$.  Hence one method to determine the quasinormal
frequencies is to use a two-dimensional shooting method, \ie solving
\reef{Eeq2} numerically with incoming wave boundary conditions at
the horizon and then tuning the (complex) frequency to find a
solution behaving as $\rho^{-2}$ asymptotically.  For small $q$ we
solved \reef{Eeq2} for various $m$ to determined the lowest
quasinormal frequency and our results for the diffusion constant are
identical to those plotted in fig.~\ref{d3d7a}.

\section{Discussion} \label{discuss}

In this paper, we used holography to investigate various aspects of
the high temperature phase of an ${\cal N}=2$ super-Yang-Mills
theory with fundamental matter. The holographic description consists
of probe D7-branes in the near-horizon background of D3-branes (in
the limit of large-$\nc$ and large-$\lam$ with fixed $\nf$). In the
high temperature phase, the D7-branes extend through the event
horizon of the AdS$_5$ black hole, which describes the theory at
finite temperature. In \cite{prl,long}, this phase was denoted as
the black hole branch since the metric induced on the worldvolume of
the D7-branes is itself a black hole. Even though the latter geometry
no longer obeys Einstein's equations, the analysis of the
hydrodynamic physics found previously for bulk fields, \eg
\cite{new}, is readily transferred to the worldvolume fields on the
D7-brane. Hence we were able to examine the spectral function for
various mesonic operators in section \ref{goldfinger}, following
\cite{Teaney:2006nc,Kovtun:2006pf}, and we calculated the diffusion
constant for the quark charge in section \ref{diffusion}, adapting
techniques from \cite{Kovtun:2003wp,Policastro:2002se}.

As reviewed in section \ref{revGeom}, the induced geometry
\reef{induce} on the D7-brane is determined by first solving for the
embedding profile from eq.~\reef{psieom} with appropriate asymptotic
boundary conditions \reef{asympD7}. Given the complexity of
eq.~\reef{psieom}, these geometries are only known in general from
numerical integration.  However, there is one particularly simple
case, namely that of zero quark mass. In this case, the embedding is
trivial, \ie $\chi=0$ everywhere, and the induced geometry
\reef{induce} is precisely that of (the direct product of) an
AdS$_5$ black hole (with a constant $S^3$). Hence our analysis of
sections \ref{vectors} and \ref{diffusion} reduces to studying a
Maxwell field in an AdS$_5$ black hole geometry and the results
precisely match those found previously for bulk gauge fields. For
example, the quark diffusion constant (with $\mq=0$) matches
precisely with the $R$-charge diffusion constant calculated in
\cite{Policastro:2002se,Kovtun:2003wp}. Further the new analytic
expression of the vector spectral function presented in section
\ref{simple} was extended to an analytic result for all of the
higher-$\ell$ modes of the worldvolume vector.

As the quark mass is increased away from zero (or $T/\mq$ decreases
to finite values), the induced black hole geometry on the D7-brane
begins to deviate from that of the background. In particular, the
main differences arise near the event horizon where $\chi$ is
largest. For example, eq.~\reef{induce} shows that the size of the
$S^3$ and hence the induced horizon area shrinks as $\chi_0$ grows.
Hence the physical properties of the fundamental fields were seen to
depart (dramatically in some cases) from the standard results with
the growth of the quark mass. Recall that some of the most
interesting behaviour appeared as $\chi_0\rightarrow1$, \ie
approaching the critical solution for which the effective horizon
area vanishes. Hence this behaviour can be seen as a precursor of
the phase transition to the low temperature phase or the Minkowski
branch, in which the D7-brane smoothly closes off above the event
horizon.

In the low temperature phase (and in the limit of large $\nc$), the
spectrum of mesons is characterized by a discrete set of stable
states \cite{meson,long} and the spectral function is a series of
$\delta$-function peaks, as illustrated in fig.\ref{schema}a. A
derivation of the spectral function for the scalar meson at $T=0$
appears in appendix \ref{specT0}.
These mesonic states correspond to open string excitations which are
essentially living at the minimum radius of the D7-brane. Since in
the high temperature phase the D7-branes extends through the event
horizon, these states are destabilized. In this phase, the spectrum
can be characterized by a discrete set of quasinormal modes in the
effective black hole metric induced on the D7-brane. The spectral
functions calculated in section \ref{goldfinger} reveal interesting
information about this quasinormal spectrum. We focussed on three
particular operators, which are bilinears of the fundamental fields
-- the details appear in appendix \ref{diction} -- corresponding
vector, pseudoscalar and scalar channels.

The behaviours of the vector and pseudoscalar spectral functions are
very similar, as can be seen in figures \ref{gaugePlots} and
\ref{pseudoPlots}. The following physical picture emerges from these
plots: At very high temperatures, the spectral function closely
resembles that for a vector in ${\cal N}=4$ SYM. (Of course, as
discussed above, the bulk and worldvolume vector results are
identical for $T/\mq\rightarrow\infty$.) In this regime, the
spectral functions show essentially no structure and the
eigenfrequencies of quasinormal modes must all be deep in the
(lower) complex plane. As the temperature decreases (with fixed
quark mass), both the real and imaginary parts of a given
quasinormal frequency decrease but the formation of peaks in the
spectral function suggests that the imaginary part decreases more
rapidly. At temperatures just above the phase transition, there are
vector and pseudoscalar quasiparticles. Continuing along the black
hole branch to even lower temperatures beyond the phase transition
(\ie following the black hole line through $A_1$ in
fig.~\ref{action}), peaks grow very sharp and even more prominent
indicating that the quasinormal modes have $\textrm{Re}(\omega) \gg
\textrm{Im}(\omega)$ in this regime.

Appendix \ref{schroe} presents a complementary discussion which
reaches the same conclusion. Plots of the effective potential for
the pseudoscalar (and vector) excitations in appendix \ref{schroe}
show a finite potential barrier developing at intermediate values of
the radius as $\chi_0 \to 1$. This suggests the existence of
metastable states in the corresponding Schroedinger problem which,
as discussed in the appendix, would correspond to a quasinormal
frequency with $\Gamma \ll \Omega$ \ie the eigenfrequency approaches
the real axis in this regime. Of course, while this intuitive
picture developed from the effective potential matches the behaviour
of the spectral functions discussed above, it only gives a very
schematic picture of the quasinormal spectrum. Hence it would be
interesting to develop more detailed picture with a full calculation
of the quasinormal modes \cite{hoyos}.

If we examine the positions of the peaks in the spectral functions
more closely as the black hole embedding approaches the critical
solution, it appears that the real parts of the quasinormal
frequencies roughly match with the spectrum of the lowest (vector
and pseudoscalar) mesons on a near-critical Minkowski embedding.
Hence one notable feature of the spectral functions is that as
$\chi_0 \to 1$, the peaks are becoming sharper but also more closely
spaced and moving towards lower frequencies. For example, in figure
\ref{pseudoPlots}, the peaks in the $\chi_0=0.9999$ line are much
more closely spaced than those in the $\chi_0=0.99$ line.  This
behaviour is similar to what is seen for the $\delta \phi$ spectrum
for near-critical Minkowski embeddings: For these near-critical
embeddings, the tower of masses appears to be collapsing to the mass
of the lowest meson -- see fig.~7 in \cite{long}. As the phase
transition occurs well away from the critical solution (\ie
$\chi_0=0.94$ versus 1), the positions of the spectral peaks are not
closely matched with the corresponding meson spectrum for the
Minkowski embedding at the phase transition. Of course, both spectra
still characterized by the same general mass scale, as given in
eq.~\reef{mbar}.

We also examined the spectral functions for $\ell>0$ in vector and
pseudoscalar channels. These modes of the worldvolume fields
correspond to higher dimension operators in the field theory, which
are charged under the internal symmetry group $SO(4)=SU(2)\times
SU(2)$ -- see the discussion in appendix \ref{diction}. For these
modes, the results are qualitatively similar to those for $\ell=0$
as one approaches the critical embedding. However, the rate at which
the quasinormal frequencies approach the real axis is much slower --
that is, the spectral functions only develop pronounced peaks very
close to $\chi_0=1$. In fact, these peaks are already washed out at
the phase transition, \ie $\chi_0=0.94$. Hence the corresponding
mesons with $\ell>0$ do not survive as quasiparticles through the
phase transition. The analogous observation applies to the excited
mesons with $\ell=0$ and $n\ge1$. Examining the spectral functions
in figs.~\ref{gaugePlots} and \ref{pseudoPlots}, one finds that only
the first peak remains pronounced at $\chi_0=0.94$. Hence it seems
that only the ground state mesons (with $n=0=\ell$) can be said to
survive the phase transition as quasiparticles. However, even these
resonances have disappeared in the quark-gluon plasma by
$\chi_0\simeq0.8$ or $T\simeq 1.1 T_\mt{fund}$, where $T_\mt{fund}$ is
the temperature of the phase transition.

As shown in figures \ref{scalarLow} and \ref{scalarKink1}, the
behaviour of the scalar spectral function is qualitatively different
from that found in the vector and pseudoscalar channels, shown in
figs.~\ref{gaugePlots} and \ref{pseudoPlots}. As before, at high
temperatures, the spectral function shows no distinguished
structure, indicating that the quasinormal eigenfrequencies are all
deep in the (lower) complex plane. As the temperature decreases, a
small peak develops near the origin although it is still not very
prominent at the phase transition. However, continuing to the
D7-brane embeddings on the black hole branch for temperatures below
the phase transition, this single peak grows and becomes extremely
sharp and centred at $\omega=0$, precisely at the first kink in the
free energy, \ie $\chi_0 = 0.9621$. Beyond this point, the peak
decays and moves away from $\omega=0$. Our interpretation of this
behaviour is that the lowest (pair) of quasinormal frequencies
approaches the origin and actually crosses the real axis  at
the point $A_2$ in fig.~\ref{action}. Continuing beyond this point,
this eigenfrequency moves into the upper half plane, where it
actually corresponds to an unstable mode.

This interpretation of the behaviour of the scalar spectral function
is confirmed by the qualitative analysis of the corresponding
quasinormal modes in appendix \ref{schroe}. Examining the effective
potential for the scalar excitations shows that a negative potential
well develops and grows as $\chi_0 \to 1$. As discussed in the
appendix, when this well is large enough, it can support long-lived
`bound' states for which (the real part of) the effective energy is
negative. These modes are distinguished since $\Gamma^2>\Omega^2$
and further $\Gamma<0$, so that these bound states correspond to
instabilities of the D7-brane. The spike (or pole) at $\omega=0$ in
the scalar spectral function discussed above results from the
formation of the first bound state where the eigenfrequency crosses
the real axis. Ref.~\cite{hoyos} studied the quasinormal modes in
this channel and developed a qualitative picture which is in
agreement with our results.

Recall that the thermodynamic discussion of section
\ref{thermoBrane} predicted that the system should become unstable
at the point $A_2$ in fig.~\ref{action} because the specific heat of
the black hole branch becomes negative there. Hence our analysis
above is in precise agreement with this result and it shows that the
instability corresponds to unstable `quasinormal' modes appearing on
the D7-branes.

In fact, we found the scalar spectral function also displays a spike
at $\omega=0$ at the second kink in the free energy, \ie $\chi_0 =
0.99973885$. Hence it appears that the second lowest quasinormal
mode becomes unstable at this point. It is natural to conjecture
then that each time the free energy turns, a new `tachyonic' mode
appears in the scalar spectrum. In fact, the spectrum of scalar
mesons on the Minkowski branch was found to display precisely this
behaviour \cite{long}.

Again the spectral functions of mesonic operators which we
calculated exhibited a number of interesting features, which had a
clear interpretation in terms of the spectrum of quasinormal modes.
It would, of course, be interesting to confirm these behaviours by a
detailed investigation of the quasinormal modes, as was begun in
\cite{hoyos}. In the present paper, the spectral functions were only
calculated for zero spatial momentum for computational simplicity.
So another natural extension of this work is to consider the
behaviour at nonvanishing spatial momentum. In particular, the
spectral functions for general time-like and light-like
four-momentum can be used to calculate photon and dilepton
production rates, respectively \cite{Caron-Huot:2006te}. An analysis
of these results for the present ${\cal N}=2$ gauge theory has been
made recently \cite{davidphoto}.

The other main result of this paper was the calculation of the
diffusion constant for the quark charge. We used a number of
techniques developed for bulk black holes in the calculation of the
$R$-charge diffusion constant in the ${\cal N}=4$ theory: the
membrane paradigm method \cite{Kovtun:2003wp}, the Green-Kubo
formula \cite{Policastro:2002se} (which for us relied on previous
studies of this system at finite quark density \cite{findens}), and
the lowest quasinormal frequency \cite{Kovtun:2003wp}. It is
gratifying that as demanded by the internal consistency of the
holographic framework, all three of these independent calculations
yield the same results \cite{andrei2}, which are shown in figure
\ref{d3d7a}.

At very high temperatures (\ie $T/\mbar\gg1$), the diffusion
constant approaches $2\pi D\,T=1$, which as discussed above matches
the $R$-charge diffusion constant for the ${\cal N}=4$
super-Yang-Mills theory \cite{Kovtun:2003wp}. As $T/\mbar$
decreases, the product $D\,T$ decreases. Intuitively, we might
understand this result as the rate of diffusion decreasing at a
fixed temperature when the quark mass is increased. In fact, the
decrease is remarkably small at first, \eg $2\pi D\,T= 0.9$ at
$T/\mbar\simeq1.5$. However, figure \ref{d3d7a} then shows a
dramatic decline as we approach the phase transition at
$T/\mbar=0.7658$. At the phase transition, $2\pi D\,T\simeq0.226$,
but if we continue following the black hole branch it appears that
$D\,T$ continues to decrease and would vanish at the critical
embedding. The membrane paradigm approach to calculating $D$, as
described in section \ref{membrane}, provides a straightforward
understanding of this vanishing from the bulk perspective. Examining
the expression in eq.~\reef{membraneD}, we see that in the critical
limit, the integral remains finite but the prefactor vanishes
because $\sqrt{-g}\to0$. Hence the dominant effect which causes $D$
to decrease is the reduction of the effective horizon area of the
brane geometry \reef{induce} as we approach the critical embedding.
That is, $(1-\chi_0^2)\to0$ in advancing toward the critical
embedding. Further, this area vanishes precisely at the critical
embedding confirming that $D$ should vanish there.

We have considered the ${\cal N}=2$ gauge theory to have $\nf$
flavours of quarks. Hence it is worth noting that the results for
the quark diffusion constant in the present holographic framework
are independent of both $\nf$ and $\nc$. Of course, the same
independence of $\nc$ was seen with the $R$-charge
\cite{Kovtun:2003wp}. This must certainly arise because we are
working in the limit of large $\nc$ and large $\lambda$.

There has also been a great deal of interest in the diffusion of
`heavy quarks' in holographic theories recently
\cite{herzog,teen,first} -- see also \cite{flurry} and the
references therein. In the present context, this simply refers to
the quark diffusion constant in the low temperature phase where the
D7-branes only approach to some finite distance away from the black
hole horizon. In the low temperature or Minkowski embedding phase, a
quark is represented by a fundamental string stretching between the
D7-brane and the horizon. As such, a heavy quark is holographically
represented by a macroscopic object (on a similar footing with the
probe D7-branes) and classically this object will remain at rest
(\ie it does not `diffuse'). It is only when semiclassical effects
are taken into account that the heavy quarks diffuse through the
appearance of Hawking radiation from the effective black hole metric
induced on the string worldsheet \cite{teen}. This process should be
contrasted with the diffusion process in the high temperature phase
which we have been considering here. As stressed above, in this
phase the induced metric on the D7-brane is a black hole metric and
so if quark number is injected into the system, the holographic
description of diffusion is simply the classical process of the
corresponding worldvolume excitations falling towards the event
horizon. Given these two disparate descriptions, it is not
surprising that the diffusion constant takes a qualitatively and
quantitatively different form in the two phases. In the low
temperature phase, the diffusion constant is governed by the heavy
quark result\cite{herzog,teen}\footnote{Note we present this result
with the same normalisation for the 't Hooft coupling used
throughout this paper. Ref.~\cite{herzog} uses a convention such
that $\tilde{\lambda}=2\lambda$. This difference arises from the
implicit normalisation of the $U(\nc)$ generators:
Tr($T_a\,T_b)=d\,\delta_{ab}$. The standard field theory convention
used in \cite{herzog} is $d=1/2$ while our choice is $d=1$, as is
prevalent in the D-brane literature.}
\beq 2\pi\,D\,T=\sqrt{8/\lambda} \,.\labell{heavy}\eeq
On the other hand, after the transition to the high temperature phase, we
found above that $2\pi\,D\,T=O(1)$. Note that the holographic
analysis applies for strong 't Hooft coupling (\ie $\lambda\gg1$)
and so these results show the expected suppression of quark
diffusion in the low temperature phase.

We comment on a possible puzzle with the above description of the
diffusion of heavy quarks as due to semiclassical Hawking radiation.
As such, the diffusion constant would be expected to vanish in the
limit $\hbar\to0$ for the bulk theory. Now the standard AdS/CFT
dictionary would associate Planck's constant $\hbar$ in the bulk
supergravity with $1/\nc$ in the dual gauge theory \cite{bigRev}.
However, above, we see the result \reef{heavy} is seen to be
independent of $\nc$ and $D$ certainly does not vanish in the limit
$\nc\to\infty$. The resolution of this puzzle is that we have
misidentified the correct `semiclassical' nature of the diffusion
process here. Above we observed that the heavy quarks diffuse
because of the appearance of Hawking radiation in the effective
field theory on the string worldsheet dual to such heavy quark. That
is, the `fields' on the worldsheet are the transverse coordinates
describing the embedding of the string and so fluctuations in these
fields (arising from Hawking radiation) corresponds to fluctuations
in the position of the quark (\ie diffusion of the quark). Now as
usual, $\hbar$ for the worldsheet theory is identified with the
inverse string tension, $\alpha'=\ls^2$. More correctly, the
dimensionless $\hbar$ in the nonlinear sigma model one the
worldsheet is identified with the ratio of $\alpha'$ and the
background curvature scale, \ie
$\hbar_{ws}\simeq\ls^2/L^2\simeq1/\sqrt{\lambda}$. Now we see this
physical picture matches precisely with the calculated result
\reef{heavy} and the limit $\lambda\to\infty$ is the semiclassical
limit on the string worldsheet.

The results for the diffusion constant in both the high and low
temperature phases are combined in figure \ref{invDTvsLam}, where
$2\pi D\,T$ is shown as a function of $\lambda$ (for fixed $\mq/T$).
The thick black curve shows a canonical result for the ${\cal N}=2$
gauge theory, which we are displaying for $\mq/T=200$. The diffusion
constant starts at very large values on the left in nearly
perturbative results. The dashed `perturbative' line is the
extrapolation of the perturbative calculation of \cite{pert} for
${\cal N}=4$ super-Yang-Mills at large $\nc$. As the curve shows, we
expect $D\,T$ to make a transition to the $\lambda^{-1/2}$ behaviour
of heavy quarks in the low temperature phase. If the quark mass was
infinite, this behaviour would extend out to infinite 't Hooft
coupling. However, for a finite mass, increasing $\lambda$
eventually takes the system to the high temperature phase. It would be
interesting to understand the corrections to the heavy quark result
\reef{heavy}, as we approach the phase transition. For $\mq/T=200$,
the latter occurs at $\lambda\simeq9.4\times10^4$ \cite{prl,long}.
At this first order phase transition, $2\pi\,D\,T$ jumps
discontinuously up to the `light quark' curve at $2\pi\,D\,T =0.226$ and it
quickly reaches the asymptotic value of 1 as $\lambda$ continues to
increase.
\FIGURE{
  \includegraphics[width =  0.8 \textwidth]{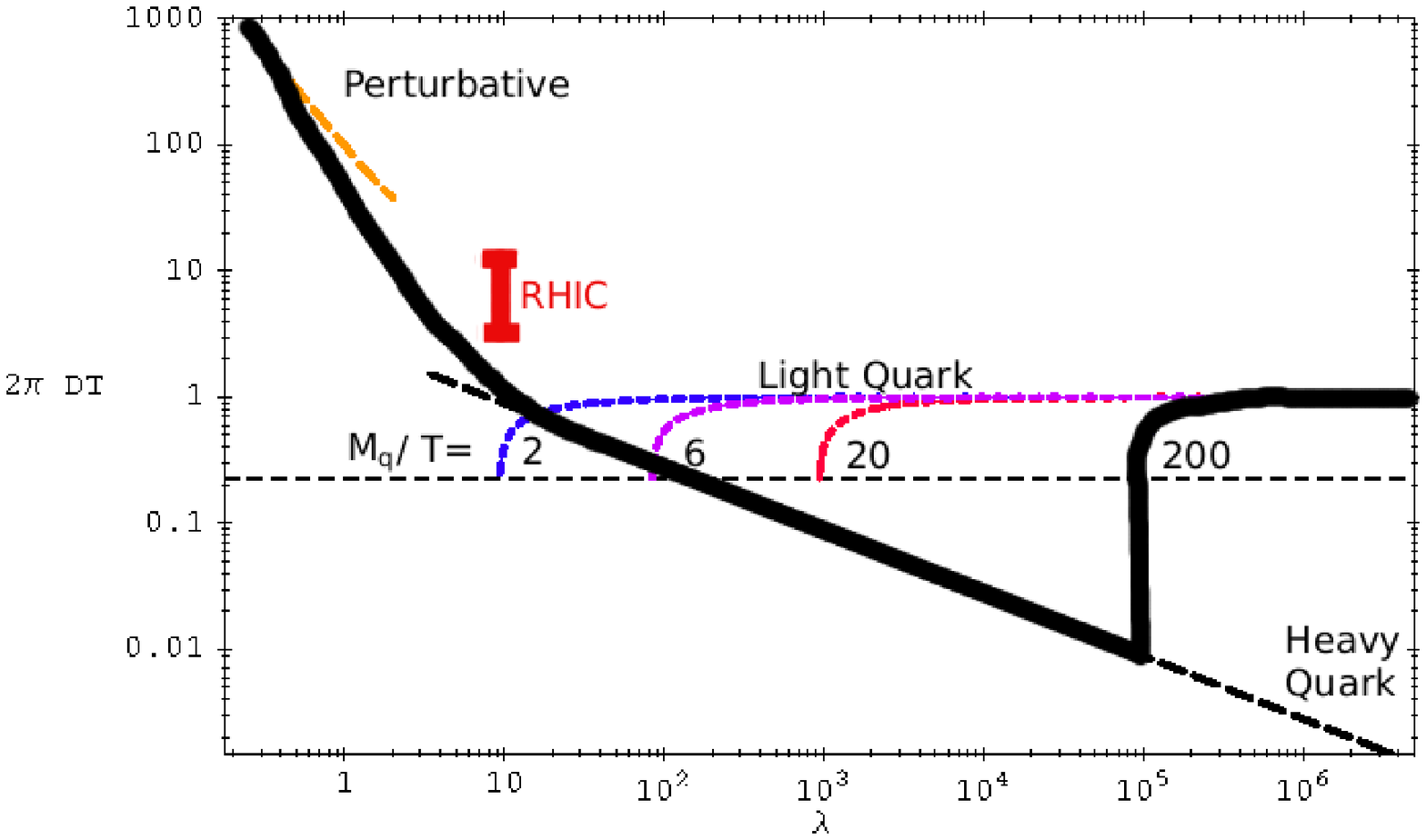}
\caption{Sketch of $2\pi\,D\,T$ versus $\lambda$ for a canonical
${\cal N}=2$ gauge theory with $\mq/T=200$. To the far left, we
have the perturbative regime \cite{pert}, which crosses over to the
`heavy' quark behaviour \reef{heavy}. Above the phase transition to
the black hole phase at $\lambda\simeq9.4\times10^4$, the curve
reflects the `light' quark behaviour of fig.~\ref{d3d7a}. The light
quark curves are also shown for $\mq/T=2$, 6, and 20. The horizontal
dotted line marks the value of $2\pi\,D\,T$ for the `light' quarks
at the phase transition.} \label{invDTvsLam} }

Much of the recent interest in holographic calculations of the
diffusion constant was generated by the possibility to compare these
strong coupling results with experimental results for the QCD
quark-gluon plasma measured at RHIC \cite{data}. It is interesting
then to place the RHIC results in the context of the phenomenology
of the ${\cal N}=2$ gauge theory studied here.  The bar labeled RHIC
corresponds to $\alpha_{strong}=0.5$ (or $\lambda=3\pi$) and the
range $2\pi\,D\,T=3\sim12$, which is used in fitting the observed
nuclear modification factor and elliptic flow amplitude for heavy
(charm) quarks -- see \cite{data} for details. This value of the 't Hooft coupling is well away
from the light quark behaviour of our canonical theory but also lies
in an intermediate regime between the heavy quark and perturbative
regimes. Hence a direct comparison of either approach for the ${\cal
N}=2$ theory to the experimental data for QCD is difficult
\cite{pert,comp}. We emphasize that in this intermediate region our
canonical curve is simply an `artistic' impression of the cross-over
between these two regimes.

For the canonical theory, we chose $\mq/T=200$ so that the phase
transition between the low and high temperature regimes took place
for a value of $\lambda$ that we could confidently characterize as
strong coupling. So we extend our discussion of RHIC results by
noting that the charm quark has a mass of roughly $1500$ MeV while
the temperatures achieved in the RHIC collisions are in the regime
$250$ MeV. Hence in these experiments, we are considering
$\mq/T\simeq 6$. As illustrated in figure \ref{invDTvsLam}, the
effect of reducing this ratio is to slide the light quark curve to
the left. That is, the phase transition occurs at smaller and
smaller values of $\lambda$, \eg $\lam_c\simeq85$ for $\mq/T\simeq
6$. However, it seems that this critical value is still well away
from the value of the coupling relevant for RHIC. Further it seems
that the same will still be true for charm quarks even at the higher
temperatures that might be achieved in the future at the LHC. Thus it is unlikely that a dramatic jump in
the diffusion constant, such as that seen for our canonical theory,
will appear in these experiments. It is noteworthy that in any
event, the experimentally favoured values of the heavy quark
diffusion constant are in fact above those calculated in the high
temperature phase. Hence it would seem that a QCD phase transition
would be characterized by a sudden decrease rather than a sudden
increase in the diffusion constant.

While our results for the diffusion constant do not seem relevant
for `heavy quarks', they might be considered as that for `light
quarks' in QCD. Note that our holographic model gave
$2\pi\,D\,T\le1$ where the effect of a finite quark mass was to give
a slight (less than order one) reduction below the asymptotic value
of 1. Appendix \ref{app-diffuse} extends the computation of the
diffusion constant described in section \ref{membrane} to more
general holographic theories. In particular, figure \ref{d4d6a}
shows the results for the D4/D6-brane system, which is the basis for
the construction of one holographic model which mimics QCD at large
$\nc$ \cite{QCDN}. The results are similar to those above with
$2\pi\,D\,T\le3/2$ with finite mass effects giving a small reduction
from the asymptotic value. Another interesting holographic model of
a QCD-like theory comes from introducing D8 and anti-D8 probe branes
in a D4-brane background \cite{sugimoto}. The resulting diffusion
constant is simply $2\pi\,D\,T=1$.  In this model, the current quark
mass is fixed vanish and so no finite mass effects appear. We might
also recall the results for the $R$-charge diffusion constants
$2\pi\,D\,T=1$ and 3/4, for the near-extremal D3- and D4-brane
backgrounds \cite{Kovtun:2003wp}. Hence in all these cases then, we
find that the calculations yield $2\pi\,D\,T=O(1)$. This might
suggest that in QCD, the diffusion constant associated with the
light quarks falls dramatically at strong coupling, as compared to
the perturbative results \cite{blast}, but that they should saturate
around this level. One might note then that these diffusion
constants would be smaller than for heavy quarks at presently
accessible energies but not much smaller. However, we must recall
that these calculations are all performed at large $\lambda$ and
large $\nc$ (with $\nf/\nc\ll1$) and so it would of course be
interesting to understand the corrections to these results at finite
$\lambda$ and finite $\nc$.

In the present study, we focussed on calculating the diffusion
constant for the overall quark charge. Of course, with $\nf$
flavours of quarks with identical masses, the ${\cal N}=2$ gauge
theory has a global $U(\nf)$ flavour symmetry and in the dual
gravity description, the worldvolume theory of the D7-brane contains
a nonabelian $U(\nf)$ gauge field. Our calculations have only
considered the diagonal $U(1)$ component of this gauge field.
However, as noted in appendix \ref{diction}, one can easily examine
the nonabelian $SU(\nf)$ flavour currents with the corresponding
components of the worldvolume gauge field  -- see, \eg
eq.~\reef{charge2}. Now in principle, we would describe the
diffusion of the full set of flavour currents with a diffusion
matrix $D_{ab}$, rather than a single constant. However, our
calculations of the diffusion constant in section \ref{diffusion}
only relied on knowing the quadratic action for the dual gauge field
and hence the nonabelian character of the gauge fields would play no
role. Hence the diffusion matrix is, in fact, diagonal: $D_{ab} =
D\, \delta_{ab}$ where $D$ is precisely the constant determined for
the $U(1)$ charge. On general grounds, this is of course the
expected result for the ${\cal N}=2$ gauge theory in the absence of
any chemical potentials. With a nonvanishing chemical potential, the
diffusion matrix will not take this simple form
\cite{Erdmenger:2007ap}. Similar comments to those above also apply
for extending our calculations of the spectral function to operators
that are no longer $SU(\nf)$ singlets.

\acknowledgments We thank Alex Buchel, Jaume Gomis, Pavel Kovtun,
Peter Langfelder and David Mateos for helpful discussions and
comments. Research at the Perimeter Institute is supported in part
by the Government of Canada through NSERC and by the Province of
Ontario through MRI. We also acknowledge support from an NSERC
Discovery grant (RCM), NSERC Canada Graduate Scholarship (RMT) and
the Canadian Institute for Advanced Research (RCM,RMT). Research at
the KITP was supported in part by the NSF under Grant No.
PHY05-51164. RCM would like to thank the organizers of Strings 2006
and String Phenomenology 2006 for the opportunity to present some of
the results discussed here.

\appendix

\section{Holographic dictionary}\label{diction}

In sections \ref{vectors} and \ref{scalars}, we focussed on
three worldvolume fields: the gauge field $A_\mu$ and the embedding
coordinates $\chi$ and $\phi$. We made use of the fact that the
asymptotic behaviour of these D7-brane fields has a direct
translation in terms of hypermultiplet operators in the gauge
theory. In this appendix, we elucidate this
holographic dictionary in more detail. Our discussion provides more
detail on the pseudoscalar operator since the other two operators
have already been considered in some detail in \cite{long,findens}.

Let us remind the reader that a hypermultiplet consists of two Weyl
fermions $\psi,\tilde{\psi}$ and two complex scalars $q,\tilde{q}$.
Of these, $\psi$ and $q$ transform in the fundamental of the
$SU(\nc)$ gauge group, while $\tilde{\psi}$ and $\tilde{q}$
transform in the antifundamental. Further, with $\nf$ flavours (of
equal mass), the hypermultiplets transform under a global $U(\nf)
\simeq SU(\nf) \times U(1)_\mt{q}$ symmetry. The charges of the
fields under the diagonal $U(1)_\mt{q}$ are +1 for $\psi$ and $q$
and --1 for $\tilde{\psi}$ and $\tilde{q}$. Hence the $U(1)_\mt{q}$
charge naturally counts the net number of quarks in a given state.
Here and below, we follow the notation of \cite{hassan}.

The operators dual to $A_\mu$, $\chi$ and $\phi$ can be determined
by considering the interactions of the open strings on the D3/D7
array \reef{array} before the decoupling limit \cite{joebook}, in
analogy with the closed strings. Such an exercise leads to the
following operators:
\beqa A_\mu\ &\leftrightarrow&\ J^\mu_\mt{q}=
\bar{\psi}\bar{\sigma}^\mu\psi+\tilde{\psi}\sigma^\mu\tilde{\psi}^\dagger
-i\left({q}^\dagger{\cal D}^\mu {q}-({\cal D}^\mu{q})^\dagger
{q}\right)- i\left(\tilde{q}\,({\cal D}^\mu \tilde{q})^\dagger-{\cal
D}^\mu\tilde{q}\,\tilde{q}^\dagger\right) \,,
\labell{charge}\\
\chi\ &\leftrightarrow&\ {\cal O}_\mt{m}=
i\tilde{\psi}\psi+\tilde{q}(\mq+\sqrt{2}\Phi^1) \tilde{q}^\dagger +
{q}^\dagger(\mq+\sqrt{2}\Phi^1) {q} + h.c.\,,\labell{mass}\\
\phi\ &\leftrightarrow&\ {\cal O}_{\phi}=
\tilde{\psi}\psi+i\sqrt{2}\,\tilde{q}\,\Phi^1\,\tilde{q}^\dagger +
i\sqrt{2}\,{q}^\dagger\,\Phi^1\, {q} + h.c.\,.\labell{pseudo2}
\eeqa
Note that $\Phi^1$, a complex scalar in the ${\cal N}=4$
supermultiplet, as well as $M_q$, appear in the scalar terms after
solving for the auxiliary field constraints within the full coupled
theory. Note that both these operators have conformal
dimension\footnote{This dimension applies in the UV where the
effects of quark mass are negligible and the theory becomes
conformal.} $\Delta=3$, which matches the standard prescription for
the powers of $\rho$ appearing in the asymptotic behaviour of the
fields -- see below.

Let us comment on the derivation of ${\cal O}_{\phi}$. The mass term
for the hypermultiplet fields originates from the following
superpotential term
\beq i\sqrt{2}\,\int d^2\theta\,\left(\tilde{Q}\,
\Phi_{7,7}\,Q-h.c.\right)\labell{sumass}\eeq
where $\tilde{Q}$ and $Q$ are chiral superfields containing
$(\tilde{q},\tilde{\psi})$ and $(q,\psi)$, respectively. These
hypermultiplet fields appear as ground states of the 3-7 and 7-3
strings while, as the subscript indicates, the superfield
$\Phi_{7,7}$ describes a particular set of massless modes in the 7-7
string sector. In particular, the lowest component of $\Phi_{7,7}$
is a complex scalar describing the transverse fluctuations of the
D7-brane position, \ie
\beq \Phi_{7,7} =
\frac{1}{\sqrt{2}}\left(\frac{X^8+i\,X^9}{2\pi\ell_s^2}\right)+\cdots\
. \labell{displace}\eeq
for the orientation in eq.~\reef{array}. After the decoupling limit,
this is no longer a dynamical field in the gauge theory but its
expectation value sets the mass of the hypermultiplets, \ie $\langle
\Phi_{7,7}\rangle=\mq/\sqrt{2}$. One sees from eq.~\reef{displace}
that the geometric angle $\phi$ appearing in our construction of the
D7-brane embeddings in section \ref{revGeom} corresponds precisely
to the phase of the complex field $\Phi_{7,7}$. Hence in deriving
${\cal O}_{\phi}$, we consider a phase rotation with the given
expectation value for $\Phi_{7,7}$ in the Lagrangian \reef{sumass}.
The resulting variation yields $\delta{\cal
L}=\delta\phi\,\mq\,{\cal O}_{\phi}$ and so we have divided by the
factor of $\mq$ in defining the operator given in
eq.~\reef{pseudo2}.

Recall that the full flavour symmetry is $U(\nf)$, which of course
matches the worldvolume gauge symmetry of the D7-branes. The current
$J^\mu_\mt{q}$ is the conserved current corresponding to the
diagonal $U(1)_\mt{q}$ of this global symmetry, \ie $J^t_\mt{q}$ is
the quark charge density. Our discussion can easily extended to the
$SU(\nf)$ symmetry by considering the nonAbelian gauge fields on the
D7-brane. The corresponding flavour currents would be
\beqa A^a_\mu\ &\leftrightarrow&\ (J^a)^\mu=
\left(T^a\right)_i{}^j\left[
\bar{\psi}^i\bar{\sigma}^\mu\psi_j+\tilde{\psi}^i\sigma^\mu
\tilde{\psi}^\dagger_j -i\left({q}^{\dagger i}{\cal D}^\mu
{q_j}-({\cal
D}^\mu{q})^{\dagger i} {q}_j\right)\right.\labell{charge2}\\
&&\left.\qquad\qquad\qquad\qquad\qquad\qquad\qquad\qquad\qquad -
i\left(\tilde{q}^{\,i}\,({\cal D}^\mu \tilde{q})^\dagger_j-{\cal
D}^\mu\tilde{q}^{\,i}\,\tilde{q}^\dagger_j\right)\right] \,,
\nonumber \eeqa
where $T^a$ are (Hermitian) generators of $SU(\nf)$ and we have
restored explicit flavour indices on the fields. With the operators
${\cal O}_\mt{m}$ and ${\cal O}_\phi$, we have also focussed on the
$SU(\nf)$ neutral terms but it would be straightforward to extend
these to a more general discussion. For example, in general
$\Phi_{7,7}$ transforms in the adjoint of $U(\nf)$ and so one can
easily chose a more elaborate mass matrix rather than one
proportional to the identity as above.

Recall the equation of motion \reef{psieom} for $\chi$ implies that
asymptotically
\beq \chi = \frac{m}{\rho} + \frac{c}{\rho^3}+\cdots \,.
\labell{asympD7a} \eeq
The dimensionless constants $m$ and $c$ are related to the quark
mass and condensate as \cite{long}:
\beqa \mq &= &
\frac{1}{2}\sqrt{\lambda}\,T\,m\,, \labell{Mm}\\
\langle{\cal O}_\mt{m}\rangle&=&
-\frac{1}{8}\sqrt{\lambda}\,\nf\,\nc T^3\, c\, . \labell{Oc} \eeqa
As presented here, this dictionary was established for a constant
coefficient (\ie the mass) and uniform expectation value of the
operator ${\cal O}_\mt{m}$. However, the same relationships also
apply in considering correlators where $\mq$ is shifted with a
general space- and time-dependent coefficient or source $\mu(x)$. In
section \ref{scalars}, we express the relevant correlators in terms
of variations $\delta\theta(x,\rho)$ (or its Fourier transform along
the $x^\mu$ directions). Given that $\chi\equiv\cos\theta$, we have
$\delta\theta\simeq-\delta\chi$ asymptotically where $\chi$
approaches zero. Hence, as confirmed from eq.~\reef{thetaeom},
$\delta\theta$ has the following the asymptotic behaviour:
\beq \delta\theta(x,\rho)=\frac{\delta\theta_0(x)}{\rho}+
\frac{\Theta(x)}{\rho^3}+\cdots\,,\labell{asymdtheta}\eeq
The source term is related to the first coefficient above as in
eq.~\reef{Mm} (up to a sign): $\mu(x)=- \frac{1}{2} \sqrt{\lambda}
\,T \,\delta\theta_0(x)$. Similarly the induced expectation value
and $\Theta(x)$ are related as in eq.~\reef{Oc}, again up to an
overall sign.

Ref.~\cite{findens} investigated the present holographic theory at
finite quark density and so established a similar dictionary for the
(time component of the) worldvolume gauge field. With the asymptotic
behaviour
\beq A_t=\mu-\frac{\tilde{\sigma}}{\rho}+\cdots\,, \labell{asympAt}
\eeq
it was found that $\mu$ is precisely the chemical potential or the
coefficient of the charge density operator $J_q^t$ and the quark
density was given by
\beq n_q\equiv\langle\, J_q^t\,\rangle=
\frac{1}{4}\nf\nc\,T^2\,\tilde{\sigma} \,. \labell{denseq} \eeq
Again these relations were originally established for a constant
chemical potential and uniform quark density but they still apply
for more general configurations. In particular for the correlators
of section \ref{vectors}, we consider more general gauge field
configurations with asymptotic behaviour
\beq A_\mu=\Sigma_\mu(x)+\frac{\sigma_\mu(x)}{\rho}+\cdots\,.
\labell{asympAmu} \eeq
In this case, $\Sigma_\mu$ corresponds to the (space- and
time-dependent) coefficient of the current operator $J_q^\mu$ and
the induced expectation values are given by
\beq\langle J_q^\mu(x)\rangle= \frac{1}{4}\nf\nc\,T^2\,\sigma^\mu(x)
\,. \labell{denseJ} \eeq
Note that here the index on $\sigma^\mu$ is raised with
$\eta^{\mu\nu}$, the inverse metric in the CFT.\footnote{Note that
while the calculation of the spectral function in section
\ref{vectors} is presented in terms of the gauge-invariant field
strengths $F_{\mu\nu}$, this was simply choice of convenience and
the final spectral function corresponds to that for current in
eq.~\reef{charge}.}

Now we would like to turn to the holographic dictionary for the
$\delta\phi$ modes. From the equation of motion \reef{phieom} and
the asymptotic behaviour of $\chi$ given in eq.~\reef{asympD7a}, we
can determine that $\delta\phi$ has the following asymptotic
behaviour
\beq
\delta\phi(x,\rho)=\delta\phi_0(x)+\frac{\Phi(x)}{\rho^2}+\cdots\,.
\labell{asymphi} \eeq
From eq.~\reef{displace}, we saw that the geometric angle $\phi$
appearing in the D7-brane embeddings corresponds precisely to the
phase of the complex field $\Phi_{7,7}$. Hence $\delta\phi_0$
corresponds to a fluctuation in the phase of the hypermultiplet mass
term. As usual, the dimensionless constant $\Phi$ is proportional to
the induced expectation value of ${\cal O}_\phi$ but we would still
need establish the precise constant of proportionality.

To establish the latter relationship, it is natural to frame the
discussion in terms of the thermal partition function -- see
\cite{findens}, for example.\footnote{In principle then this
discussion only concerns space- but not time-dependent sources.
However, the following results apply for the general case including
time dependence.} The source potential $\delta\phi_0$ enters the
partition function as
\be \exp\left[-\beta\,F(\beta,\delta\phi_0)\right]\equiv
\sum\,\exp\left[-\beta\int d^3x\,\left({\cal H}- \delta\phi_0\,
\mq\,{\cal O}_\phi\right) \right]\labell{parti}\ee
where a sum over all states is denoted on the right hand side and
$\beta$ denotes the inverse temperature. Of course,
$F(\beta,\delta\phi_0)$ and ${\cal H}$ are the free energy and
Hamiltonian densities, respectively. To begin, we note that, as can
be seen from eq.~\reef{parti},
\beq \frac{\delta F}{\delta(\delta\phi_0)} = -\mq\,\langle\, {\cal
O}_\phi\, \rangle\,. \labell{wow2}\eeq
To compare to the string description, we turn to the semiclassical
analysis of the Euclidean supergravity path integral -- see
\cite{long,findens}, for example. Here the on-shell action gives the
leading contribution to the free energy, \ie $\ide=\beta\,F$. Hence
to compare to eq.~\reef{wow2}, we need to evaluate the change of the
on-shell D7-brane action induced by a variation $\delta\phi_0$. The
background solution for this field is simply $\phi=0$ and so to
linear order the action is invariant. Hence we can focus on the
appropriate Euclidean version of the quadratic action
\reef{quadLphi} and the variation yields a boundary term, as in
eq.~\reef{boundphi},
\beqa \delta F &=& \frac{\pi^2}{4}\nf T_\mt{D7} \om^4\,\int d^3x \, \delta
\phi_0\, \left[\frac{\rho^5 f \tilde{f} (1-\chi^2)^2 \chi^2}{
\sqrt{1-\chi^2+\rho^2 \dot{\chi}^2}}   \,
\partial_\rho \delta \phi \right]_{\rho \to \infty} \nonumber\\
&=&-\frac{\pi^2}{2}\nf T_\mt{D7} \om^4m^2\,\int d^3x \,\Phi(x)\, \delta
\phi_0\,.\labell{vary}\eeqa
where the last expression uses the asymptotic behaviour of both
$\chi$ and $\delta\phi$ from eqs.~\reef{asympD7a} and
\reef{asymphi}, respectively. Comparing eqs.~\reef{wow2} and
\reef{vary}, we find
\beq \langle\, {\cal O}_\phi\, \rangle=\frac{\pi^2}{2\mq}\nf T_\mt{D7}
\om^4m^2 \,\Phi(x)=\frac{1}{8} \nf\nc\,\mq T^2\,\Phi(x)\,,
\labell{for}\eeq
which completes the holographic dictionary for the $\delta\phi$
modes.

Above we have identified the $\ell=0$ modes of the worldvolume
fields with operators in the dual field theory. Of course, a similar
dictionary relates the higher-$\ell$ modes to dimension $\ell+3$
operators $\mathcal{O}^\ell$ in the gauge theory. Qualitatively, we
may say that the latter are constructed with products of $\ell$ adjoint
scalar fields `sandwiched' between two fundamental fields in the
above operators -- see \eg \cite{meson, DeWolfe:2001pq}. For
example, the expression \reef{sumass} for $\mathcal{O}_\phi$ would
be generalized to
\beq \mathcal{O}^\ell_\phi \sim\int d^2\theta\,\tilde{Q}\,\,
\Phi_{7,7}\,\left(\Phi_{3,3}\right)^\ell\,Q\labell{sumore}\eeq
where $\left(\Phi_{3,3}\right)^\ell$ represents a traceless
combination of scalar superfields in the adjoint
hypermultiplet.\footnote{Hence in this expression \reef{sumore}, the
fundamental fields have an implicit sum of over the global $U(\nf)$
indices with $\Phi_{7,7}$ and the gauge $U(\nc)$ indices with
$\left(\Phi_{3,3}\right)^\ell$.} As is often the case in the AdS/CFT
correspondence and its generalisations, the precise matching between
normalisations in the field theory and string theory is difficult,
mainly due to the fact that one would need the full D3/D7 brane
action before taking the decoupling limit to determine the couplings
of the bulk fields to field theory operators.

However, to study the spectral functions overall numerical constants
need not concern us and we simply choose a convenient normalisation
which is consistent with holography. However, we must ensure that
the spectral function has the proper scaling dimension
\reef{hightail}. The three equations of motion, \reef{Eeq2} for the
vector, \reef{phieom2} for the pseudoscalar, and \reef{thetaeom2}
for the scalar imply the asymptotic behaviour
\beq
\Psi_\ell = A_\ell \, \rho^\ell + B_\ell \, \rho^{-\ell-2} \, , \labell{asymtYay}
\eeq
for some constants $A_\ell, \,B_\ell$ where $\Psi_\ell = E_x^\ell,
\, \delta \phi_\ell, \, \rho \delta \theta_\ell$. In order to obtain
the correct scaling of the spectral function, we change to the
standard dimensionful coordinates $z=L^2/r=\sqrt{2}/\pi T \rho$ used
in the usual AdS/CFT prescriptions for computing correlation
functions -- see, \eg \cite{bigRev}.  In these coordinates,
\reef{asymtYay} becomes
\beq
\delta \Psi_\ell =  \tilde{A}_\ell\, z^{-\ell} + \tilde{B}_\ell\, z^{\ell+2} \, .
\eeq
Hence, for $z \to 0$ (the boundary), we expect the leading behaviour
$\Psi \sim z^{-\ell}$.  Taking a cutoff at small $z = \epsilon$,
this implies that we should take
\beq \Psi_\ell (\rho, k) = \Psi_\ell ^0 (k) \epsilon^{-\ell}
\frac{\Phi_{\ell}(z)}{\Phi_{\ell}(\epsilon)} = \Psi_\ell ^0 (k)
\frac{(\pi T \,\rho_\infty)^\ell}{2^{\ell/2}}
\frac{\Phi_{\ell}(\rho)}{\Phi_{\ell}(\rho_\infty)} \, ,
\labell{asymptYay2} \eeq
where $\Phi_\ell$ represents  $E_{\ell,k}$ for the vector,
$\mathcal{P}_{\ell,k}$ for the pseudoscalar, and $\rho
\mathcal{R}_{\ell,k}$ for the scalar. As we will see in the next
section, eq.~\reef{asymptYay2} ensures that the spectral functions
have the correct asymptotic scaling \reef{hightail} appropriate for
an operator of dimension $\Delta=\ell+3$.

\section{Spectral function for scalar meson at $T=0$}\label{specT0}

In the low temperature phase of the D3/D7 brane theory (in the
large-$\nc$ limit), we expect the scalar spectral function to be a
series of $\delta$-functions positioned at mass eigenvalues of the
corresponding mesons. In this appendix, we explicitly demonstrate
this expectation by calculating the scalar spectral function at zero
temperature. The mesons and their spectrum in this theory at $T=0$
were studied in detail in ref.~\cite{meson}.

The background geometry corresponding to the field theory at $T=0$
is that of \reef{D3geom-r} with $r_0=0$ so that $f(r)=1$, \ie the
background no longer contains a black hole.  We embed $\nf$
D7-branes in this geometry as described by the array \reef{array},
at a distance $\mo=2\pi \ls^2 \mq$ from the D3-branes.  The
resulting configuration is supersymmetric with eight supercharges
preserved. Using spherical polar coordinates in the 4567-space with
radial coordinate $\bar{\varrho}=\mo \varrho$, the induced metric on
the D7-branes is
\beq ds^2 = \frac{\mo^2}{L^2}(1+\varrho^2)\left( -dt^2+d{\bf
x}^2\right)+ \frac{L^2}{1+\varrho^2} \left(d\varrho^2 +\varrho^2
d\Omega_3^2 \right) \, . \labell{induceGeomT0} \eeq

Following the same procedure as described in section \ref{scalars},
we consider small scalar fluctuations $\delta R $ of the D7-branes
about this fiducial embedding. That is, taking polar coordinates
$\phi$ and $\bar{R}=\mo R$ in the 89-directions, we consider
$\phi=0$ and $\bar{R} = \mo(1+ \delta R)$. Expanding the DBI action
to quadratic order in $\delta R$, we find
\beq S = -\frac{L^2}{2}\,T_\mt{D7}\,\nf \int d^8 \sigma
\frac{\sqrt{-g}}{1+\varrho^2}\, g^{ab} \, \partial_a \delta R
\,\partial_b \delta R\, . \labell{acta} \eeq
The corresponding equation of motion is
\beq
\partial_a \left[ \frac{\sqrt{-g}}{1+\varrho^2} g^{ab}\,
\partial_b \delta R\right]=0 \, ,\labell{eomYa}
\eeq
where $g$ is the induced metric \reef{induceGeomT0}.

We take the fluctuations to be constant on the internal $S^3$. Then
integrating over the $S^3$ and integrating by parts, the action
\reef{acta} becomes
\beq S = - \pi^2{\mo^4}\,T_\mt{D7}\, \nf \int d^4 x \left[ \varrho^3
\delta R \, \partial_\varrho \delta R\right]_{\varrho \to \infty}\,
. \eeq
Expanding the fluctuation in terms of Fourier modes in the
worldvolume directions, we take $\delta R(k,\varrho) = \delta R_0
(k) \mathcal{R}_k(\varrho)/\mathcal{R}_k(\varrho_\infty)$ as the
solution of  eq.~\reef{eomYa} regular at $\varrho =0$ and normalized
to $\delta R_0 (k)$ at $\varrho=\varrho_\infty$. The correlation
function is given by
\beq G = 2\pi^2\mo^4\, T_\mt{D7}\, \nf \left[\varrho^3
\frac{\mathcal{R}_{-k}(\varrho)\,\partial_\varrho
\mathcal{R}_k(\varrho)}{\mathcal{R}_{-k}(\varrho_\infty)\mathcal{R}_k(\varrho_\infty)}
\right]_{\varrho \to \infty} = 2\pi^2\mo^4\, T_\mt{D7}\,\nf
 \left[\varrho^3
\frac{\partial_\varrho\mathcal{R}_k(\varrho)}{\mathcal{R}_k(\varrho)}
\right]_{\varrho \to \infty}. \labell{fluff} \eeq
Changing the radial coordinate from $\varrho$ to
$\zb=\varrho^2/(1+\varrho^2)$ and setting $k^\mu=(\omega,0,0,0)$,
the equation of motion \reef{eomYa} for a fluctuation with
vanishing spatial momentum can be written as
\beq
\partial_{\zb} ^2 \mathcal{R}_k (\zb)+
\frac{2}{\zb} \partial_{\zb} \mathcal{R}_k (\zb)+
 \frac{\bw^2}{4\zb(1-\zb)}\mathcal{R}_k(\zb)=0\,,
\eeq
with $\bw^2 = \omega^2 L^4/\mo^2$. The solution regular at $\zb =0$
is given by a hypergeometric function
\beq
\mathcal{R}_k (\zb) = {}_2F_1\left( \frac{1}{2}+\frac{1}{2}\sqrt{1+\bw^2},
 \, \frac{1}{2}-\frac{1}{2}\sqrt{1+\bw^2} ; \, 2;\,\zb\right)\, .
\labell{housefire}\eeq
In \cite{meson}, the meson spectrum was determined by adjusting
$\bw$ such the this radial wavefunction vanishes asymptotically as
$\zb\to1$. Hence the correlator \reef{fluff} has a series of poles
positioned precisely at the mass eigenvalues of the meson spectrum.
The imaginary part of the correlator will then come from deforming
the integration contour for $\omega$ into the complex plane, in the
usual way. The difference between various types of correlators (\eg
retarded, advanced, Feynman) then comes from the precise choice of
this contour.

With the solution in eq.~\reef{housefire}, the correlator
\reef{fluff} becomes\footnote{In order
 to obtain the appropriate normalization we must multiply
 $G$ by $1/\mq^2$ -- see section \ref{scalars}.}
\beq
G =  \frac{2\, \pi^2\, \mo^4 \, T_\mt{D7}\, \nf }{\mq^2}
 \,  \bw^2\, \left[  \psi \left(\frac{1}{2} +
\frac{1}{2}\sqrt{1+\bw^2}\right) -\frac{\pi}{2}
\tan{\frac{\pi\sqrt{1+\bw^2}}{2} }\, \right]\,,
 \label{hryuk}
\eeq where we dropped contact terms. As indicated above, the
correlator \reef{hryuk} has poles at $\omega_n^2 = 4 n (n+1)
\mo^2/L^4$, $n=1,2,...$, corresponding to the meson spectrum found
in \cite{meson}. Using the expansion
$$
\frac{\pi}{4 a} \tan{\frac{\pi a}{2}} = \sum_{n=0}^\infty
\frac{1}{(2 n+1)^2 - a^2}
$$
and Sokhotsky's formula
$$
\lim_{\epsilon \rightarrow 0}
 \frac{1}{x\pm i \epsilon} = \mp i \pi
\delta (x) + {\cal P}
\left( \frac{1}{x}\right)\,,
$$
we find
\beq
\rn  = - 2 \,\textrm{Im } G =  \frac{\nf \nc}{\pi}\,
 \sum_{n=1}^\infty \,  \omega^2_n \, \sqrt{1+\bw^2_n} \,  \delta
 \left(\bw^2-\bw^2_n\right)\,.
\eeq
Since
\beq \delta \left(\bw^2-\bw^2_n\right) = \frac{\delta
\left(\bw-\bw_n\right) + \delta \left(\bw+\bw_n\right)}{2 \bw_n} \,
, \eeq
the spectral function  for $\omega \geq 0$
can be expressed as
\beq
\rn  = \frac{\nf \nc}{2 \pi}\,
 \sum_{n=1}^\infty \,  {\omega_n^2} \, \sqrt{1+\frac{1}{\bw^2_n}} \,  \delta
 \left(\bw-\bw_n\right) \, .\labell{endgame}
\eeq
This expression confirms our expectations for the spectral function
in the low temperature phase of the theory, illustrated in
fig.~\ref{schema}a:  the spectral function is a sum of
$\delta$-functions positioned at the meson mass eigenvalues. Of
course, the same is true in the vector and pseudoscalar channels (as
well as for $\ell>0$ in any of the channels). Here we have
explicitly calculated the spectral function only for $T=0$ because
we can analytically solve for the radial profile of the
fluctuations. However, for the general case with $0<T<\tf$, the
calculation is similar and the spectral functions again would take
the form of a sum of $\delta$-functions positioned at the meson
masses.

For large values of $\omega$ (at which large $n$ is relevant), the
above result \reef{endgame} can be written as
\beq
\rn  \sim \frac{\nf \nc \omega^2}{4 \pi}\,
 \sum_{n} \,   \, \Delta \bw_n \,  \delta
 \left(\bw-\bw_n\right) \, .
\labell{gamagama}\eeq
where $\Delta \bw_n =\bw_{n+1} - \bw_n\simeq 2 $ is the spacing
between delta functions for large $n$.   Note that the factor $\nf
\nc \omega^2/4 \pi$ is precisely the high frequency asymptotics
found for the $\ell=0$ spectral functions in appendix \ref{high} --
see eq.~\reef{hah}. With the appropriate choice of coordinates, the
radial equation in appendix \ref{high} takes the same form as here,
however, the boundary conditions chosen there for small radii assume
a black hole embedding and naturally produce a complex result for
the radial profile. This can be contrasted with the (naively) real
result \reef{housefire} for the Minkowski embeddings here. However,
eq.~\reef{gamagama} shows that if we examine the low temperature
spectral function \reef{endgame} with low resolution on the scale of
the spacing $\Delta \bw_n$, the $\delta$-functions in the spectral
function are averaged out and we reproduce the same high frequency
tail as for the black hole embeddings.

\section{Spectral function high frequency asymptotics}\label{high}

In this section we find expressions for the vector, pseudoscalar and
scalar spectral functions in the high frequency limit, \ie $\omega$
much larger than all scales: $\omega \gg T,\mq$.  Note that in a
sense this limit is equivalent to taking the limit of zero
temperature and quark mass.

The spectral functions, $\tilde{\rn}_{\theta_\ell}(\omega)$,
$\tilde{\rn}_{\phi_\ell}(\omega)$, ${\rn}_{\ell}(\omega)$, are
collectively denoted by $\rn_s^\ell(\omega)$ here and are defined as
\beq \rn_s ^\ell = - \frac{\pi^{2\ell}}{2^{\ell+2}} \nf \nc
T^{2\ell+2}\ \textrm{Im} \left[ \rho^{3+2 \ell} \frac{\partial _\rho
\Phi_\ell}{\Phi_\ell}\right]_{\rho \to \infty} \, , \eeq
where $\Phi_\ell = \rho {\cal R}_{\ell,k}, {\cal P}_{\ell,k},
E_{\ell,k}$ for the scalar, pseudoscalar, and vector fluctuations,
respectively.  The desired limit is achieved by considering both
$\rho \to \infty$ and $\tom^2-\tk^2 \to +\infty$. In this limit, the
embedding function is $\chi \sim m/\rho$, and the three equations of
motion, \reef{Eeq22} for $E_k$, \reef{thetaeom2} for ${\cal R}_k$,
and \reef{phieom2} for ${\cal P}_k$, reduce to
\beq
\partial_\rho^2 \Phi_\ell + \frac{3}{\rho}  \partial_\rho \Phi_\ell -
\left[\frac{\ell(\ell+2)}{\rho^2} -\frac{8 (\tom^2 -\tk^2)}{\rho^4}
\right] \Phi_\ell = 0. \labell{approx} \eeq

Changing variables first from $\rho$ to the `standard' radial
coordinate $r$ using \reef{rhor} (which becomes $r = \rho
\om/\sqrt{2}$ asymptotically) and then to $z=L^2/r$, we obtain
\beq \Phi_\ell''(z)-\frac{1}{z}\Phi_\ell'(z)-\left(\frac{\ell(\ell+2)}{z^2}
+k^2 \right) \Phi_\ell(z) =0\, , \eeq
where $k^2 = -\omega^2 + q^2$.  For timelike momenta, the solution
satisfying the incoming wave boundary condition at $z=\infty$ (the
horizon in the zero temperature limit) can be written in terms of
the Hankel function of the first kind,
\beq \Phi = z H_{\ell+1}^{(1)}(|k|z) \,, \labell{laugh3}\eeq
assuming $\omega>0$ \cite{Son:2002sd}. The spectral function then
becomes
\beq \rn_s = \frac{\nf \nc}{2\pi^2} \, \textrm{Im}\!\left[
\lim_{\epsilon \to 0} \frac{\Phi'(\epsilon)}{\epsilon^{2
\ell+1}\Phi(\epsilon)} \right]\, . \labell{laugh2}\eeq

For $\ell=0$, this is
\beq \rn_s = \frac{\nf\nc}{4\pi} \, (\omega^2 -q^2)
\theta(\omega^2 -q^2)\, \textrm{sgn} \,\omega \eeq
which for vanishing three-momentum $q=0$ reduces to
\beq \rn_s = \frac{\nf \nc \omega^2}{4 \pi} \labell{hah}\eeq
which coincides with the high frequency asymptotics which we found
for all three $\ell=0$ spectral functions in section
\ref{goldfinger}. Of course, this asymptotic behaviour precisely
matches that of the analytic solution \reef{lfal} with $\ell=0$
found for the vector modes in the massless quark limit.

For $\ell=1$ and vanishing three-momentum $q$, the spectral function
is
\beq \rn_s^{\ell=1} = \frac{\nf \nc \omega^4}{16\pi } \,. \eeq
Again, this matches the asymptotic behaviour of the analytic vector
solution \reef{lfal} for $\mq=0$ with $\ell=1$. In fact, given that
these asymptotics are independent of $\mq$ and that we have found a
common expression for all three channels, we can use eq.~\reef{lfal}
to write the general asymptotic behaviour with $q=0$ as
\beq \rn^\ell_s(\omega,q=0) = \frac{\nf
\nc}{2^{2\ell+2}\pi(\ell!)^2}
 \omega^{2\ell+2}\,.\labell{laugh}\eeq
Hence we have produced the expected asymptotic behaviour
\reef{hightail} for an operator of scaling dimension
$\Delta=\ell+3$. Of course, the behaviour for general $q$ is in
principle given by combining the expressions \reef{laugh3} and
\reef{laugh2}.

\section{Effective potentials and quasinormal modes}\label{schroe}

In this section, we rewrite the equations of motion for the
pseudoscalar and scalar fluctuations of the D7-brane in the form of
the Schroedinger equation. This can be considered a first step
towards calculating the spectrum of quasinormal modes for these
fields -- see, \eg \cite{hoyos}.  The effective
potential in each of these effective Schroedinger problems allows us
to infer certain aspects of the quasinormal spectra. In particular,
we argue that tachyonic modes appear in the scalar spectrum
sufficiently close to the critical solution. The same analysis for
the vector modes gives results that are essentially identical to
those found for the pseudoscalar.

\subsection{Pseudoscalar} \label{psoodo}

Considering fluctuations of the pseudoscalar, we take the general
ansatz
\beq\delta \phi \sim e^{ikx}\,{\cal P}(\rho)\, {\cal Y}_\ell (S^3)\,
,\labell{ansszz}\eeq
where ${\cal Y}_\ell (S^3)$ are spherical harmonics on the $S^3$ and
$k_\mu = (-\omega, q,0,0)$. Then the pseudoscalar's wave equation
\reef{phieom} can be written as
\beq -{H_0\over H_1}\,\partial_\rho\!\left[\vphantom{\tk^2}
H_1\,\partial_\rho\mathcal{P}\right]+\left[
\tk^2\,H_2+L^2\,H_3\right]\mathcal{P}=\tom^2\,\mathcal{P}\ ,
\labell{phieom4} \eeq
with:
\beq
\begin{array}{lll}
 H_0 \equiv {\rho^4f^2\over
8\tilde{f}}{(1-\chi^2)\over1-\chi^2+\rho^2 \dot{\chi}^2} \, , \quad
& H_1 \equiv {\rho^5 f \tilde{f} \chi^2(1-\chi^2)^2 \over
\sqrt{1-\chi^2+\rho^2\dot{\chi}^2}} \, , \quad &   \quad\\
& &  \\
H_2 \equiv {f^2\over\tilde{f}^2} \, , & H_3\equiv {\rho^2 f^2\over
8\tilde{f} (1-\chi^2)} \, , & L^2 \equiv \ell(\ell+2) \, .
\end{array} \labell{notaPhi}
\eeq
Then, with the substitution $\mathcal{P}=h \psi$, eq.~\reef{phieom4}
becomes
\beq
-H_0\,\ddot{\psi}-H_0\left(2{\dot{h}\over h}+{\dot{H}_1\over
H_1}\right)\,\dot{\psi} +\left[ \tk^2\,H_2+L^2\,H_3-H_0
\left({\ddot{h}\over h}+{\dot{H}_1\over H_1}\,{\dot{h}\over
h}\right)\right]\,\psi=\tom^2\,\psi\  \labell{phieom5}
\eeq
where, as usual, the dot denotes a derivative with respect to
$\rho$. The first term above can be rewritten as
\beq
-H_0\,\ddot{\psi}=-\sqrt{H_0}\partial_\rho\left(\sqrt{H_0}\partial_\rho
\psi\right)+{1\over2}\dot{H}_0\dot{\psi}=-\partial^2_{R^*}\psi+{1\over2}\dot{H}_0\,
\dot{\psi} \labell{kinetic}\eeq
where
\beq
R^*=\int_\rho^\infty{d\tilde{\rho}\over\sqrt{H_0(\tilde{\rho})}}\ .
\labell{radius} \eeq
In terms of this new radial coordinate, the second derivative term
takes the simple form found in a one-dimensional Schroedinger
equation. Combining eqs.~\reef{phieom5} and \reef{kinetic}, all of
the terms involving $\dot{\psi}$ are eliminated if we choose $h$ as
\beq h={H_0^{1/4}\over H_1^{1/2}}\ . \labell{hsol} \eeq
Hence the radial equation reduces to
\beq -\partial^2_{R^*}\psi+V_{eff}\,\psi=\cE\,\psi \labell{schroeEq}
\eeq
where the effective energy and potential are given by
\beq\cE=\tom^2\,,\qquad
 V_{eff} =
\tk^2\,H_2+L^2\,H_3-H_0\left[{\ddot{h}\over h}+{\dot{H}_1\over
H_1}\,{\dot{h}\over h}\right] \, . \labell{potSclr} \eeq

Let us comment on the new radial coordinate. In some sense, this
coordinate is like the `tortoise' radial coordinate introduced in
the analysis of physics in the Schwarzschild geometry \cite{wald}.
Approaching the event horizon, \ie as $\rho\rightarrow 1$,
$H_0\simeq (\rho-1)^2$ and so
$R^*\propto-\log(\rho-1)\rightarrow+\infty$. For large $\rho$,
$H_0\simeq\rho^4$ and so $R^*\simeq 1/\rho\rightarrow0$. Note that
given the definition in eq.~\reef{notaPhi}, $H_0$ is positive
everywhere on the range $\rho\in(1,\infty)$. Hence we are assured
that $R^*$ is a monotonic function of $\rho$.

Although $R^*$ is the appropriate coordinate to analyse the
effective Schroedinger equation \reef{schroeEq}, we can still gain
some intuition  for the problem by plotting the effective potential
\reef{potSclr} as a function of $\rho$ for various different
D7-brane embeddings. A summary of results (with $q,\ell=0$) is given
in fig.~\ref{summPlotPhi}. Note that for all embeddings, the
effective potential exhibits a large barrier in the asymptotic
region, as expected for an asymptotically AdS geometry. Further, in
all cases, the effective potential vanishes at the horizon. For
small $\chi_0$, the potential is monotonically increasing with
$\rho$. For larger $\chi_0$, a small potential barrier develops at
intermediate values of the radius. Intuitively, the latter might
give rise to metastable states in the effective Schroedinger
problem.
\FIGURE{
\begin{tabular}{cc}
\includegraphics[width=0.5 \textwidth]{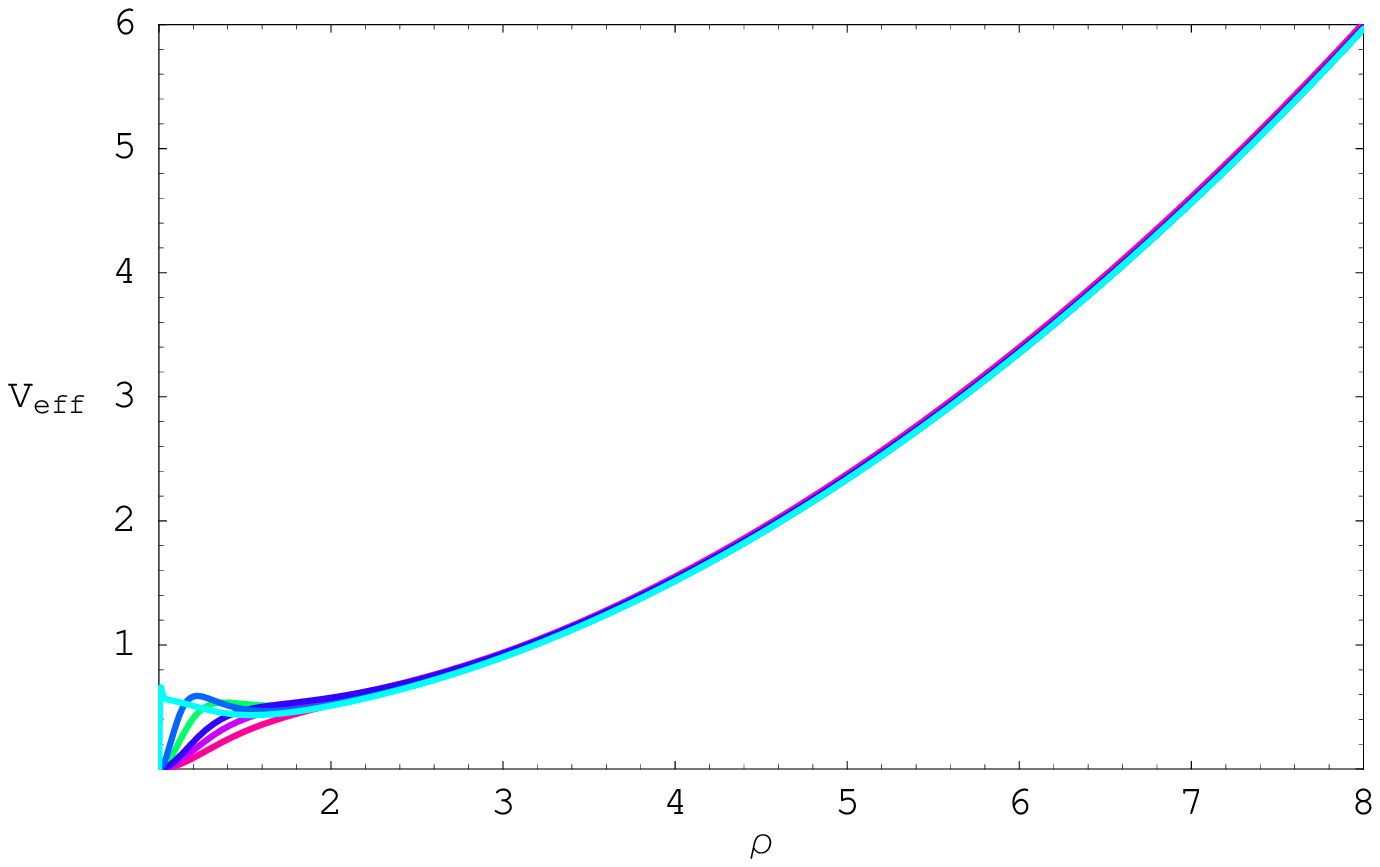}
&\includegraphics[width=0.5 \textwidth]{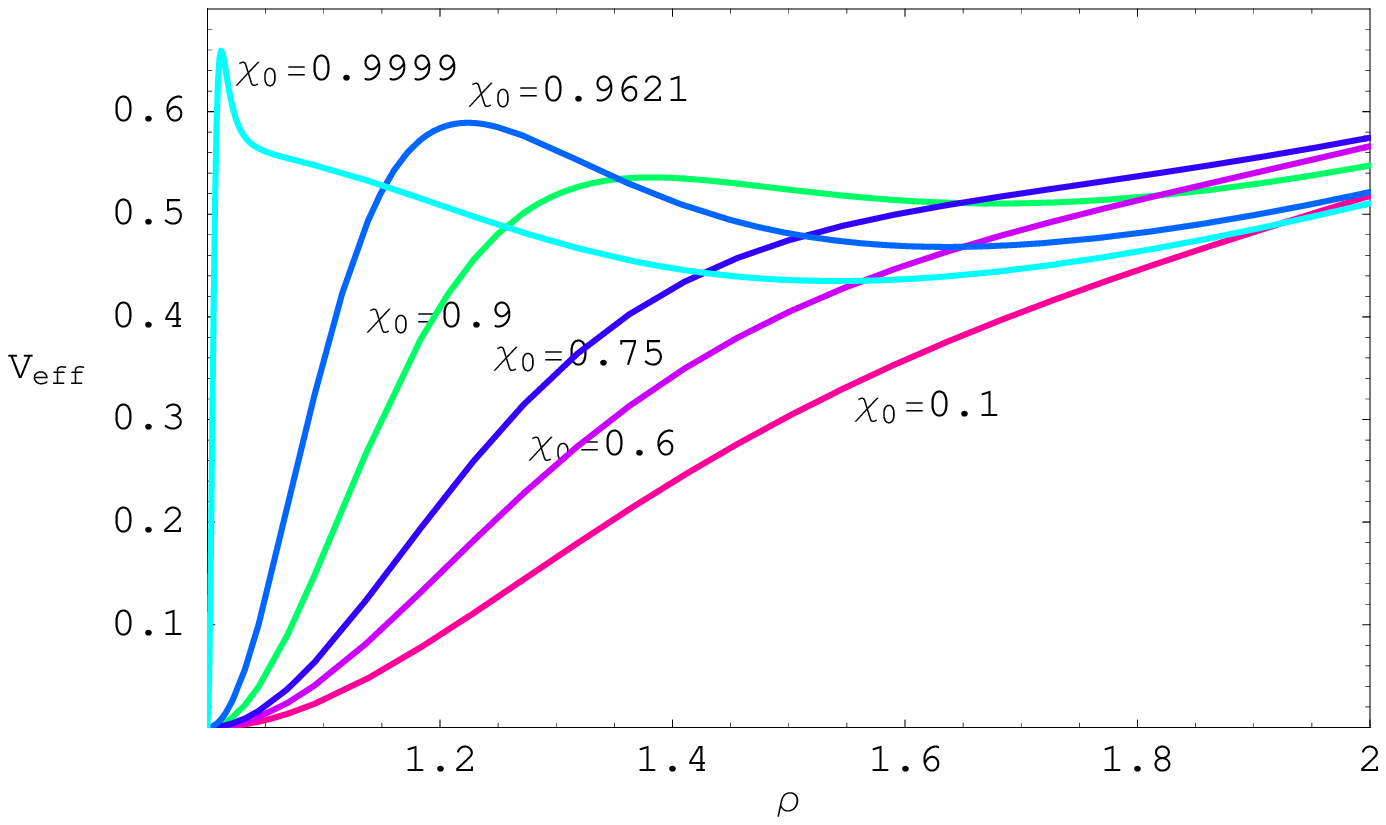}
 \end{tabular}
\caption{The effective potential (for $\delta \phi$ fluctuations)
versus $\rho$ for various $\chi_0$ with $\ell=0$ and $q=0$.
}\label{summPlotPhi}}

For $\ell>0$ nonzero and $q\not =0$ the results are qualitatively
similar to those depicted in fig.~\ref{summPlotPhi}.  The effective
potential vanishes at the horizon and grows as $\rho \to \infty$.
The potential barrier grows most quickly for $\ell \not= 0$, due to
the term proportional to $L^2$ in \reef{potSclr} which is roughly
$\ell (\ell+2) \rho^2 /8$ for large $\rho$.   For values of $\chi_0$
near 1, a small potential barrier develops for intermediate values
of the radius.  Note that while a small potential barrier is already
evident for $\chi_0=0.9$ (with $\ell=0$ and $q=0$) in
fig.~\ref{summPlotPhi}, the potential barrier only develops for
larger values of $\chi_0$, \eg $\chi_0 >0.99$ for $\ell > 0$ and/or
$q\not=0$.

The quasinormal modes of the pseudoscalar can be found by solving
the Schroedinger problem constructed above, with the appropriate
boundary conditions. One of the boundary
conditions is that the pseudoscalar should have only an incoming
wave component at the horizon, \ie $\rho\rightarrow1$ or
$R^*\rightarrow\infty$. Given the ansatz \reef{ansszz}, we are thus
looking for solutions with\footnote{Note that taking the limit
$\rho\rightarrow1$ in eq.~\reef{hsol} yields a simple constant for
$h$ and so we have ${\cal P}\propto\psi$ as we approach the
horizon.} $\psi\propto \exp(i\tom \,R^*)$. For large $\rho$ or small
$R^*$ where the effective potential diverges, we demand that the
wavefunction vanish. Generically, these boundary conditions lead
to complex eigenvalues for the effective energy $\cE$, which is in
accord with our expectation that the quasinormal frequencies have
the form \cite{garyver,revQNM}
\beq \tom = \pm \Omega - i\,\Gamma \, ,\label{tom9}\eeq
with $\Omega,\Gamma>0$. Note that $\Gamma>0$ ensures that the
quasinormal excitations decay in time, as can be seen from the
ansatz \reef{ansszz}.   However, given that
\beq \cE=\tom^2=(\Omega^2-\Gamma^2)\mp 2i\,\Omega\,\Gamma\,,
\label{tom8}\eeq
some translation is required to use our intuition for the
Schroedinger problem to infer general characteristics of the
quasinormal spectrum. Note that as the sign of $Im(\cE)$ is not
fixed, we are implicitly admitting energy eigenvalues which would
correspond to both decaying and growing wavefunctions in the
effective Schroedinger problem. However, this Schroedinger problem
is purely an auxiliary tool and so one should not ascribe any
physical significance to this observation.

Above, we observed that for small $\chi_0$, the effective potential
rises monotonically from zero as we move away from the horizon
towards larger $\rho$. Hence we would infer that $Re(\cE)>0$ or
$\Omega>\Gamma$. Further we should not expect any suppression of
$Im(\cE)$, \ie $Im(\cE)\sim Re(\cE)$, which means that we should
still expect to be $\Omega$ and $\Gamma$ to be the same order of
magnitude in this regime. This intuition would then suggest the
absence of any interesting structure in the corresponding spectral
functions, as the quasinormal frequencies should be far from the
real axis. However, as noted above, a small potential barrier
appears at intermediate values of $R^*$ (or $rho$) as $\chi_0$
approaches one. Intuitively, then one might expect to find
long-lived states with $Re(\cE)\gg Im(\cE)$ and $Re(\cE)\sim
V_{eff}$(well), \ie $Re(\cE)$ would be roughly given by the height
of the potential in this intermediate potential well. From
eq.~\reef{tom8}, this requires $\Omega\gg\Gamma$ with $\Omega$
finite and so would correspond to quasinormal frequencies
approaching the real axis. Hence we would expect the formation of
peaks in the spectral function in this regime, as discussed in
section \ref{prelude}. Of course, this intuitive picture developed
from the effective potential matches the behaviour of the spectral
functions found in section \ref{pseudo}. We emphasize, however, that
this intuition only gives a very rough picture of the quasinormal
spectrum and it would be interesting to develop more detailed
picture with a full calculation.

As mentioned above, the results for the effective potential for the
vector fluctuations are essentially the same as for the pseudoscalar
and hence the quasinormal spectrum should also be similar. We saw in section \ref{vectors} that the behaviour of the
vector spectral functions is very similar to those for the
pseudoscalar.

\subsection{Scalar} \label{skalars}

With the ansatz $\delta \theta \sim e^{ikx}\,{\cal R}(\rho)\, {\cal
Y}_\ell (S^3)$), the scalar wave equation \reef{thetaeom} can be
written as
\beq -{H_0\over H_1}\,\partial_\rho\!\left[\vphantom{\tilde{\bf
k}^2} H_1\,\partial_\rho {\cal R} \right]+\left[ \tk^2\,
H_2+(\ell+3)(\ell-1)\,H_3\right]{\cal R}=\tom^2\,{\cal R}\ .
\labell{thetaeom4} \eeq
where $H_0$, $H_2$ and $H_3$ are the same as defined in
eq.~\reef{notaPhi}, while we must redefine the following:
\beq
 H_1 \equiv {\rho^5 f \tilde{f}
(1-\chi^2)^3 \over (1-\chi^2+\rho^2 \dot{\chi}^2)^{3/2}} \, ,  L^2
\equiv (\ell+3)(\ell-1) \, . \labell{nota} \eeq

Since eq.~\reef{thetaeom4} has the same form as eq.~\reef{phieom4},
we can use precisely the same steps as above to cast this equation
for the scalar fluctuations into the form of a Schroedinger
equation. That is, taking ${\cal R}=h\,\psi$ with $h =
H_0^{1/4}/H_1^{1/2}$ and defining the radial coordinate
\reef{radius}, eq.~\reef{thetaeom4} takes the form of the
Schroedinger equation \reef{schroeEq} with the effective energy and
potential given by \reef{potSclr}. Examining the effective potential
to gain some intuition for the physics, we find: Again for all
embeddings, there is a large potential barrier in the asymptotic
region and the effective potential vanishes at the horizon. For
embeddings of the D7-branes with $ 0 \leq \chi_0 < 0.7$,
fig.~\ref{summPlot} shows that the effective potential is a
monotonically increasing function of $\rho$.  For $\ell=0$ and
$q=0$, once $\chi_0 \gtrsim 0.7$, the potential develops a negative
well near the horizon. As $\chi_0$ increases towards 1, this well
near $\rho \simeq 1$ becomes deeper and wider. Note that $\chi_0
\simeq 0.9621$ and $\chi_0 \simeq 0.99973885$ correspond to the
first and second kinks, respectively, in a plot of the free energy
versus temperature -- see fig.~\ref{action}. For any modes with
$\ell>0$, one finds that there is never a region where the
effective potential becomes negative, however, for $\chi_0 $ near 1
the potential develops a small barrier near the horizon.  For
nonvanishing spatial momentum ($q \not= 0$), the effective potential
exhibits a negative well near the horizon for values of $\chi_0$
near 1. However, the well is neither as deep nor as wide as that for
$q=0$.

\FIGURE{
\begin{tabular}{cc}
\includegraphics[width=0.5 \textwidth]{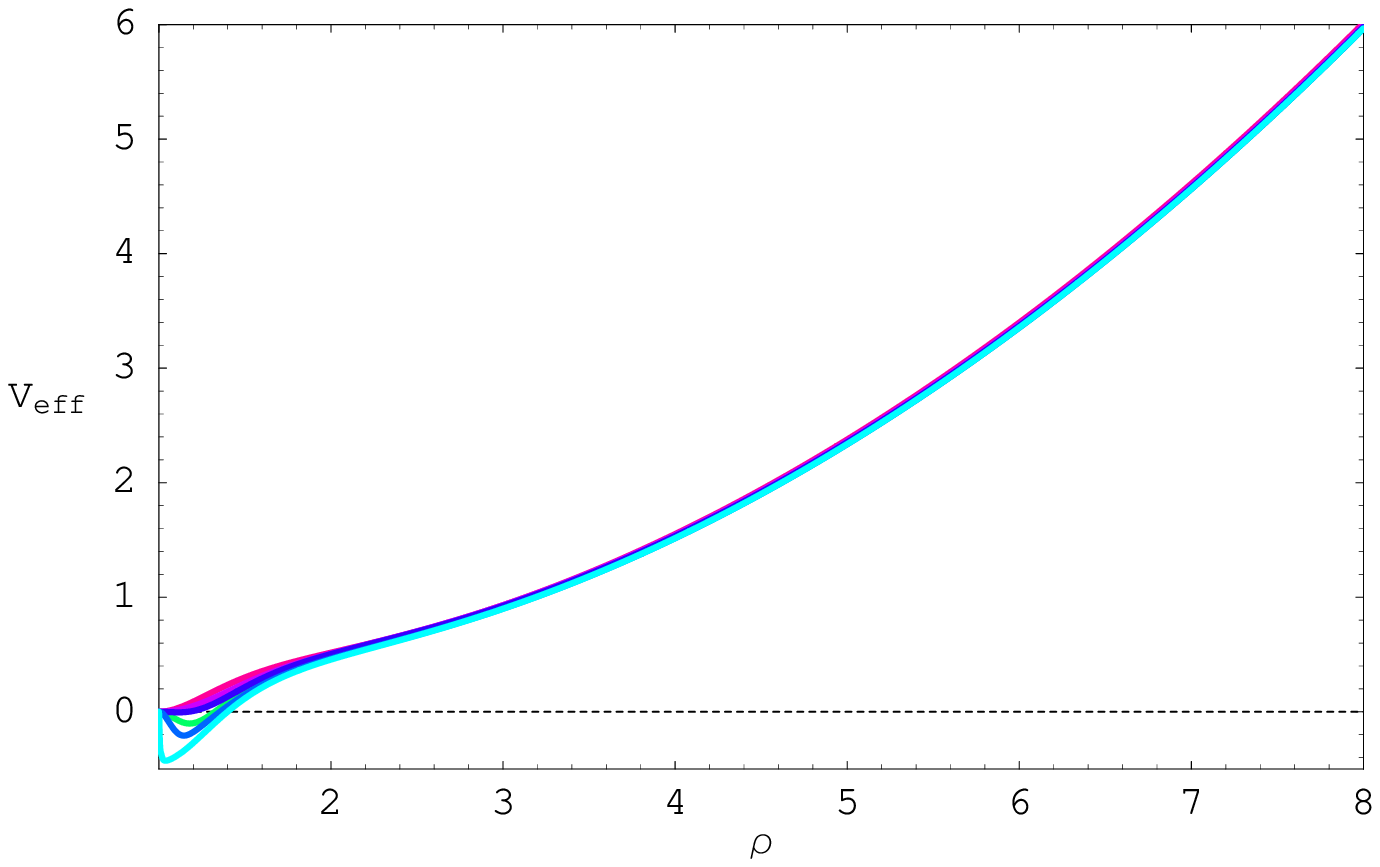}
&\includegraphics[width=0.5 \textwidth]{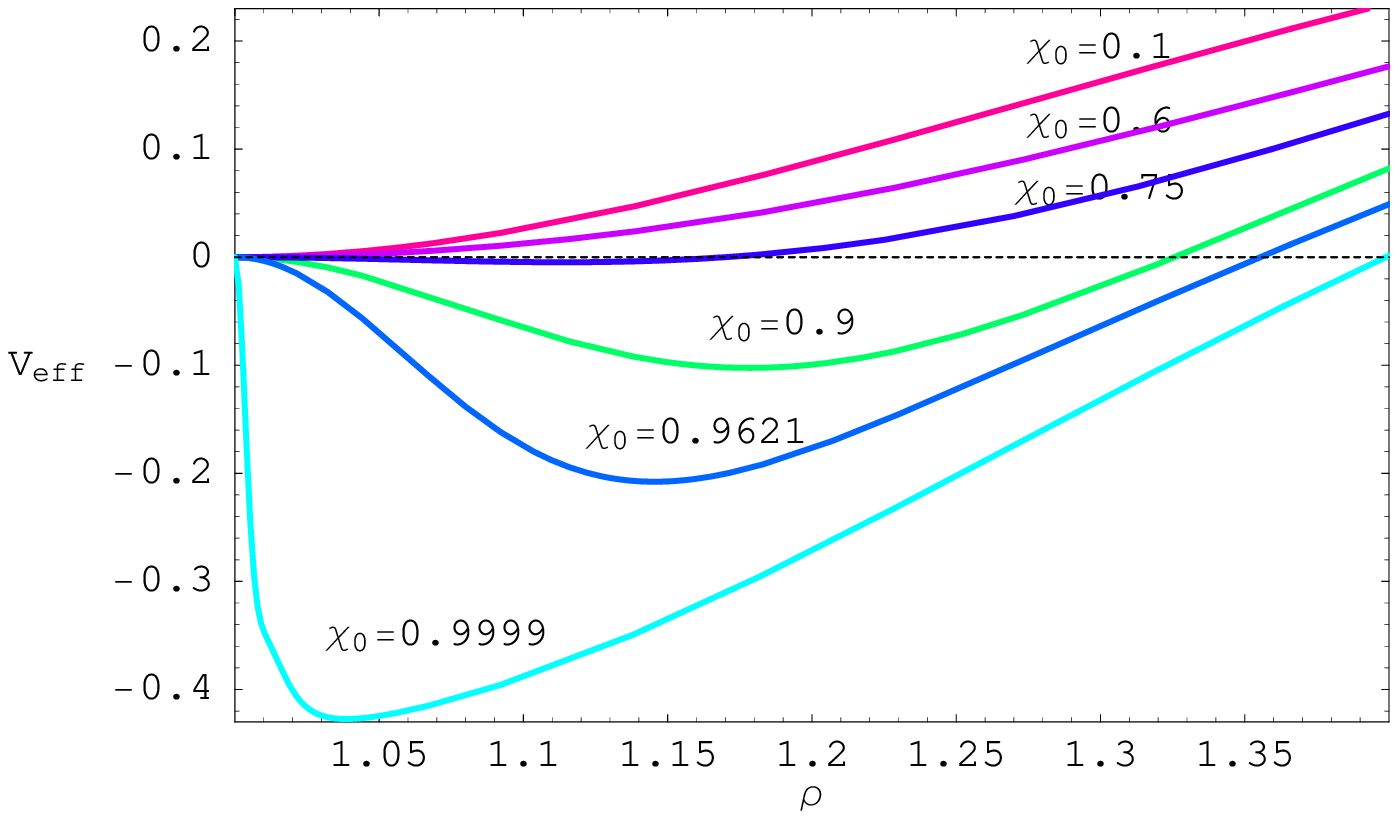}
\end{tabular}
\caption{The effective potential for the scalar field ($\delta
\theta$)  versus $\rho$ for various $\chi_0$ with $\ell=0$ and
$k=0$.}\label{summPlot}}

Certainly, the most interesting feature of the effective potential
for the scalar is the negative potential well which develops and
grows as $\chi_0 \to 1$. One would expect that if this negative well
grows large enough, it will be able to support a `bound' state with
$\cE <0$. Actually, since such a state would still see a finite
potential barrier between the center of the well and the horizon, it
would still be a long-lived state. Using a WKB approximation, we can
estimate that a (zero-energy) bound state will appear for \cite{wkbStandard,wkb}
\beqa
\left(n-\frac{1}{2}\right)\pi & =&
\int_{R_0}^{\infty} dR^\ast \sqrt{-V_{eff}(R^\ast)} \\
&= & \int_1^{\rho_0} \frac{d\rho}{\sqrt{H_0}} \sqrt{-V_{eff}(\rho)}
\equiv I \eeqa
where $n$ is a positive integer and the integration is over the
values of $\rho$ for which the potential is negative.  A plot of
$I/\pi+1/2$ is given in figure \ref{wkbFig} (for $\ell, q=0$).  This
quantity reaches 1 for $\chi_0 \simeq 0.98297$ and 2 for $\chi_0
\simeq 0.99986$, and so we expect that the first two bound states
form at roughly these values of $\chi_0$. Below we will argue that
the appearance of these bound states can be associated with the
dramatic spikes that were observed in the scalar spectral functions
in section \ref{scalar}. Not then that our results here do not
(quite) match the values of $\chi_0$ corresponding to the first two
kinks (on the black hole branch) of the free energy -- recall that
the latter correspond to $\chi_0=.9621$ and $.99973885$,
respectively. However, we expect this discrepancy is likely due to
the approximations inherent in the WKB calculation.
\FIGURE{
\includegraphics[width=0.6 \textwidth]{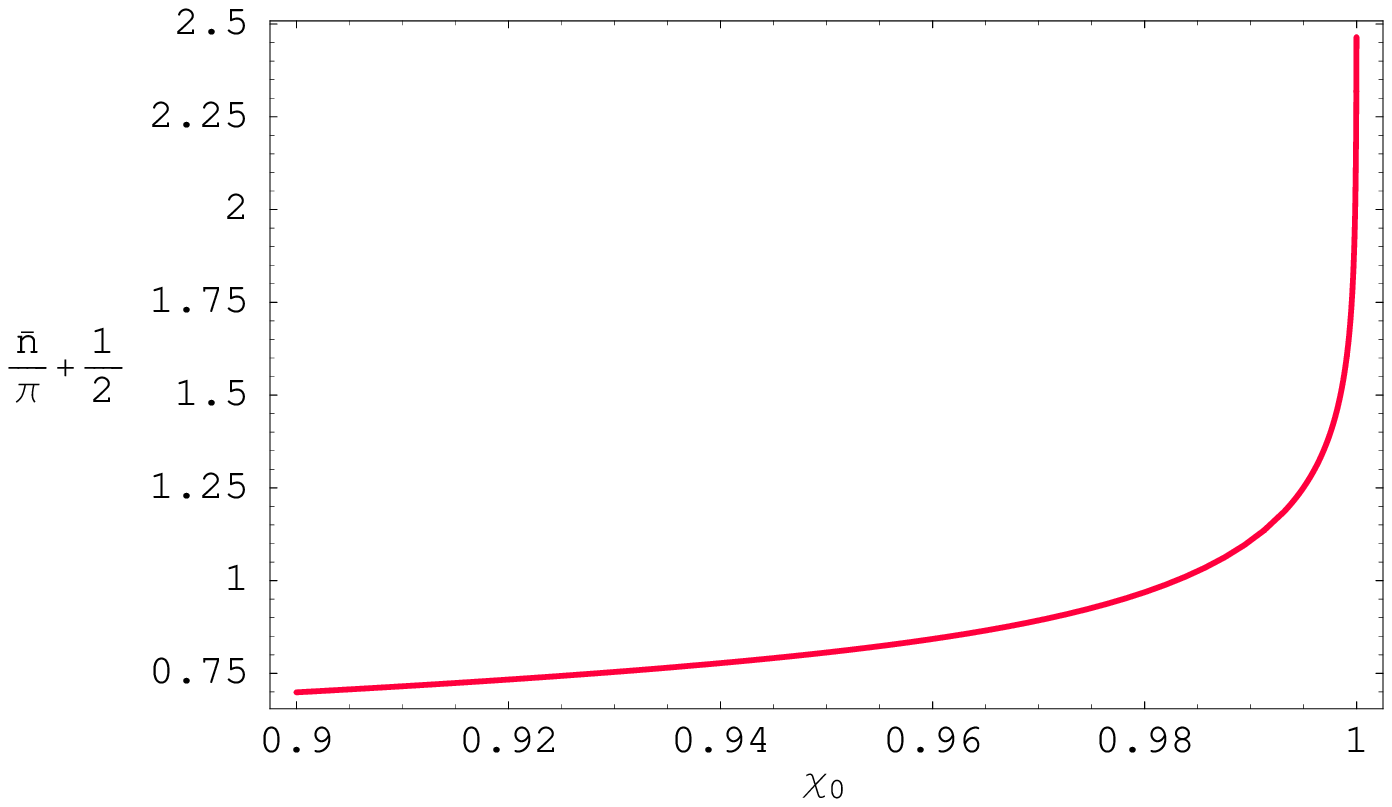}
\caption{Plot of $I/\pi+1/2$ versus $\chi_0$ for $\ell,q=0$. } \label{wkbFig} }

As the effective potential is positive and monotonically increasing
in the small $\chi_0$ regime, we expect the eigenfrequencies
\reef{tom9} in quasinormal spectrum will again have $\Omega$ and
$\Gamma$ with the same order of magnitude. Of course, the regime
with large $\chi_0$ is more interesting because of the appearance of
bound states. These modes with $Re(\cE)<0$ are distinguished since
$\Gamma^2>\Omega^2$, as seen in eq.~\reef{tom8}. Further as we
alluded to above, the corresponding wavefunctions are below a
potential barrier as $R^* \rightarrow\infty$ and so must have the
form $\psi\sim \exp(-|\Gamma|\,R^*)$ to avoid a divergence at the
horizon. Given the boundary condition there, \ie $\psi\propto
\exp(i\tom \,R^*)$, this requires that $\Gamma<0$ for these modes.
Further then, it follows that these exceptional modes grow rather
than decay in time and so these bound states really represent an
instability of the system.

The above discussion indicates that these bound states appear when a
quasinormal frequency crosses the real axis and so their appearance
should be signalled by a pole appearing in the scalar spectral
function calculated in section \ref{scalar}. Further, however, we
argued that as the eigenfrequency crosses the real axis, it moves
from a regime where $\Omega^2>\Gamma^2$ for $\Gamma>0$ to
$\Omega^2<\Gamma^2$ for $\Gamma<0$. Hence at the point that
$\Gamma=0$, we must also have $\Omega=0$.\footnote{Essentially this
we are just saying that $Re(\cE)=0$ just as the bound states form.}
Hence we see that the quasinormal frequencies must be cross the real
axis by passing through the origin. This is, of course, precisely
what was observed in section \ref{scalar}, where the poles in the
spectral function appeared at precisely $\omega=0$. Further, the
lack of much structure in the spectral function aside from these
poles would indicate that the quasinormal frequencies approach the
origin uniformly so that we never find eigenfreqencies with
$\Omega\gg\Gamma$. We reiterate that this discussion only gives a
schematic picture of the quasinormal spectrum and it would be
interesting to develop a more detailed picture with a full calculation
\cite{hoyos}.

\section{Diffusion constants for Dp/Dq systems} \label{app-diffuse}

This appendix extends the computation of the diffusion constant
using the membrane paradigm \cite{Kovtun:2003wp} described in
section \ref{membrane} to that for the gauge theory dual to the
supergravity configuration of a Dq-brane probe in the near-horizon
black Dp-brane geometry.

The background geometry \reef{D3geom-r} is generalised to the near-horizon
black Dp-brane metric (in the string-frame) \cite{itz}:
\beq
 ds^2 = H^{-\frac{1}{2}} \left( -f
dt^2 +  d{\bf x}_{\it p}^2 \right) + H^{\frac{1}{2}} \left( \frac{dr^2}{f}
+ r^2 d\Omega_{\it 8-p}^2 \right) \,, \labell{metric} \eeq
where now $H(r) = (L/r)^{7-p}$ and $f(r) =1-(r_0/r)^{7-p}$. The
background also includes non-trivial dilaton and RR fields: $e^\phi
= H^{(3-p)/4}$,  $C_{01\ldots p} = H^{-1}$. The Hawking temperature
associated with the horizon at $r=r_0$ is given by
\beq
 T = \frac{7-p}{4\pi  L} \left( \frac{r_0}{L}
\right)^{\frac{5-p}{2}} \,. \labell{beta} \eeq
According to the gauge/gravity correspondence, string theory on this
background is dual to a supersymmetric $(p+1)$-dimensional gauge
theory at temperature $T$.

Consider placing a probe Dq-brane in the above geometry such that
the probe has $d$ spatial directions parallel and $n+1$ transverse
to the background Dp-branes, so that $q=d+n+2$ and such that it
intersects the horizon at $r=r_0$.  In analogy to eq.~\reef{rhor},
it is useful to introduce a new (dimensionless) radial coordinate
$\rho$ related to $r$ via
\beq \left( r_0 \rho \right)^{(7-p)/{2}} = r^{{(7-p)}/{2}} +
\sqrt{ r^{7-p} - r_0^{7-p}} \,. \labell{rho} \eeq
The horizon is now positioned at $\rho=1$. Implicitly, we will
assume in the following that the Dp/Dq system under consideration is
T-dual to the D3/D7 one described by the array \reef{array}. This
choice ensures that the brane configuration is supersymmetric at
zero temperature and the probe brane embeddings should be stable in
the finite temperature background \reef{metric} \cite{long}.

With the coordinate \reef{rho}, the metric and the dilaton may be
written as:
\beqar ds^2 &=& \frac{1}{2} \left(\frac{r_0 \rho}{L}
\right)^{(7-p)/2} \left[-\frac{f^2}{\tilde{f}}dt^2+\tilde{f} dx_p^2
\right] + \tilde{h}(\rho)\,\left[d\rho^2 + \rho^2 \left(d\theta^2 + \sin^2
\theta d\Omega_n +\cos^2 \theta d\Omega_{7-p-n}
\right) \right] \nonumber\\
e^{\phi}& =& \left(\frac{\tilde{f}}{2}\right)^{(p-3)/2} \left( \frac{r_0 \rho}{L}
\right)^{(7-p)(p-3)/4} ,  \eeqar
where
\[ f(\rho) = 1- {1}/{\rho^{7-p}}, \quad \tilde{f} (\rho) = {1}/{\rho^{7-p}},
\quad \tilde{h}(\rho)= r_0^2\left({L}/{r_0 \rho} \right)^{(7-p)/2}
\left({\tilde{f}}/{2}\right)^{(p-3)/(7-p)}.\]

Describing the probe brane profile using $\chi (\rho) = \cos
\theta(\rho)$, the induced metric on the Dq-brane may be written as
$ ds^2 (g) = ds^2 (\tg) + Z(\rho) d\Omega_n^2 $,
where
\beqar ds^2 (\tg) &=& \frac{1}{2} \left(\frac{r_0 \rho}{L}
\right)^{(7-p)/2} \left[-\frac{f^2}{\tilde{f}}dt^2+\tilde{f} dx_d^2
\right] +\tilde{h}(\rho)\,\frac{1-\chi^2+\rho^2
\dot{\chi}^2}{1-\chi^2}d\rho^2\, , \\
Z(\rho) &=& \tilde{h}(\rho)\,\rho^2(1-\chi^2)\, .
\eeqar
Using the DBI action and expanding the gauge fields to quadratic
order, the relevant portion of the action for the gauge fields is
\beq I_{q,F} = -T_{q} (\pi \ls^2)^2 \Omega_n \int dt\, d^d x\, d
\rho\, \frac{\sqrt{-\tg}}{g_{eff}^2} F^2  , \quad  g_{eff}^2 = e^{\phi} Z^{-n/2}\, . \eeq

We are now in a position to evaluate the diffusion constant using
eq.~(2.27) from \cite{Kovtun:2003wp}:
\beqa D &=& \left.{\sqrt{-\tilde{g}} \over \tilde{g}_{xx}\, g_{eff}^2 \,\sqrt{-\tilde{g}_{tt} \,
\tilde{g}_{\rho \rho}}}\right|_{\rho=1} \int _1 ^\infty d\rho {-\tilde{g}_{tt} \,
 \tilde{g}_{\rho \rho} \, g_{eff}^2 \over \sqrt{-\tilde{g}} } \labell{genMem} \\
&=& \frac{(7-p)}{2 \pi T} 2^{\alpha} (1-\chi_0^2)^{n/2} \int_1^\infty
d\rho \frac{f \, \rho^{\beta}}{\tilde{f}^{\gamma}}\frac{\sqrt{1-\chi^2+\rho^2
\dot{\chi}^2}}{(1-\chi^2)^{(n+1)/2}}  \labell{ddudd}
\eeqa
where
\[\alpha=\frac{d-p}{2}+\frac{(n-1)(p-3)}{2(7-p)}\, ,\quad \beta=
\frac{(7-p)(p+n-3-d)}{4}-n\, , \quad
\gamma=\frac{(n-1)(p-3)}{2(7-p)}+\frac{d+4-p}{2}.\] One may check
that for $p=3=n=d$ this result reduces to that for the D3/D7 case
given in eq.~\reef{membraneD}.

We have also evaluated $DT$ numerically for the D4/D6 case ($p=4,\
n=2,\ d=3$) and the results are plotted in figure \ref{d4d6a}. The
horizontal axis is labelled by the ratio of the temperature to the
natural mass scale of the problem:
\beq \bar{M} = {3\over 4 \pi L} \left({2 \pi \ls^2 M_q  \over L}
\right)^{1/2} \simeq {M_q \over g_{eff}(M_q)} \, . \eeq
Asymptotically, $DT$ approaches $3/4\pi$ for large temperatures. As
the temperature is reduced, $DT$ decreases dramatically near the
phase transition. The value at the phase transition is
$DT=.125\simeq .785/2\pi$.
If we continue following the black hole
embeddings beyond the phase transition, $DT$ continues to fall and
it also becomes a multi-valued function of temperature, as was seen
in fig.~\ref{d3d7b} for the D3/D7 system. Again, this simply
reflects the fact that multiple embeddings can be found for a single
temperature in the vicinity of the critical solution.
\FIGURE{
  \includegraphics[width = 0.7 \textwidth]{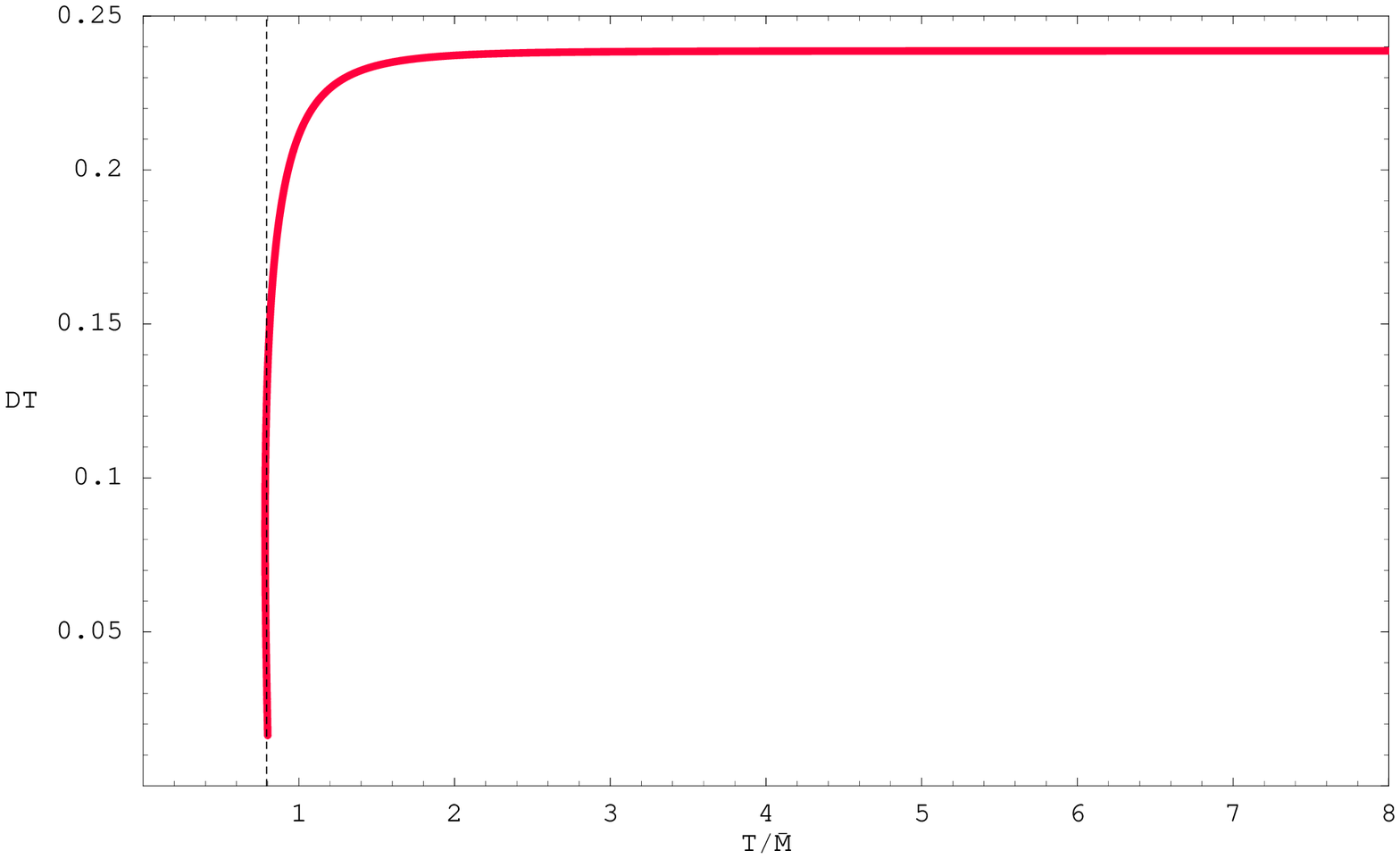}
      \caption{The diffusion constant $D$ times the temperature
      $T$ versus temperature $T/\bar{M}$ for a D6-brane probe in the black
      D4-brane geometry. For large $T$ we have $DT\simeq 3/4\pi$.}  \label{d4d6a}
}

Note that the asymptotic value for $DT$ does not match that for the $R$-charge
diffusion constant calculated for a near-extremal D4-brane result:
$3/8\pi$ \cite{Kovtun:2003wp}. However, there is no
reason that these two quantities should be equal since the D6-brane
does not fill the entire D4-brane throat, \ie one of the D4-brane
worldvolume directions is transverse to the D6 probe.

The D4/D6-brane system considered above is the basis for the
construction of one holographic model which mimics QCD at large
$\nc$ \cite{QCDN}. Another interesting holographic model of a
QCD-like theory comes from introducing D8 and anti-D8 probe branes
in a D4-brane background \cite{sugimoto}. This system displays an
interesting phase transition related to chiral symmetry breaking
\cite{ss-temp}.\footnote{These models generalise to a broad family
of models displaying a similar pattern of chiral symmetry breaking
\cite{ss-general}.} One can again calculate the diffusion constant
for the quark charge in the high temperature phase along the lines
described above. In this case, the D8-brane wraps the entire $S^4$
of the D4 background but otherwise fills the same directions as the
D6-branes above. After the phase transition the embeddings are much
simpler since there is no non-trivial radial profile to be
considered and, further, the embedding is temperature independent.
The diffusion constant may be determined from eq.~\reef{ddudd}
setting $\chi=0=\chi_0$, $p=4=n$ and $d=3$. The result for the
calculation is $DT=1/2\pi$.

\end{document}